\documentclass{cimento}

\usepackage[utf8]{inputenc}
\usepackage{amsmath,amssymb,amsfonts}

\usepackage{graphicx}

\usepackage{dcolumn}
\usepackage{bm}
\usepackage{color}
\usepackage{bbold}
\usepackage{url}

\usepackage{slashed}
\usepackage[svgnames]{xcolor}
\usepackage[english]{babel}
\usepackage{blindtext}
\usepackage{microtype}
\usepackage{tikz}

\usepackage{soul}
\usepackage{tabularx}
\usepackage{bigstrut}
\usepackage{slashbox}
\newcommand{\phiH}{\phi _h}
\newcommand{\phiS}{\phi _S}

\newcommand{\xzpt}{(x,z,\Phperp)}
\newcommand{\xkts}{(x,\kperp^2)}
\newcommand{\zpts}{(z,\pperp^2)}

\newcommand{\bfhp}{\hat{\bm h}}

\newcommand{\Phperp}{P_{T}}
\newcommand{\FUU}{F_{UU,T}}

\newcommand{\FLL}{F_{LL}}

\newcommand{\kt}{{\bf k}_T}
\newcommand{\pt}{{\bf p}_T}
\newcommand{\zh}{z}
\newcommand{\xbj}{x} 
\newcommand{\ph}{\phi}
\newcommand{\la}{\langle}
\newcommand{\ra}{\rangle}
\newcommand{\bq}{\begin{eqnarray}}
\newcommand{\eq}{\end{eqnarray}}
\newcommand{\be}{\begin{equation}}
\newcommand{\ee}{\end{equation}}

\def\la{\langle}
\def\ra{\rangle}
\def\ba{\begin{eqnarray}}
\def\ea{\end{eqnarray}}
\def\PT{{\bf P}_{hT}}

\def\pperp{\Phperp}
\def\kperp{k_T}
\def\pdperp{p^2_T}
\def\kdperp{k^2_T}
\def\Phperp{P_{hT}}

\def\phperp{P_{hT}}
\def\apperp{\langle \pperp^2 \rangle}
\def\akperp{\langle \kperp^2 \rangle}

\DOI{10.1393/ncc/i2019-10155-3}

\title{Spin Orbit Correlations and the Structure of the Nucleon}
\author{H.~Avakian~\from{ins:jlab}~\thanks{e-mail:avakian@jlab.org},
B.~Parsamyan~\from{ins:bp}~\thanks{e-mail:bakur.parsamyan@cern.ch},
A.~Prokudin~\from{ins:psu}~\from{ins:jlab}~\thanks{e-mail:prokudin@jlab.org}}

\instlist{\inst{ins:jlab} Thomas Jefferson National Accelerator Facility, Newport News, Virginia 23606
\inst{ins:bp} CERN, CH-1211 Geneva 23, Switzerland
\inst{ins:psu} Penn State Berks, Reading, Pennsylvania 19610, USA}

\begin{document}

\maketitle

\begin{abstract}
Extensive experimental measurements of spin and azimuthal asymmetries in various processes have stimulated theoretical interest and progress in studies of the nucleon structure. Interpretation of experimental data in terms of parton distribution functions, generalized to describe transverse momentum and spatial parton distributions, is one of the main remaining challenges of modern nuclear physics.
These new parton distribution and fragmentation functions encode the motion and the position of partons and are often referred to as three-dimensional distributions describing the three-dimensional (3D) structure of the nucleon. Understanding of the production mechanism and performing phenomenological studies compatible with factorization theorems using minimal model assumptions are goals of analysis of the experimental data.
HERMES and COMPASS Collaborations and experiments at Jefferson Lab have collected a wealth of polarized and unpolarized Semi-Inclusive Deep Inelastic Scattering (SIDIS) data. These data play a crucial role in current understanding of nucleon spin-phenomena as they cover a broad kinematical range.
The Jefferson Lab 12 GeV upgrade data on polarized and unpolarized SIDIS
will have remarkably higher precision at large parton fractional momentum $x$ compared to the existing data. We argue that both experimental and phenomenological communities will benefit from development of a comprehensive extraction framework that will facilitate extraction of 3D nucleon structure, help understand various assumptions in extraction and data analysis, help to insure the model independence of the experimental data and validate the extracted functions.
In this review we present the latest developments
in the field of the spin asymmetries with emphasis on observables beyond the leading twist in SIDIS, indispensable for studies of the complex 3D nucleon structure, and discuss
different components involved in precision extraction of 3D partonic distribution and fragmentation functions.
\end{abstract}

\section{Introduction}
\label{Sec:introduction}

\begin{figure}[!htb]
\centering
\includegraphics[width=0.55\textwidth]{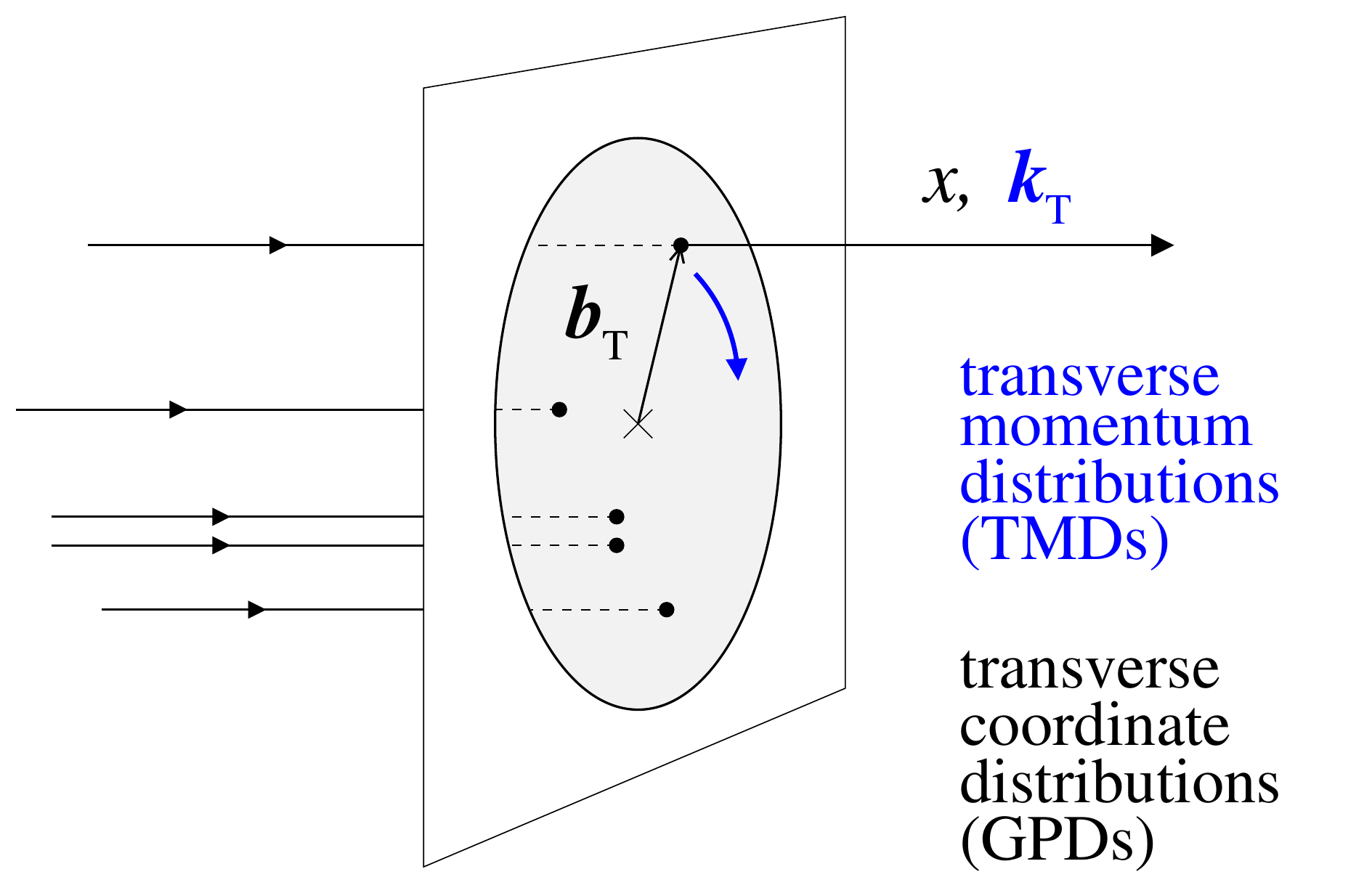}
\caption{
Three--dimensional structure of a fast--moving
nucleon. The distribution of partons (quarks, gluons) is
characterized by the longitudinal momentum fraction $x$ and the
transverse spatial coordinate $\bm{b}_T$
through the impact parameter GPDs~\cite{Burkardt:2000za}. In addition, the
partons are distributed over transverse momenta $\kt$,
reflecting their orbital motion and interactions in the system (TMDs)~\cite{Kotzinian:1994dv,Mulders:1995dh}.
Polarization distorts both the spatial and momentum distributions.
Figure from Refs.~\cite{Accardi:2011mz,Aschenauer:2014twa}.
\label{fig:1-3Dproton}
}
\end{figure}

The quantum chromodynamics (QCD) is an established theory of strong interactions between quarks and gluons, the fundamental building blocks of the proton and the neutron. The experimental studies in the last several decades of Deep Inelastic Scattering (DIS) and Drell-Yan (DY) processes were pivotal in our tests of QCD, in explaining the behavior of the strong coupling constant, and the collinear parton densities. The spin structure of the hadrons always poses additional complications, continuously challenging the theory, as quark-gluon interactions, and their correlations with the spin of partons and hadrons, are very significant and often are not easily understood in a simple picture of static partons in the nucleon. Relatively recent experimental explorations of the spin dependent observables in SIDIS, DY, $e^+e^-$, and hadron-hadron scattering spurred extensive theoretical and phenomenological studies opening a new era of exploration of the 3D structure of the nucleon. Together with rapid advances of precision of lattice QCD calculations and detailed predictions from theory and phenomenology, new precise experimental data are needed. In the process of moving from ``testing QCD'' to actually understanding it in its full complexity, SIDIS
has emerged as a powerful tool to probe the dynamics of strong interactions. Transition from simple, one dimensional description using collinear parton distributions that depend on nucleon's longitudinal momentum fraction, $x$, to more complex
nucleon picture with interacting and orbiting quarks, lead to a generalization of partonic distributions, to include also
the transverse parton momentum, $\kt$, and to introduction of Transverse Momentum Dependent (TMD) partonic distributions, see Fig. \ref{fig:1-3Dproton}. SIDIS provides access to TMD partonic distributions through measurements of spin and azimuthal asymmetries.
Studies of spin-azimuthal asymmetries in semi-inclusive and hard exclusive production of photons and hadrons have been widely recognized as key objectives of the Jefferson Lab 12 GeV~\cite{Prokudin:2013xx,Dudek:2012vr} upgrade and one of the driving forces for the future Electron Ion Collider~\cite{Accardi:2012qut,Aschenauer:2014twa,Abeyratne:2012ah,Accardi:2011mz,Anselmino:2011ay}.

%
%

\begin{figure}[htp]
\centering
\includegraphics[width=0.5\textwidth]{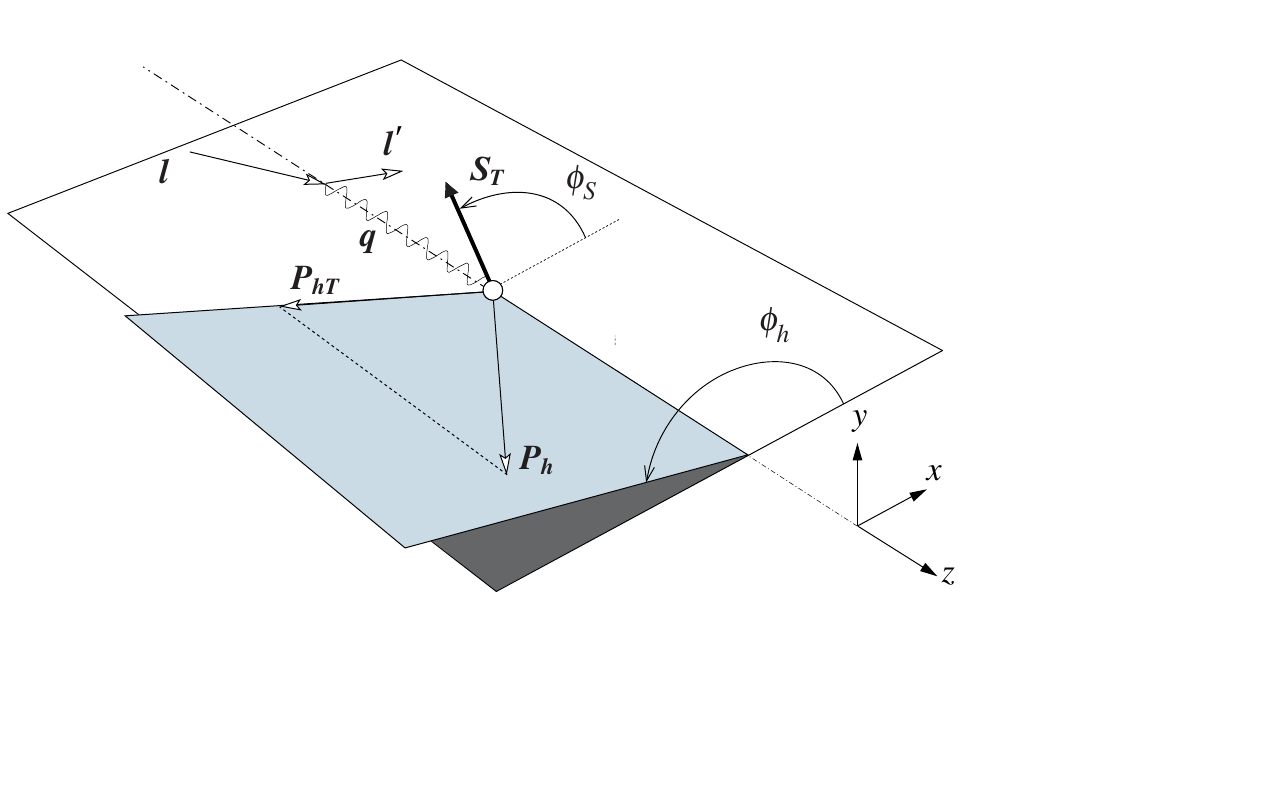}
\caption{(color online) SIDIS kinematical plane with definitions of transverse variables}
\label{fig:sidis-kin}
\end{figure}

SIDIS ($\ell (k) + N(P) \rightarrow \ell^{\prime} (k^{\prime} ) + h(P_{h}) + X(P_{X})$) reaction is such that a beam lepton $\ell$ with the 4-momenta $k$, scatters off of a target nucleon, $N$ with four momentum $P$, and the scattered lepton $\ell^{\prime}$ with four momentum $k^{\prime}$ is detected along with a single hadron, $h$ with four momentum $P_{h}$; all other produced particles in the final state, $X$, are not detected, see Fig.~\ref{fig:sidis-kin}. Assuming a single photon exchange, the SIDIS cross-section can be decomposed into a sum of various azimuthal modulations coupled to
corresponding structure functions. SIDIS cross section has following form ~\cite{Kotzinian:1994dv,Mulders:1995dh,Bacchetta:2006tn}:
\begin{eqnarray} \nonumber
\centering
\hspace*{0.cm}\frac{{d\sigma }}{{dxdydzd\Phperp^{2}d{\phiH}d{\phiS}}} =
\hat{\sigma}_{U} &&
\Bigg\{ 1 +
\varepsilon A_{UU}^{\cos 2{\phi _h}}\cos {2{\phiH}} + \sqrt {2\varepsilon \left( {1 + \varepsilon } \right)} A_{UU}^{\cos {\phi _h}}\cos {\phiH}
\\ \nonumber
&&\hspace*{-3.55cm} + \lambda_{\ell} \sqrt {2\varepsilon \left( {1 - \varepsilon } \right)} A_{LU}^{\sin {\phi _h}}\sin {\phiH} \\ \nonumber
&&\hspace*{-3.55cm}+\,{S_{||}}\Big[ \sqrt {2\varepsilon \left( {1 + \varepsilon } \right)}A_{UL}^{\sin {\phiH}}\sin {\phiH}
+\,\varepsilon A_{UL}^{\sin {2{\phiH}}}\sin {2{\phiH}}\Big]\\ \nonumber
&&\hspace*{-3.55cm}+\,{S_{||}}\lambda_{\ell} \Big[ \sqrt {1 - {\varepsilon ^2}} A_{LL}
+\,\sqrt {2\varepsilon \left( {1 - \varepsilon } \right)} A_{LL}^{\cos {\phiH}} \cos {\phiH}\Big]\\ \nonumber
&&\hspace*{-4.25cm}+\,{{S}_{\bot}}\Big[
A_{UT}^{\sin \left( {{\phiH} - {\phiS}} \right)}\sin \left( {{\phiH} - {\phiS}} \right)
+ \varepsilon A_{UT}^{\sin \left( {{\phiH} + {\phiS}} \right)}\sin \left( {{\phiH} + {\phiS}} \right)
+ \varepsilon A_{UT}^{\sin \left( {3{\phiH} - {\phiS}} \right)}\sin \left( {3{\phiH} - {\phiS}} \right)\\ \nonumber
&&\hspace*{-3.5cm}+\,\sqrt {2\varepsilon \left( {1 + \varepsilon } \right)} A_{UT}^{\sin {\phiS}}\sin {\phiS}
+ \sqrt {2\varepsilon \left( {1 + \varepsilon } \right)} A_{UT}^{\sin \left( {2{\phiH} - {\phiS}} \right)}\sin \left( {2{\phiH} - {\phiS}} \right)\Big]\\ \nonumber
&&\hspace*{-4.25cm}+\,{{S}_{\bot}}\lambda_{\ell} \Big[\sqrt {\left( {1 - {\varepsilon ^2}} \right)} A_{LT}^{\cos \left( {{\phiH} - {\phiS}} \right)}\cos \left( {{\phiH} - {\phiS}} \right)\\ \nonumber
&&\hspace*{-3.5cm}+\,\sqrt {2\varepsilon \left( {1 - \varepsilon } \right)} A_{LT}^{\cos {\phiS}}\cos {\phiS}
+ \sqrt {2\varepsilon \left( {1 - \varepsilon } \right)} A_{LT}^{\cos \left( {2{\phiH} - {\phiS}} \right)}\cos \left( {2{\phiH} - {\phiS}} \right)\Big]
\Bigg\} \label{eq:x_secSIDIS}
\\
\end{eqnarray}
where asymmetries $A_{\dots}^{\dots}$~\cite{Bacchetta:2006tn} depend on kinematical variables $x,Q^2,z,\Phperp$ and correspond to azimuthal modulations of the cross section in the azimuthal angle $\phiS$ of transverse spin and/or azimuthal angle $\phiH$ of the produced hadron, both defined in the $\gamma^* N$ cm frame (see Fig.~\ref{fig:sidis-kin}). The first and second subscripts denote respectively the lepton and target nucleon polarizations, while the superscript indicates the corresponding azimuthal modulation. Asymmetries are defined as ratios of corresponding polarized structure functions $F_{\dots}^{\dots}$ and unpolarized structure function $F_{UU}$ The unpolarized structure function, $F_{UU}$, or more precisely combination of structure functions corresponding to transverse and longitudinal polarization of the virtual photon $F_{UU,T}+\varepsilon F_{UU,L}$, is included in the definition of $\hat{\sigma}_{U}$.
We use the usual SIDIS kinematical variables $\xbj$, $y$, and $z$ defined as:
$\xbj = Q^2/{2(P\cdot q)}$, $y={(P \cdot q)/(P \cdot k)}$, $\zh=(P_h \cdot P)/(P \cdot q)$,
where $Q^2=-q^2=-(k-k^\prime)^2$ is the negative four-momentum squared
of the virtual photon, and $\Phperp$ is the transverse momentum of the
detected hadron. The ratio $\varepsilon$ of the longitudinal and transverse photon flux is given by: $\varepsilon=\frac{1-y-\gamma^2y^2/4}{1-y+y^2/2+\gamma^2y^2/4}$, where $\gamma=2M\xbj /Q$,
and $M$ is the mass of the nucleon.

In the kinematical region, where TMD description of SIDIS is appropriate, namely in the beam fragmentation region, $\Phperp/z \ll Q$, the transverse momentum of the produced hadron $\Phperp$ is generated by intrinsic momenta of the parton in the nucleon $\kt$ and the transverse momentum of the produced hadron with respect to the fragmenting parton $\pt$, such that
the structure functions become convolutions of TMD parton distribution functions (PDFs),
and TMD fragmentation functions (FFs). The convolution integral, for a given combination of TMD PDF $f$ and FF $D$ reads~\cite{Bacchetta:2006tn}
\begin{eqnarray}
\hspace*{0.5cm} {\cal C}[w f D] &=& x \sum_q e_q^2 \int d^2\kt \, d^2\pt \delta^{(2)}(\pt + z\kt - \PT) w(\kt, \pt) f^{q}\xkts D^{q}\zpts, \label{Eq:convo}
\end{eqnarray}
where $w$ is a kinematical factor, and the sum goes over all flavors of quarks and anti-quarks.
Well known SIDIS structure functions $F_{UU,T}$ and $F_{LL}$ will be, thus, described by convolutions
of $f_1$ and $g_1$ TMD PDFs and $D_1$ the unpolarized TMD fragmentation function, with $\FUU =  {\cal{C}}\left[ f_1 D_1 \right], \FLL  =  {\cal{C}}\left[ g_1 D_1 \right]$. The full list of TMD PDFs accessible in SIDIS is given in Table.\ref{tab:TMD-tables}. The TMDs depend on polarization state of the quark (rows) and polarization state of the nucleon (columns). The corresponding tables for TMD FFs can be found in Review~\cite{Metz:2016swz}.

Apart from $Q^2$ dependence of the elementary lepton-quark cross section $\propto Q^{-4}$, some structure functions appear in the cross section either unsuppressed or suppressed
by an additional power of the hard scale $Q$. Accordingly, structure functions can be classified as subleading-twist $Q^{-1}$ (twist-3) effects and
leading-twist $Q^0$ (twist-2) effects ~\cite{Jaffe:1996zw}. Higher twist structure functions will include convolutions of higher twist TMD functions. Several collinear higher twist distributions, like $g_T$, for instance, are
accessible in DIS where collinear factorization applies, and bear physics of significant interest~\cite{Jaffe:1989xx}.
While the definition of twist itself is problematic~\cite{Gamberg:2006ru,Bacchetta:2008xw} in TMD formalism, one can characterize twist expansion rigorously using collinear operator product expansion (OPE) of the scattering amplitude, such that
the twist classifies operators and isolates the leading
contributions, at twist-$2$, from the sub-leading higher-twist contributions,
power suppressed as $~Q^{2-\rm twist}$. The leading and higher twist non-perturbative functions
describe various spin-spin and spin-orbit correlations as corresponding operators include additional gluon and/or quark fields in the matrix element.

Many structure functions involve transversely polarized quarks. For example the $F^{\cos 2\ph}_{UU}$, at leading twist is interpreted as a convolution of Boer-Mulders distribution function, $h_1^\perp$ that encodes the correlation between the transverse motion
of a quark and its own transverse spin~\cite{Boer:1997nt}, and the Collins fragmentation function, $H_1^{\perp}$~\cite{Collins:1992kk}, that describe fragmentation of transversely polarized quarks into an unpolarized hadron.

\begin{table}[h!]
\centering
\caption{Tables for leading (left) and sub-leading (right) TMDs, with columns indicating the quark, and rows the nucleon polarization~\cite{Mulders:1995dh,Bacchetta:2006tn,Chen:2015tca}.}
\label{tab:TMD-tables}
%
\resizebox{0.421\linewidth}{!}{
\begin{tabularx}{0.46865\linewidth}{|c|c|c|c|} 
\hline
\backslashbox{nucleon}{quark}& {U}& {L}& {T}\bigstrut \\ \hline
{U} &$f_{1}$& & $h_{1}^{\bot}$\bigstrut \\ \hline
{L} & &$g_{1L}$&$h_{1L}^{\bot}$\bigstrut \\ \hline
{T} &$f_{1T}^{\bot}$ & $g_{1T}$ & $h_{1}$, $h_{1T}^{\bot}$ \bigstrut \\ \hline
\end{tabularx}
}
\resizebox{0.55\linewidth}{!}{
\begin{tabularx}{0.61765\linewidth}{|c|c|c|c|} 
\hline
\backslashbox{nucleon}{quark}
      & {U}    & {L}  & {T}    \bigstrut \\ \hline
{U} & $f_{\bot}$, $e$ &  $g^{\bot}$     & $h$  \bigstrut \\ \hline
{L} & $f_{L}^{\bot}$  & $g_{L}^{\bot}$ &  $h_{L}$, $e_{L}$ \bigstrut \\ \hline
{T} & $f_{T}$, $f_{T}^{\bot}$, $e_{T}^{\bot}$ & $g_{T}$, $g_{T}^{\bot}$, $e_{T}$  & $h_{T}$, $h_{T}^{\bot}$ \bigstrut \\ \hline
\end{tabularx}
}
\end{table}

Measurements of flavor asymmetries in sea quark distributions performed in DY experiments, indicate very significant non-perturbative effects at large Bjorken-$x$, where the valence quarks are relevant~\cite{Alberg:2017ijg}.
In perturbative QCD $q\bar{q}$ pairs are created from the gluon splitting. Since the masses of $u$ and $d$ quarks are small, the gluon splitting
is not expected to generate quark flavor asymmetries. Older measurements by NMC~\cite{Amaudruz:1991at} indicated that integrated
$\bar{d}$ is more than integrated $\bar{u}$. The measurements by E866 collaboration~\cite{Garvey:2001yq}, and more recently by SeaQuest~\cite{Nagai:2017dhp} suggest that $\bar{d}$ is significantly larger than $\bar{u}$ in the full accessible $x$-range.
The non-perturbative $q\bar{q}$ pairs, are also correlated with spins and play a crucial role in spin orbit correlations, and in particular, single-spin asymmetries measured by various experiments in last few decades.

Interpretation of leading-twist structure functions in terms of convolutions of TMD distributions arise from factorization theorems, see Ref.~\cite{Collins:2011zzd} and references therein. Subleading structure functions require a proof of validity of TMD factorization at higher twist and the proof is not yet available, however studies of
sub-leading twists are also important, as they may affect significantly the extraction of leading-twist moments, as the radiative effects, and complicated acceptances of wide angle spectrometers may introduce cross-talk between different azimuthal moments in the spin dependent and spin-independent moments. For instance $\cos \phi$ modulation asymmetry known as the Cahn effect~\cite{Cahn:1978se}), is significant $\sim 20\div30\%$ and dominating in the $\Phperp \sim $1 GeV range, even though it is suppressed by $\Phperp/Q$ with respect to leading twist asymmetries. Additional contributions to $\cos \phi$ and $\cos 2\phi$ moments coming from processes when the final meson
is produced at short distances via hard-gluon exchange~\cite{Berger:1979xz} may also be
significant in the kinematic regime where the ejected meson carries most of
the virtual photon momentum ($z$ approaching 1).

Even though we do not have enough direct information about higher twist TMDs from experimental data one may hope that quark models give estimates for higher-twist TMDs, and whether the modeling can be useful
for phenomenological studies. The applicability of quark models to TMDs beyond leading twist still remains debatable, however the additional information on higher twist TMDs from models may become very important for phenomenology and for experimental event generators.

The projected accuracy of Jefferson lab 12 GeV upgrade data for $\phiH$-dependent multiplicities is comparable to effects arising from the fine structure constant, $\alpha_{em}\approx$1/137, and therefore neglecting electromagnetic corrections of the order ${\cal O} (\alpha_{em})$ may lead to significant mis-interpretation of the data. Extension and further development of the theoretical and computational approaches to higher-order QED effects for electron scattering is an outstanding problem for the upcoming physics analysis effort at Jefferson Lab.
Disentangling the cross section $\sigma_0^{ehX}(x,y,z,\Phperp,\phiH,\phiS)$ from measurements is a very important task, complicated by the fact that radiative corrections to cross section introduce significant corrections to contributions and may introduce additional azimuthal moments in the cross section~\cite{Akushevich:1999hz,Akushevich:2007jc}:
\begin{eqnarray}
 &\hspace*{-8.0cm}\sigma_{Rad}^{ehX}(x,y,z,\Phperp,\phi,\phiS) \rightarrow \\ \nonumber &\hspace*{2.0cm}\sigma_0^{ehX}(x,y,z,\Phperp,\phiH,\phiS)\times R(x,y,z,\Phperp,\phiH)+R_A(x,y,z,\Phperp,\phiH,\phiS) \; .
\end{eqnarray}

Studies of the model dependence in Radiative Correction (RC) calculations in the full kinematical region will be important for precision studies of underlying 3D PDFs from Jefferson Lab12 to EIC and require
simultaneous extraction of all moments to account for various correlations. The methodology for accounting for radiative effects in SIDIS is currently a generalized version tested for DIS studies: typically one uses RADGEN generator~\cite{Akushevich:1998ft} combined with different full event generators, like PYTHIA, LEPTO, PEPSI~\cite{Sjostrand:2006za,Ingelman:1996mq,Mankiewicz:1991dp}. This approach, while providing some estimates for RC, is not fully consistent for SIDIS, as RADGEN itself, contains only DIS structure functions, and the LUND model based generators, at the moment do not include spin-orbit correlations in the fragmentation. Precision studies of azimuthal moments in SIDIS will require a completely new methodology for accounting of RC effects in SIDIS, taking as input a set of realistic structure functions that describe all relevant moments for specific observables under study. Thus we expect that phenomenological studies will attempt extraction of all azimuthal moments and a full set of TMDs contributing to the SIDIS cross section for a given configuration of beam and target polarizations.

One of the most important questions about the 3D structure of the nucleon is the transverse momentum dependence of the distribution and fragmentation TMDs and flavour and spin dependence of those shapes. For precision studies of TMDs it is also important to understand the role of medium, and the effects of in medium modifications of TMDs. That is crucial, since both COMPASS and JLab use nuclear targets
to study polarization effects. Another important question to address is the role of exclusive processes in studies of SIDIS.
In order to extract underlying functions and thus details of dynamics of quarks and gluons from SIDIS data one also has to have a good understanding of the underlying fragmentation process in which quark fragments into an observed hadron. Exclusive processes may shed light on the fragmentation process itself.

The structure of this mini-review is the following. After a brief introduction of experimental facilities and important measurements of SIDIS leading twist observables ( Sections \ref{Sec:experiments}-\ref{Sec:leadingtwist}), we will discuss the higher twist SIDIS observables (Section \ref{Sec:highertwist}). In the Section \ref{Sec:flavor} we discuss the relevance of different flavors of hadrons in studies of the complex nucleon structure, followed by discussion of some unique possibilities enabled by detection of di-hadron final states in Section \ref{Sec:dihadrons}. The final section \ref{Sec:precision} is devoted to challenges and possible methodology of extraction of non-perturbative partonic distribution and fragmentation functions from the wealth of the experimental data, already available and expected from future measurements at COMPASS and upgraded Jefferson Lab, as well as future Electron Ion Collider.

\section{SIDIS Experiments}
\label{Sec:experiments}

Several experiments worldwide were involved in studies of SIDIS with different hadrons produced. Most of the latest relevant data
for studies of spin-orbit correlations was coming from HERMES Collaboration at HERA, COMPASS collaboration at CERN, and measurements already performed at JLab. As for the near future, a wide spectra of high-precision measurements will be done at JLab12, while COMPASS plans to collect more SIDIS data with transversely polarized deuteron target in 2021~\cite{cmps_prop_add}.
Major advantages of different setups 
include gas target from HERMES with fast target spin flip, providing clean target
spin asymmetries with no dilution from nuclear target, high energy muon beam of COMPASS with relatively small radiative
corrections, and superior beam polarization at JLab, allowing clean measurements of beam-spin asymmetries.
Wider angle coverage of CLAS12 detector allows measurements in a wide range of $\Phperp$ (up to 1.5 GeV), and $Q^2$ (up to 10 GeV$^2$), while the SoLID detect would allow measurements of all kind of polarization asymmetries at large Bjorken-$x$ with superior precision.

%
\begin{table}[h]
\caption{Main characteristics of SIDIS detectors.}
\centering
\resizebox{13cm}{!}{%
\begin{tabular}{c|c|c|c|c|c}
Experiment & Beam     & Target     & Energy GeV & Lumi      & Polarization \\[0.3em] \hline \\[-0.3em]
HERMES   & $e^+e^-$     & H$_2$,D$_2$,N,C   & 27.5   & $10^{32},10^{33}$ & U/L/T\\
COMPASS  & $\mu^+$ & NH$_3$,6LiD    & 160, 200    & $ 10^{33}$    &U/L/T\\
JLab CLAS  &$e^-$      & H$_2$,D$_2$,NH$_3$,ND$_3$  & 6, 11   & $10^{34}-10^{35}$   &U/L/T\\
JLab Hall-A &$e^-$      & $^3$He       & 6, 11   & $10^{37}$   &U/L/T\\
JLab Hall-C &$e^-$      & H$_2$,D$_2$      & 6, 11   & $10^{38}$   &U\\
JLab Solid &$e^-$      & $^3$He,NH$_3$    & 6, 11   &  $10^{36}-10^{37}$  &U/L/T\\
EIC~\cite{EIC_JLEIC}     &$e^-$       & p,d,A      & $3-12(e)/20-400(p)$   &  $0.5-5.10^{34} $    &U/L/T\\ \hline 
\end{tabular}}
\label{tab1}
\end{table}

The Table~\ref{tab1} shows main characteristics of SIDIS experiments involved in TMD studies.
At JLab all 3 halls are involved in 3D structure studies~\cite{Dudek:2012vr} including the HMS and Super HMS at Hall C~\cite{E12-06-104,E12-09-017,E12-13-007},
the BigBite and Super BigBite, as well as, the SoLID detector at Hall A~\cite{E12-09-018,E12-10-006,E12-11-007}, and
CLAS12 at Hall-B~\cite{E12-06-112,E12-07-107}. Several experiments
are already approved to study in details the azimuthal modulations in SIDIS for
different hadron types, targets, and polarizations in a broad kinematic range~\cite{E12-06-112,E12-07-107,E12-09-008,E12-09-009,E12-10-006,E12-09-017,E12-09-018,E12-11-007}.
The experimental investigation of
medium modification of quark fragmentation and spin-orbit correlation will be
also extensively pursued at the upgraded Jefferson Lab facility, for which several
related experimental proposal already exist~\cite{LOI12-14-004,E12-14-001}.

\section{Leading twist observables}
\label{Sec:leadingtwist}
Correlations of quark transverse momenta with their own spin or spin of the parent hadron manifest
in different spin dependent azimuthal moments in the cross section, generated either by correlations in the distribution of quarks or in the fragmentation process. The most known correlations are often referred to as
Sivers type~\cite{Sivers:1989cc} and Collins type~\cite{Collins:1992kk}, respectively, see Review~\cite{Boglione:2015zyc}.
Involved structure functions factorize into TMD parton distributions and fragmentation functions, and hard parts~\cite{Ji:2004wu}.

The most prominent leading twist observable is the $\phiH$-integrated cross section described by
the $F_{UU}$ structure function. Experiments, however, prefer to measure the multiplicities of hadrons, which is the ratio of SIDIS cross sections for a given type of hadron divided by DIS cross section in a given bin in $x,Q^2$, (the advantage is that \textit{e.g.} the scattered lepton acceptance entering in the numerator and denominator cancels).
As one can see from Eq.~(\ref{Eq:convo}), in the TMD formalism the final hadron $\PT$ results from the initial quark $\kt$ and the fragmenting quark $\pt$ and
up to order ${\cal O}(\kperp/Q)$ the momentum conservation gives
$\PT=z\kt+\pt$. The structure function $F_{UU}$ is given by the convolution
integral ${\cal C}[f_1D_1]$:
\begin{eqnarray}
\label{Eq:convolution}
F_{UU} \xzpt &=& x \sum_q e_q^2 \int d^2\kt \, d^2\pt \delta^{(2)}(\pt + z\kt - \PT) f^{q}_1\xkts D^{q}_1\zpts, \label{eq:fuu}
\end{eqnarray}

Notice that Eq.~(\ref{eq:fuu}) was initially proposed in the parton model approximation, see for instance ~\cite{Bacchetta:2006tn}. The  result of factorization proof of Ref.~\cite{Collins:2011zzd} formally coincides with Eq.~(\ref{eq:fuu}), however the TMD functions have a much more intricate dependence on the scales present in the process. This dependence is governed by the evolution equations and allow to predict the change of the shape of TMDs without using the model assumptions. The complication of implementation of TMD factorization is the presence~\cite{Collins:2011zzd} of a universal non-perturbative kernel of evolution. This kernel and non-perturbative shape of TMDs should be extracted from the global fit of low and high energy experimental data.

In this review we will consider only the approximate description of TMDs in the non-perturbative region and analytical results that can be obtained using simplified assumptions.
In order to resolve the convolution integral of Eq.~(\ref{Eq:convolution}) one makes assumptions on $\kperp$-dependence of $f_1$ and $\pperp$-dependence of $D_1$ and individuates a set of parameters which would be then extracted in an analysis of multidimensional data on either $F_{UU}$ or multiplicity. For example, a common
assumption is the Gaussian ansatz for the transverse momentum dependence of
distribution and fragmentation functions~\cite{Anselmino:2013lza, Signori:2013mda} which would result in the average $\Phperp$ given by
\ba
    \la \Phperp (z)\ra &=& \frac{\sqrt{\pi}}{2}\,
    \sqrt{z^2\la \kdperp \ra+\la \pdperp \ra}\;,
\ea
where $\la \kdperp \ra$ and $\la \pdperp \ra$ (GeV$^2$) are Gaussian widths of $\kperp$-dependence of $f_1$ and $\pperp$-dependence of $D_1$.

Collinear PDFs have flavour dependence, thus it is not unexpected that also
the transverse momentum dependence may be different for the different
flavours~\cite{Signori:2013mda}. Model calculations of transverse momentum dependence of TMDs ~\cite{Pasquini:2008ax,Lu:2004au,Anselmino:2006yc,Bourrely:2010ng} and lattice QCD results~\cite{Hagler:2009mb,Musch:2010ka} suggest that the dependence of widths of TMDs on the quark polarization and flavor may be
significant. It was found, in particular, that the average
transverse momentum of antiquarks is considerably larger than that of
quarks~\cite{Wakamatsu:2009fn,Schweitzer:2012hh}.
The frequently used assumption of factorization of $x$ and $\kperp$
(or $z$ and $\pperp$) dependencies~\cite{Anselmino:2013lza} may be significantly violated (see Fig. 10 of~\cite{Bacchetta:2017gcc}). For instance the predicted average
transverse momentum square $\langle \kdperp \rangle$ of quarks and antiquarks
may depend strongly on their longitudinal momentum fraction $x$ within the
framework of the chiral quark soliton model.

In the fragmentation process, one would expect~\cite{Matevosyan:2011vjb} that the {\it dis-favored}
fragmentation of a quark into a hadron would be broader in the transverse momentum with respect the
{\it favored} fragmentation (fragmentation of a quark to a hadron that has this type of quark as a valence quark).

Production of charged pions in SIDIS has been measured from both
proton and deuteron targets, using a 5.5 GeV energy electron beam in Hall-C at
Jefferson Lab~\cite{Mkrtchyan:2007sr}. In the limited $\Phperp^2<0.2$ explored,
the $\Phperp$ dependence from the deuteron was found to be slightly weaker than
from the proton. In the context of a simple model this would suggest that
transverse momentum distributions may depend on the flavor of quarks.
Multiplicities of charged pion and kaon mesons have been measured by
HERMES using the electron beam scattering off hydrogen and deuterium
targets~\cite{Airapetian:2012ki}. Multiplicities of charged hadrons
produced in deep inelastic muon scattering off a $\rm ^6LiD$ target have been
measured at COMPASS~\cite{Adolph:2013stb}. These high-statistics data samples
have been used in phenomenological analyses~\cite{Anselmino:2013lza, Signori:2013mda,Bacchetta:2017gcc} to extract information on the flavor
dependence of unpolarized TMD distribution and fragmentation
functions. Restricting the ranges of the available data to
$Q^2>1.69\ (\mathrm{GeV}/c)^2$, $z<0.7$ and $0.2\ \mathrm{GeV}/c < \phperp <
0.9\ \mathrm{GeV}/c$, the authors of Ref.~\cite{Anselmino:2013lza}
obtained a reasonable description of the experimental data within a Gaussian
assumption for TMDs with flavour independent and constant widths, $\langle \kdperp \rangle$
and $\langle \pdperp \rangle$. Nevertheless, indications were reported that
favoured fragmentation functions into pions have smaller average transverse
momentum width than unfavoured functions and fragmentation functions into
kaons~\cite{Signori:2013mda}, consistent with
predictions based on the NJL-jet model~\cite{Matevosyan:2011vjb}.
\begin{figure}[ht!]
\centering
\includegraphics[width=0.6\textwidth]{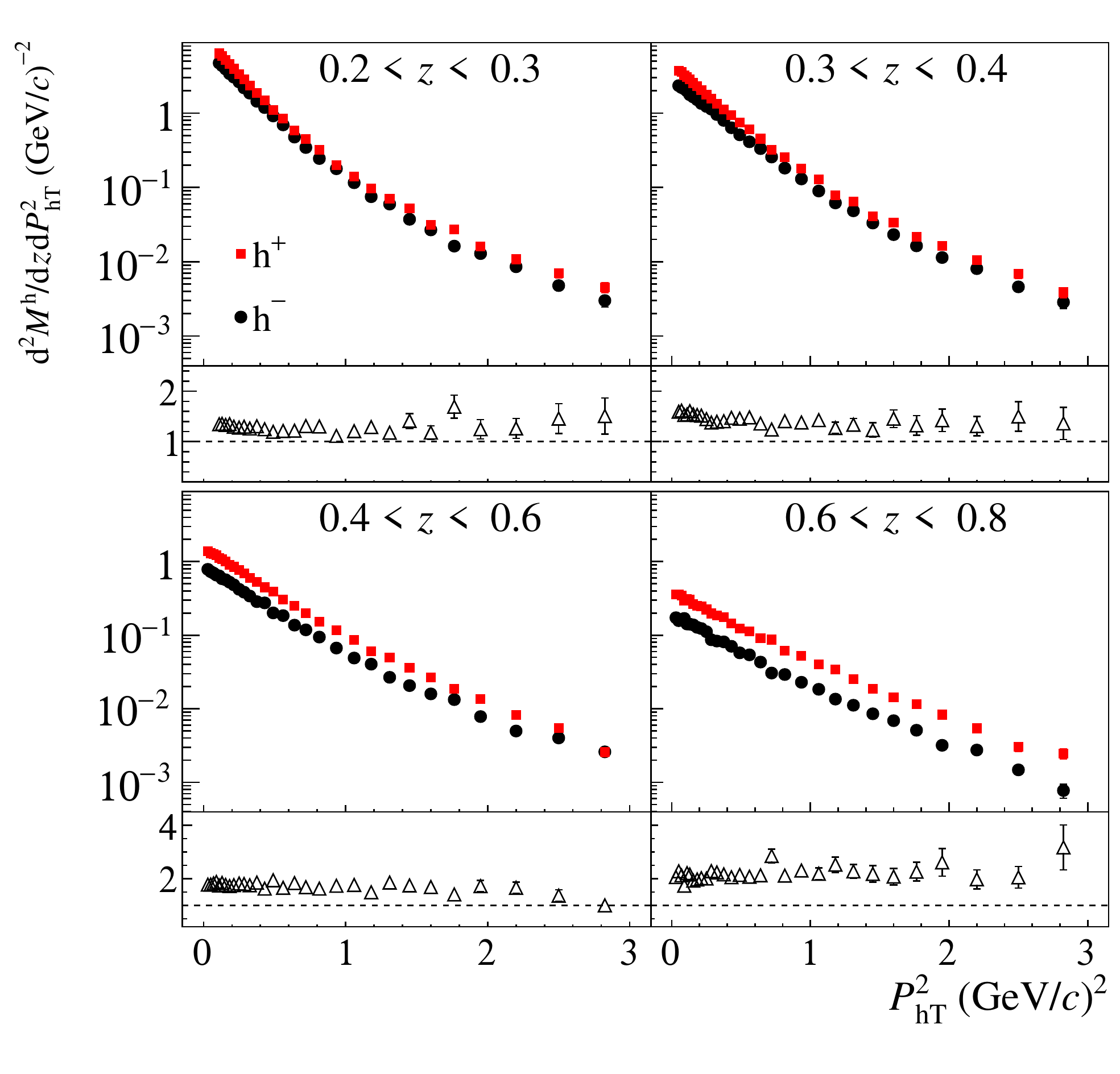}
\caption{Multiplicities of positively (full squares) and negatively (full circles) charged hadrons at COMPASS~\cite{Aghasyan:2017ctw}. Upper panels: Multiplicities of charged hadrons as a function of $P_{hT}$ in four z bins at $Q^2$ = 9.78 (GeV/c)$^2$ and $x$ = 0.149. Lower panels: Ratio of multiplicities of positively and negatively charged hadrons.}
\label{fig-compass-mult}
\end{figure}
Latest multiplicity measurements at COMPASS~\cite{Aghasyan:2017ctw} shown in Fig.~\ref{fig-compass-mult} indicate that the ratio of counts of positive and negative hadrons increases with $z$, which can be explained by the fact that, in contrast to $\pi^\pm$, $K^+$ and $p$, negative hadrons $K^-$ and $\bar{p}$ cannot be produced by the favoured fragmentation of nucleon valence quarks. In addition, the ratio tends to decrease with $\Phperp$ at relatively small $z$, where the fraction of exclusive events is not essential. Comparison of COMPASS with HERMES and JLab measurements of multiplicities performed in Ref.~\cite{Aghasyan:2017ctw} shown in Fig.~\ref{fig-cmps-hms_jlab} unveils significant differences which could be due to the
different $Q^2$ ranges covered by the experiments and supports findings of studies from Ref.~\cite{Aghasyan:2014zma} indicating that at lower energies the large values of $\Phperp$ are suppressed due to smaller phase space, in particular at large $z$.
The latter is confirmed by recent COMPASS results obtained for the $K^-$ over $K^+$ multiplicity ratio at large fraction $z$ of the virtual-photon energy~\cite{Akhunzyanov:2018ysf}.
\begin{figure}[ht!]
\centering
\includegraphics[width=0.54\textwidth]{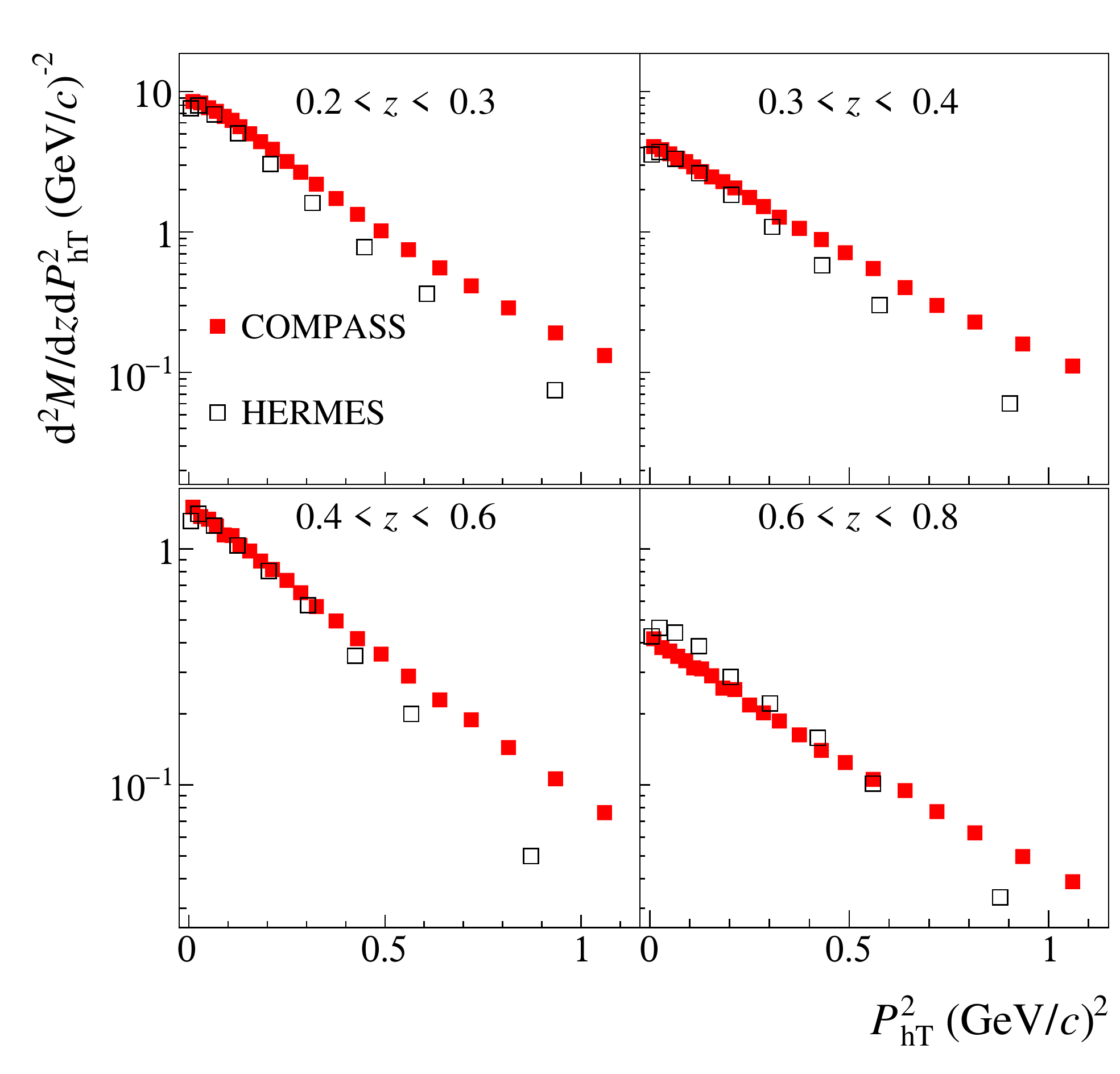}
\includegraphics[width=0.42\textwidth]{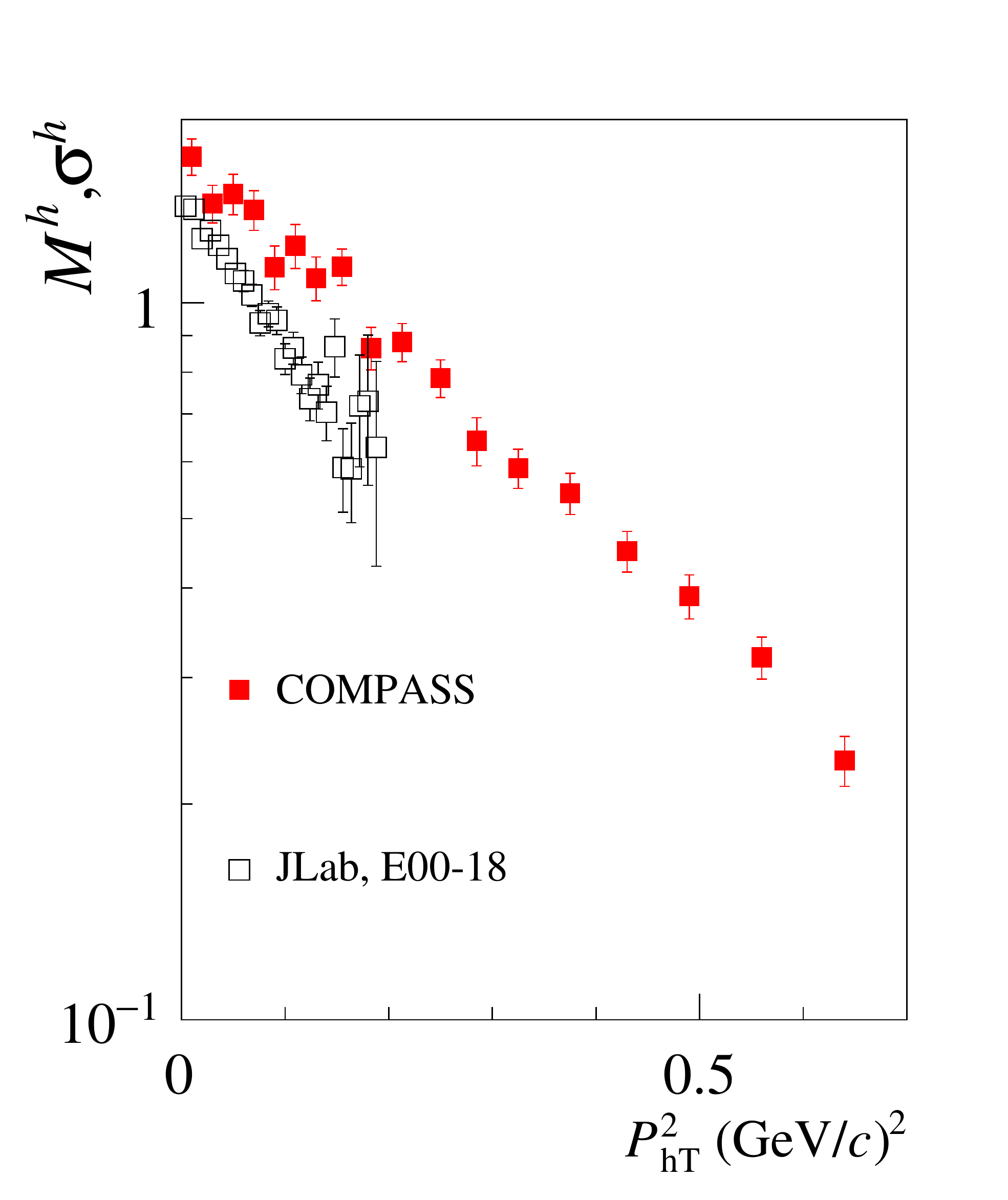}
\caption{Comparison of multiplicities of positively charged hadrons at COMPASS~\cite{Aghasyan:2017ctw} and HERMES~\cite{Airapetian:2012ki} (left panel) and at COMPASS~\cite{Aghasyan:2017ctw} and JLab~\cite{Asaturyan:2011mq} (right panel).}
\label{fig-cmps-hms_jlab}
\end{figure}

The origin of the $\cos 2\phi_h$ modulation, $F^{\cos2\phi_h}_{UU}$, due to convolution of
the Collins fragmentation function $H_1^\perp$, describing fragmentation of transversely polarized quarks, and the Boer-Mulders distribution function $h_{1}^\perp$, describing distributions of transversely polarized quarks in an unpolarized nucleon,
was first discussed by Boer and Mulders in 1998~\cite{Boer:1997nt}. The structure function reads:
\begin{eqnarray}
F_{UU}^{\rm cos2\phi_h}\xzpt & = & {\cal C} \left[\frac{2 (\bfhp\cdot\kt)(\bfhp\cdot\pt) - (\pt\cdot\kt)}{z M_N m_h}h_{1}^\perp\xkts H_1^\perp\zpts \right]
\end{eqnarray}

In addition to this, when the intrinsic transverse momenta $\kt$ of quarks inside the nucleon is taken into account a subleading-twist $Q^{-2}$ (twist-4) contribution to the $\cos2\phi$ amplitude originates from the Cahn effect~\cite{Cahn:1978se} (kinematic correction at the level of the elastic quark–lepton cross-section). This contribution is expected to dominate at small $x$.
Measurements of the $\cos2\phi$-moments have been published by different experiments.
A significant positive $\cos2\phi$ amplitudes for both positively and negatively charged
hadrons were measured at COMPASS~\cite{Akhunzyanov:2018ysf}. At HERMES~\cite{Airapetian:2012yg}, positive $\cos 2\phi$ amplitudes
are extracted for negatively charged pions, while for positively charged pions
the moments are compatible with zero, but tend to be negative in some kinematic
regions. In all the cases, the amplitudes of the cosine modulations show strong
kinematic dependencies. Comparisons between COMPASS and HERMES 
$\cos2\phi$ modulations for hadrons in the almost overlapping kinematic region
($0.02 <x <0.13$, $\langle Q^2 \rangle\simeq 4$ $(\mathrm{GeV}/c)^2$ of COMPASS
and $0.023 <x <0.145$, $\langle Q^2 \rangle\simeq 2$ $(\mathrm{GeV}/c)^2$ of
HERMES) require point-to-point correction for the so-called depolarization factor $D(y)=\epsilon$ (ratio of the longitudinal to transverse virtual photon flux).
There is some tension in the $z$-dependence between the two
experiments in the $\cos2\phi$ modulation of positive
hadrons (that show the same behaviour but have an off-set of about 0.05).
 For a detailed comparison between
results of different experiments and between results and theoretical models, a full
differential analysis, using the complete multi-dimensional information is needed~\cite{Avakian:2016pqj}.

The first observation of a Single Spin Asymmetry (SSA) in semi-inclusive DIS pion
electroproduction was made by HERMES~\cite{Airapetian:1999tv}.
The main goal of original measurements was to access distributions of transversely polarized quark in the longitudinally polarized nucleon, $h_{1L}^\perp$.
The physics of $F_{UL}^{\sin 2\phi}$, which involves the Collins fragmentation
function $H_1^\perp$ and Mulders distribution function $h_{1L}^\perp$,
was first discussed by Kotzinian and Mulders in 1996
\cite{Mulders:1995dh,Kotzinian:1994dv,Kotzinian:1995cz}.

\begin{eqnarray}
F_{UL}^{\rm \sin2\phi_h}\xzpt & = & {\cal C} \left[\frac{2 (\bfhp\cdot\kt)(\bfhp\cdot\pt) - (\pt\cdot\kt)}{z M_N m_h} h_{1L}^\perp\xkts H_1^\perp\zpts \right]
\end{eqnarray}

The same
distribution function is accessible, in particular, in double polarized Drell-Yan, where it
gives rise to the $\cos2\phi$ azimuthal moment in the cross section
\cite{Tangerman:1994eh}.
The behavior of the Mulders distribution function was subsequently studied in many models, including
 large-$x$~\cite{Brodsky:2006hj} and large $N_c$~\cite{Pobylitsa:2003ty} limits of QCD. Model calculations of Boer-Mulders functions, phenomenological analysis and predictions for JLab measurements of $\cos2\phi$ azimuthal moment were given in Ref.~\cite{Gamberg:2007wm}.

\begin{figure}[ht!]
\centering
\includegraphics[width=0.7\textwidth]{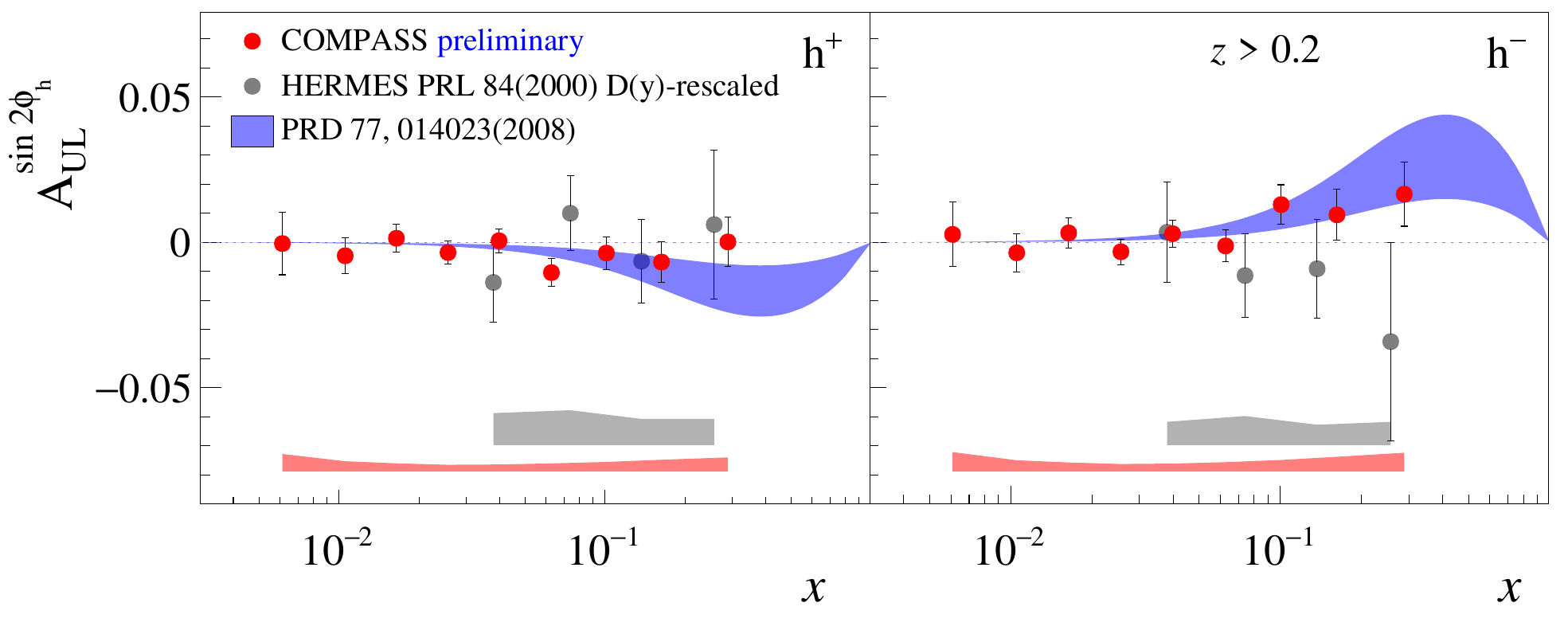}
\caption{The $A_{UL}^{\sin(2\phiH)}$  results obtained by HERMES~\cite{Airapetian:1999tv} and preliminary results by COMPASS~\cite{Parsamyan:2018ovx,Parsamyan:2018evv} and available model predictions~\cite{Avakian:2007mv}.}
\label{fig:AUL2cmps_X_th}
\end{figure}

Measurements of the $\sin2\phi$ SSA~\cite{Kotzinian:1995cz}, allows the
study of the Collins effect with no contamination from other mechanisms. Measurement of the $\sin 2\phi$ moment of $F_{UL}$ by HERMES~\cite{Airapetian:1999tv} appeared to be consistent with zero. A measurably large
asymmetry has been predicted only at large $x$ ($x>0.2$), a region
well-covered by JLab~\cite{Efremov:2002ut}.
The existing data indeed indicates that at large $x$ the $F_{UL}$ may be significant~\cite{Avakian:2010ae,Parsamyan:2018ovx,Parsamyan:2018evv}. In
Fig.~\ref{fig:AUL2cmps_X_th} the latest COMPASS measurements~\cite{Parsamyan:2018ovx,Parsamyan:2018evv} are compared with $D(y)$-rescaled HERMES points~\cite{Airapetian:1999tv} and model predictions for COMPASS kinematics~\cite{Avakian:2007mv}.

\begin{figure}[ht!]
\centering
\includegraphics[width=0.5\textwidth]{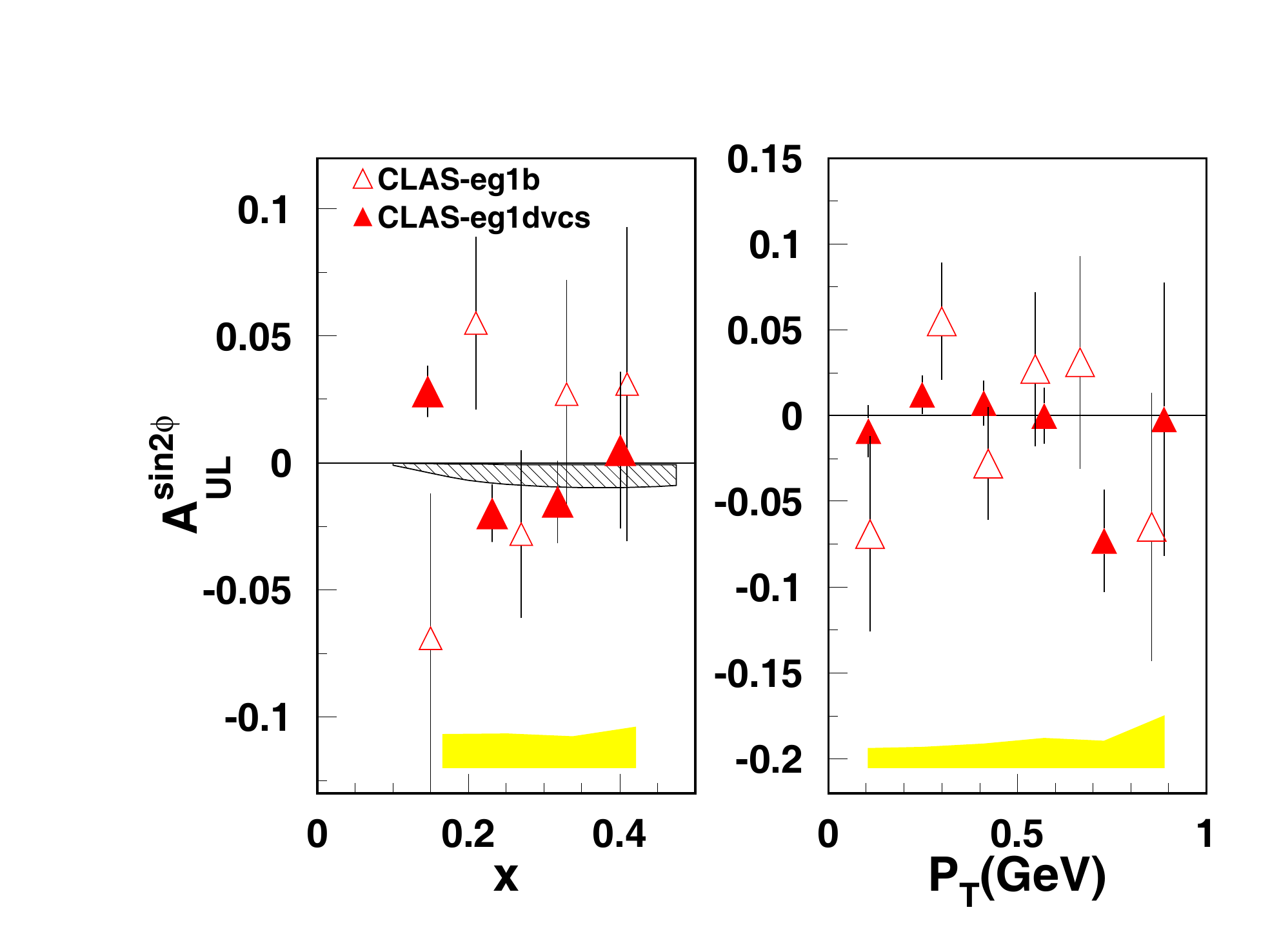}
\caption{The $\sin 2\phi_h$ moments for $A_{UL}$ plotted versus $\xbj$ (left) and $\PT$ (right)~\cite{Jawalkar:2017ube} compared to
previous CLAS measurements~\cite{Avakian:2010ae}  and theory predictions (gray band)  ~\cite{Avakian:2007mv}.
The error bars represent the statistical uncertainties, whereas the yellow bands represent the
total experimental systematic uncertainties.}
\label{fig-pi0aul-sin2}
\end{figure}

Due to opposite sign of the Collins fragmentation functions for the favored and disfavored hadrons, all kind of SSAs originating from Collins mechanism are in principle expected to be suppressed for $\pi^0$-production.
The latest data from CLAS collaboration~\cite{Jawalkar:2017ube} is consistent with previous measurements~\cite{Avakian:2010ae} indicating that the $\sin2\phi$ target spin dependent moment,
which is expected to depend on the Collins fragmentation function is much smaller for $\pi^0$ than for charged pions Fig.~\ref{fig-pi0aul-sin2}.

\begin{figure}[ht!]
\centering
\includegraphics[width=0.7\textwidth]{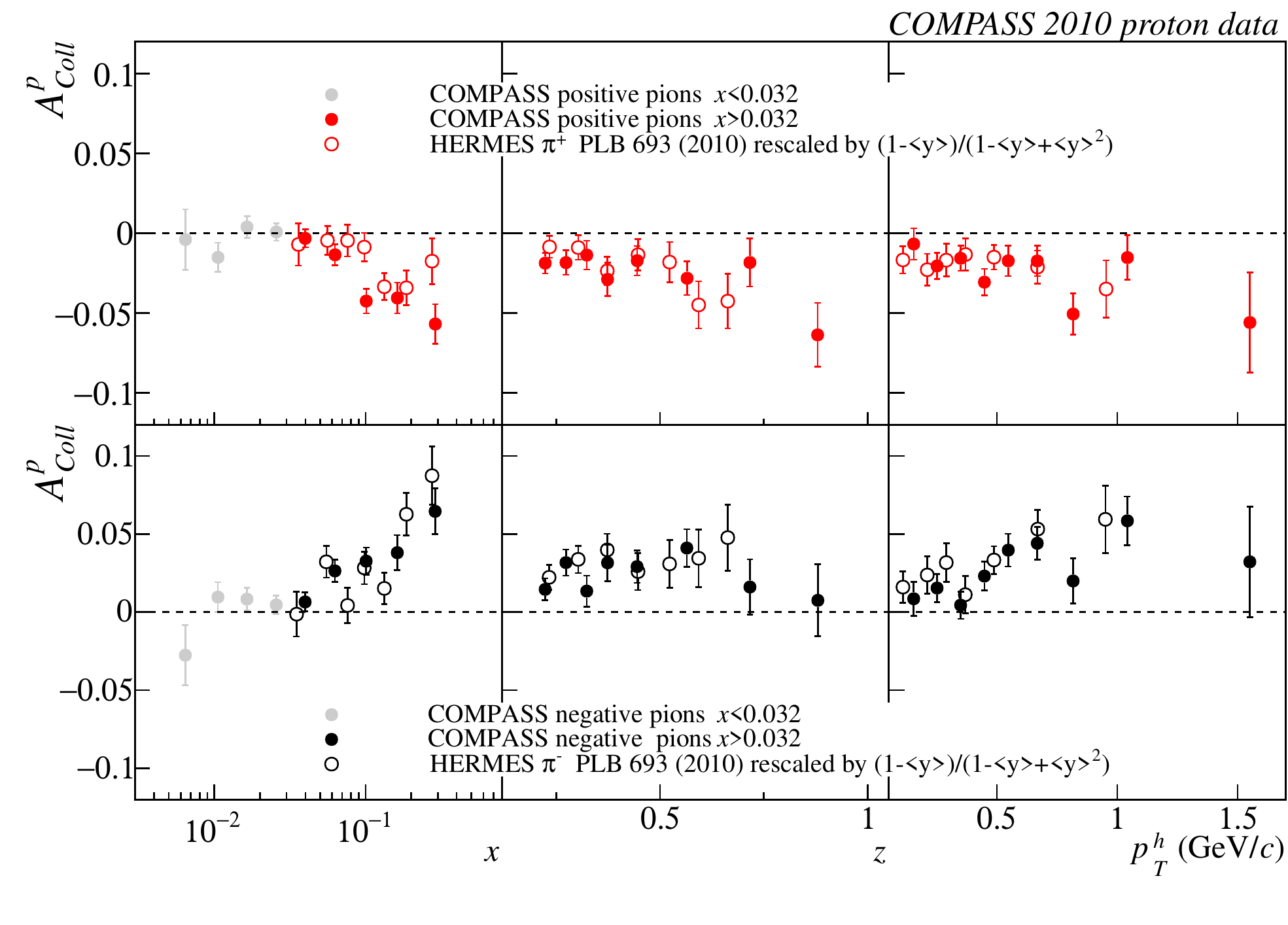}
\caption{Collins asymmetries, $A_{UT}^{\sin(\phi_h+\phi_S)}$, for positive and negative pion production on proton measured at COMPASS~\cite{Adolph:2014zba} requiring $x > 0.032$ (filled circles) are compared with HERMES proton results~\cite{Airapetian:2010ds} (empty circles).}
\label{fig-cmps-hms-col}
\end{figure}

Large $\sin\phi$ SSA measurements by HERMES~\cite{Airapetian:1999tv} spawned a number
of additional measurements of SSAs and DSAs (Double Spin Asymmetries) using polarized hydrogen and deuterium targets~\cite{Airapetian:2001eg, Airapetian:2002mf}. Most prominent are the Collins and Sivers asymmetries.
With an unpolarised beam and a transversely polarised target one can
get access to the structure function $F_{UT}^{{\rm sin}(\phi+\phi_S)}(x,z,\Phperp,Q^2)$. The latter can be written as a
convolution of $h_{1}(\xbj, \kperp, Q^2)$ and $H_1^\perp(z,\pperp,Q^2)$,
integrated over the transverse momentum of the initial, $\kperp$, and
fragmenting $\pperp$ partons, providing access to distributions of transversely polarized quarks, also known as ``transversity''
TMD and the Collins fragmentation function:
\begin{eqnarray}
F_{UT}^{\rm \sin(\phi_h+\phi_S)}\xzpt & = & {\cal C} \left[\frac{\bfhp\cdot\pt}{z m_h} h_1\xkts H_1^\perp\zpts \right]
\end{eqnarray}
The Collins asymmetries were measured by HERMES~\cite{Airapetian:2010ds} with proton target and by COMPASS with deuteron~\cite{Alexakhin:2005iw,Ageev:2006da,Alekseev:2008aa} and proton targets~\cite{Adolph:2012sn,Adolph:2014zba}.
The Collins asymmetry on deuteron was found to be small and compatible with zero within the uncertainties, while on proton it has a strong $x$ dependence, \textit{i.e.} compatible with zero in the small $x$ region accessible at COMPASS it increases up to 0.05 in the valence quark region. The asymmetry exhibits a mirror symmetry (similar amplitude, but opposite sign) with respect to the hadron charge, which is attributed to the same size and opposite sign of the favoured and unfavoured Collins FFs. In Fig.~\ref{fig-cmps-hms-col} COMPASS and HERMES proton Collins SSA results for positive and negative pion productions are compared. The asymmetries are found to be in agreement, which is a non-obvious result, taking into account that in valence region $Q^2$ value at COMPASS is as much as two to three times larger compared to that of HERMES.
Measurements of Collins asymmetries by HERMES~\cite{Airapetian:2010ds} and COMPASS~\cite{Adolph:2012sn,Adolph:2014zba} combined with the Belle~\cite{Seidl:2008xc} $e^+e^- \rightarrow \pi^+\pi^-$ data are used in global fits allowing to extract the transversity distribution, see \textit{e.g.} Ref.~\cite{Anselmino:2007fs}. In 2021 COMPASS is planning to perform one year of semi-inclusive DIS data taking with a transversely polarised deuteron target~\cite{cmps_prop_add}. This measurement will allow to considerably improve the knowledge on $d$-quark transversity distribution.

The study of the Sivers effect, describing correlations between the transverse polarization of
the nucleon and its constituent (unpolarized) parton's transverse momentum, has been the topic
of a great deal of experimental, phenomenological and theoretical effort in recent years.
The asymmetry is related to the Sivers TMD PDF ($f_{1T}^\perp\xkts$) convoluted with ordinary fragmentation function ($D_1\zpts$), corresponding structure function reads:
\begin{eqnarray}
F_{UT}^{\rm sin(\phi_h-\phi_S)}\xzpt & = & {\cal C} \bigg[-\frac{\bfhp\cdot\kt}{M_N} f_{1T}^\perp\xkts D_1\zpts \bigg]
\end{eqnarray}
The most exciting feature, predicted for the Sivers function, is that is expected to have opposite sign when measured in SIDIS on the one hand, and in DY or $W/Z$-boson production on the other~\cite{Brodsky:2002cx,Collins:2002kn}:
%
\begin{equation}
\left( f_{1T}^\perp \right)_{\mathrm{SIDIS}} = -\left( f_{1T}^\perp \right)_{\mathrm{DY}}
\end{equation}

The Sivers SSAs for proton and deuteron targets have been published by HERMES~\cite{Airapetian:2004tw,Airapetian:2010ds} and
COMPASS~\cite{Alexakhin:2005iw,Ageev:2006da,Alekseev:2008aa,Adolph:2014zba,Alekseev:2010rw,Alekseev:2010dm,Adolph:2012nw,Adolph:2012sp}, which provided the first, direct indication of significant interference terms beyond the simple s-wave ($L_z=0$) picture. The asymmetries become larger with increasing $\xbj$, suggesting that spin-orbit correlations are significant only in the region of large-$x$  ($x>0.01$), where the valence quarks or non-perturbative sea are relevant~\cite{Alberg:2017ijg}. In addition to the classical approach, COMPASS has recently measured also the $P_T/zM$-weighted Sivers asymmetries accessing directly the first moments of the Sivers functions for $u_v$ and $d_v$ quarks~\cite{Alexeev:2018zvl}.

\begin{figure}[ht!]
\centering
\includegraphics[width=0.7\textwidth]{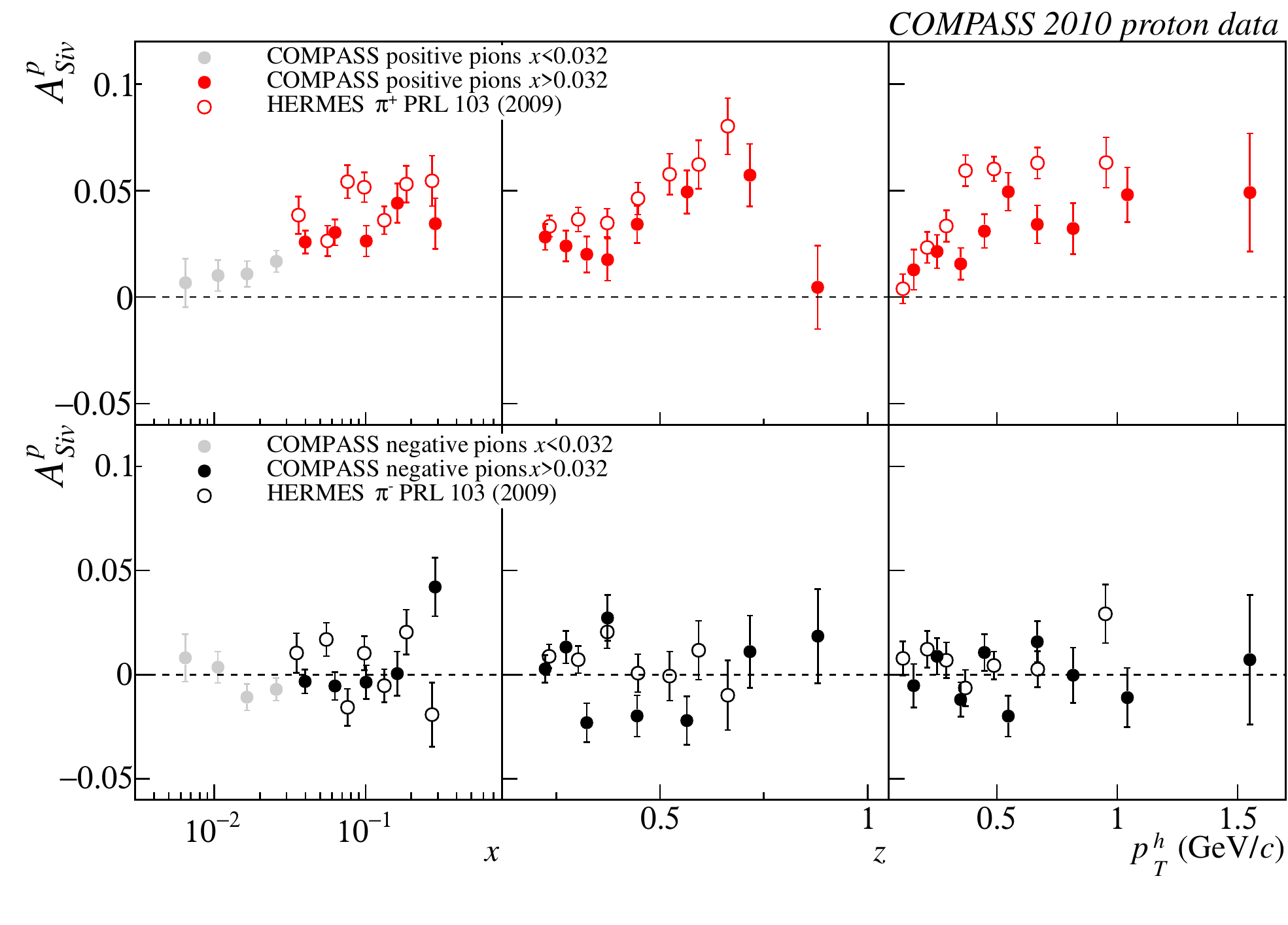}
\caption{Sivers asymmetries, $A_{UT}^{\sin(\phi_h-\phi_S)}$, for positive and negative pion production on proton measured at COMPASS~\cite{Adolph:2014zba} requiring $x > 0.032$ (filled circles) are compared with HERMES proton results~\cite{Airapetian:2009ae} (empty circles).}
\label{fig-cmps-hms-siv}
\end{figure}

\begin{figure}[ht!]
\centering
\vspace{-5cm}
\includegraphics[width=0.5\textwidth]{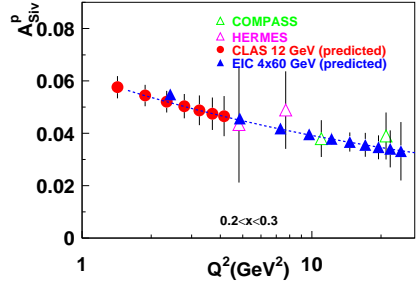}
\caption{Projections for CLAS12 measurements of $Q^2$-dependence of the Sivers asymmetry for $ep\rightarrow e^\prime\pi^+X$, compared to HERMES, COMPASS and future EIC measurements.}
\label{fig-comp-sivers-q2}
\end{figure}

From the comparison of HERMES~\cite{Airapetian:2009ae} and COMPASS~\cite{Adolph:2014zba} proton results in the overlapping kinematic region, unlike the Collins asymmetry, the Sivers effect at HERMES was found to be somewhat larger compared to that measured at COMPASS (see Fig.~\ref{fig-cmps-hms-siv}).
This observation may hint to the influence of TMD evolution effects. Present models predict for increasing $Q^2$ a mild dependence of the Sivers asymmetry when parton model approximation and  DGLAP evolution is used  and a strong or weak decrease for different TMD evolution schemes, see for instance Refs.~\cite{Anselmino:2012aa,Aybat:2011ta, Echevarria:2014xaa}. There is no theoretically compelling argument to use DGLAP evolution for TMDs, only the small-$b$ expansion \cite{Collins:2011zzd} of TMDs may be related to collinear functions that obey DGLAP evolution. TMD evolution \cite{Collins:2011zzd,Aybat:2011ge} should be quite different from DGLAP. COMPASS recently performed the first multi-differential analysis of the transverse-spin-dependent asymmetries extracting them from SIDIS data at four-five different hard scales~\cite{Adolph:2016dvl,Parsamyan:2015dfa,Parsamyan:2015nnl}. Extracted $Q^2$-dependences of the Sivers SSAs in different bins of $x$ were fitted with a linear decreasing function and a constant with a slight statistical preference for the former case.
Evolution properties of TMDs and in particular the Sivers TMD, were predicted to be very different from regular PDFs~\cite{Aybat:2011zv}.
Studies of evolution of Sivers TMD require precision measurements in different ranges of $Q^2$. A projections for $Q^2$-dependence of the Sivers effect expected from CLAS12 is shown in Fig.~\ref{fig-comp-sivers-q2}. The asymmetry, however, as other observables
which are constructed by taking ratios, are not ideal grounds for the study of TMD evolution effects, as it has additional modulations coming from the unpolarized part, making interpretation more complicated. Due to partial cancellation of evolution effects in numerator and denominator, the asymmetries themselves may exhibit only a weak $Q^2$-dependence.
It was suggested that more effort should be made towards measuring properly normalized SIDIS and $e^+e^-$, and Drell-Yan cross sections (both unpolarized and polarized).

\begin{figure}[ht!]
\centering
\includegraphics[width=0.53\textwidth]{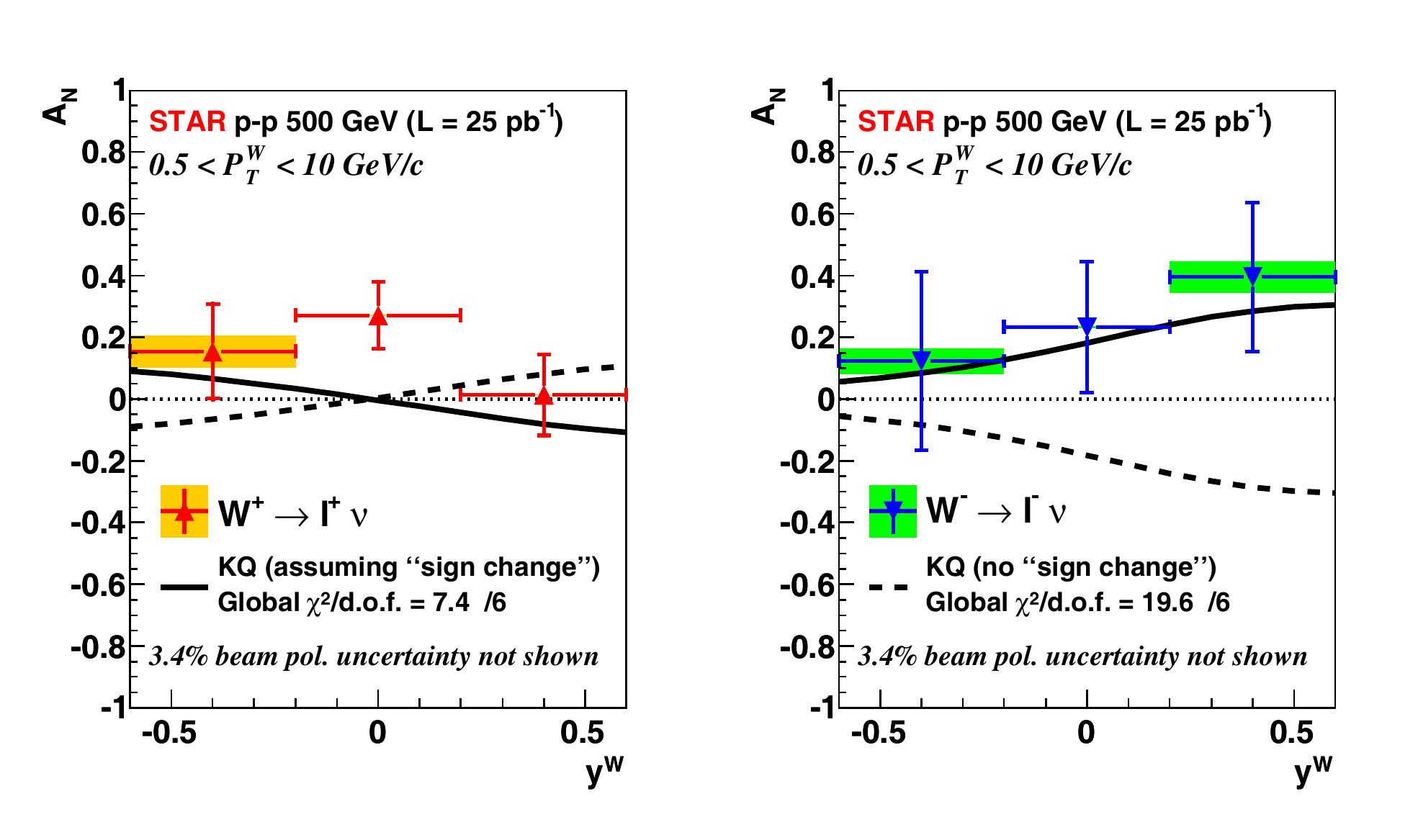}
\includegraphics[width=0.43\textwidth]{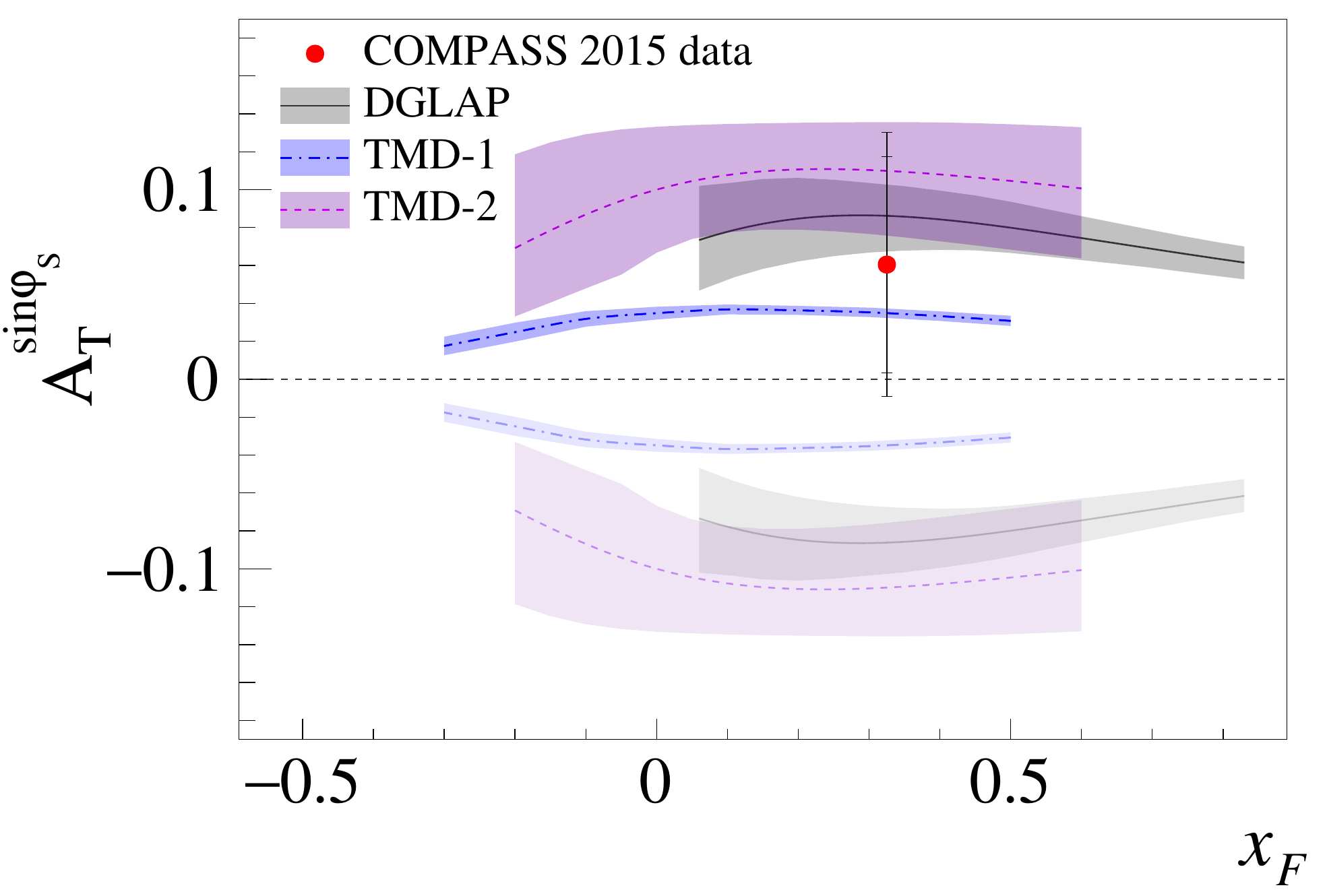}
\caption{Transverse single-spin asymmetry amplitude for $W^{+}$ (left panel) and $W^{-}$ (central panel) from STAR~\cite{Adamczyk:2015gyk} and Sivers asymmetry in Drell-Yan from COMPASS~\cite{Aghasyan:2017jop} (right panel)}.
\label{star-cmps}
\end{figure}

The first measurement of the Sivers effect in $W$ and $Z$-boson production in ${p^\uparrow} \, p \to W^\pm/Z^0 \, X$ reactions at RHIC was reported by the STAR collaboration~\cite{Adamczyk:2015gyk}, while COMPASS has recently published first ever results for Sivers asymmetry measured in the pion-induced Drell-Yan lepton-pair production off a transversely polarized proton~\cite{Aghasyan:2017jop,Parsamyan:2018zju,Parsamyan:2018evv}, see Fig.~\ref{star-cmps}. Both measurements were found to be consistent with the hypothesis of predicted change of sign for the Sivers function, but the accuracy was not enough to give a conclusive answer. Soon more precise data is expected to come from both experiments.
%
%

\begin{figure}[ht!]
\centering
\includegraphics[width=0.42\textwidth]{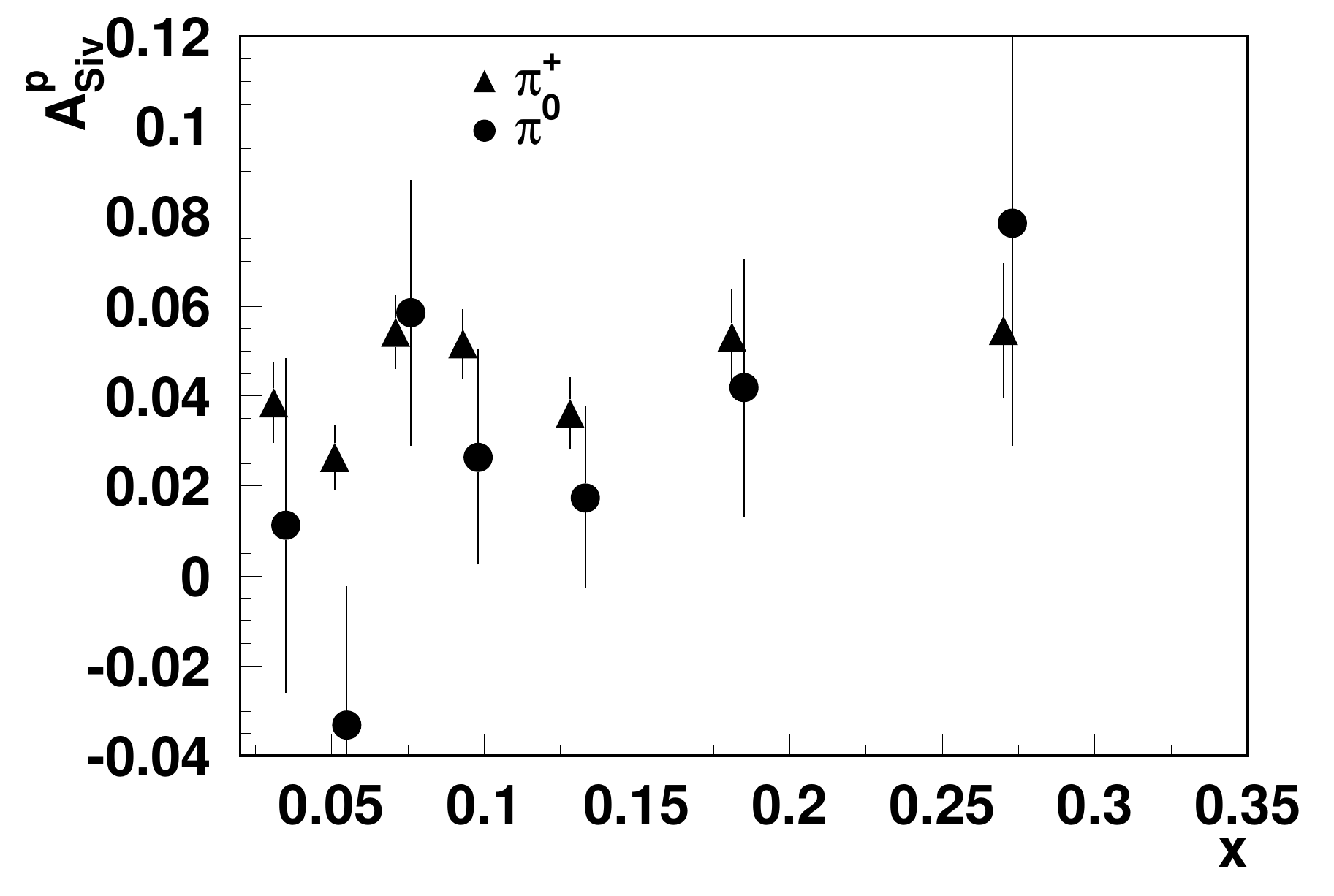}
\hspace{0.5cm}
\includegraphics[width=0.42\textwidth]{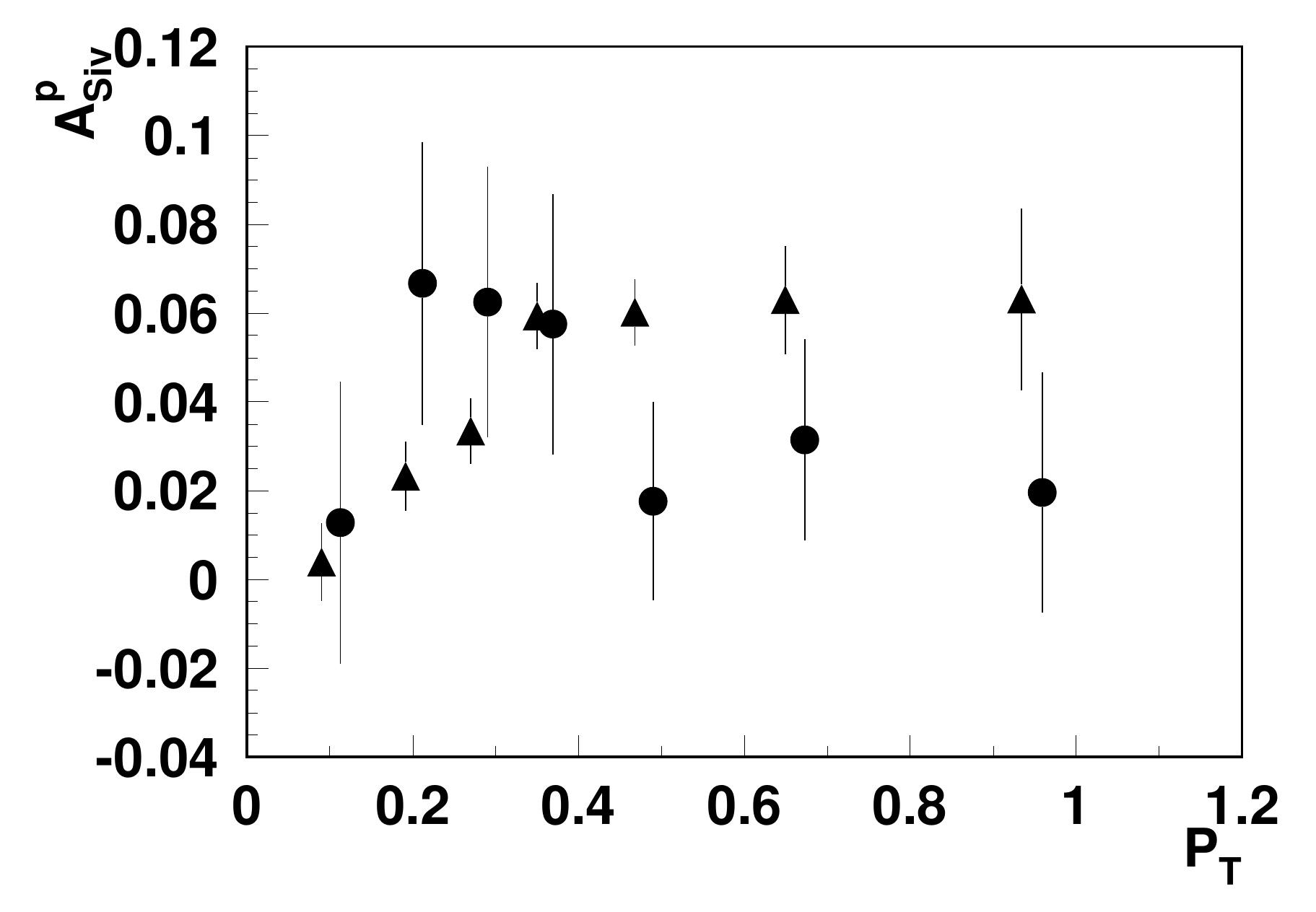}
\caption{Comparison of Sivers asymmetries, $A_{UT}^{\sin(\phi-\phi_S)}$ for neutral and positive pions measured by HERMES Collaboration~\cite{Airapetian:2009ae}.}
\label{fig-comp-sivers}
\end{figure}

The Sivers asymmetry for $\pi^+$ and $\pi^0$  productions in SIDIS appeared to be very close to each other (see Fig.\ref{fig-comp-sivers}.), a feature showing up also for many kinds of higher twist modulations.
Measurements of SSAs at JLab, performed with transversely polarised $^3$He
\cite{Qian:2011py,Huang:2011bc,Zhao:2014qvx,Zhang:2013dow,Zhao:2015wva}, indicate that spin
orbit correlations may be significant for certain combinations of spins of
quarks and nucleons and transverse momentum of scattered quarks.
%
%

\begin{figure}[ht!]
\begin{center}
\vspace{0.0cm}
\includegraphics[width=0.42\textwidth]{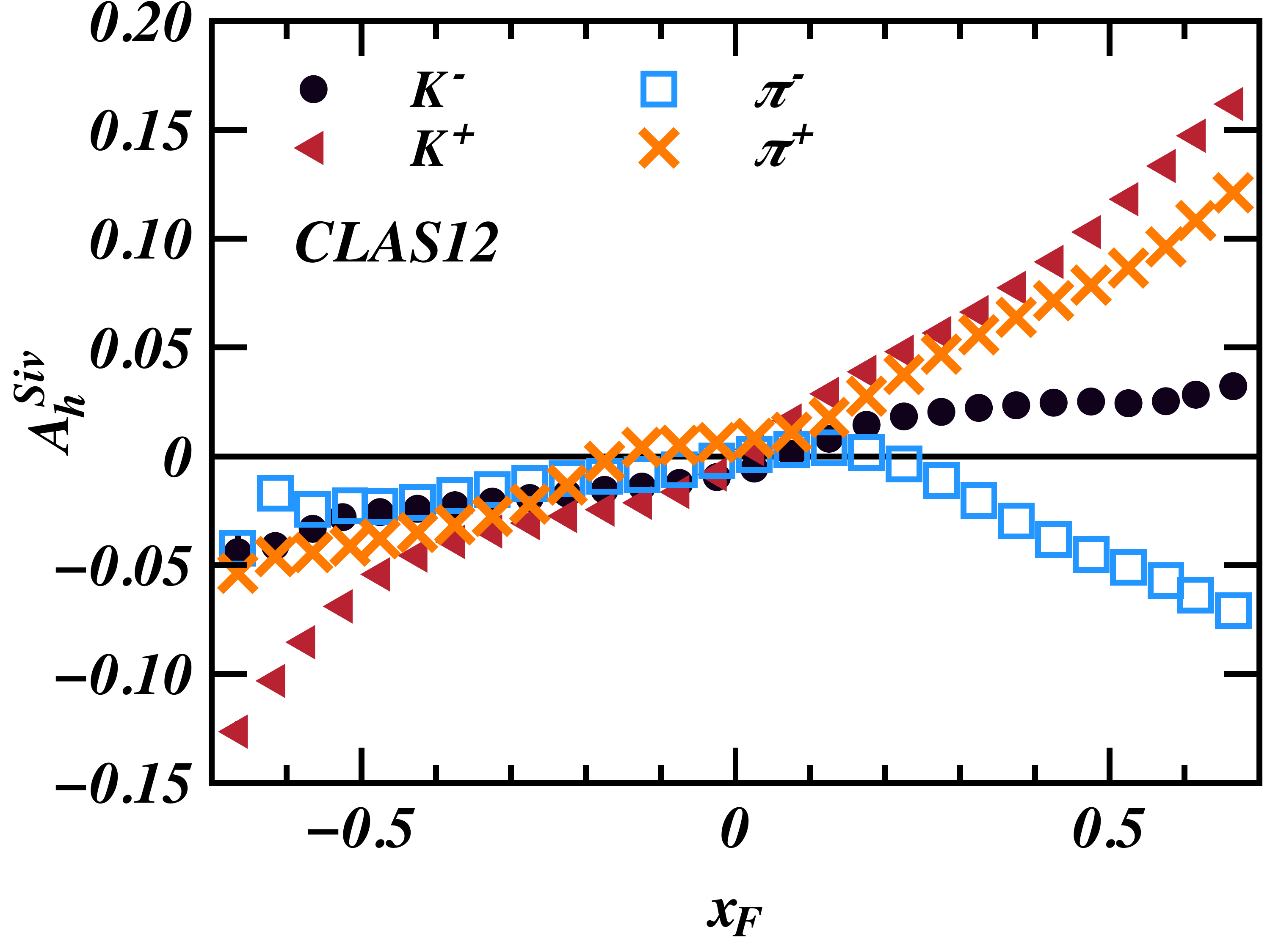}
\hspace{1.0cm}
\includegraphics[width=0.42\textwidth]{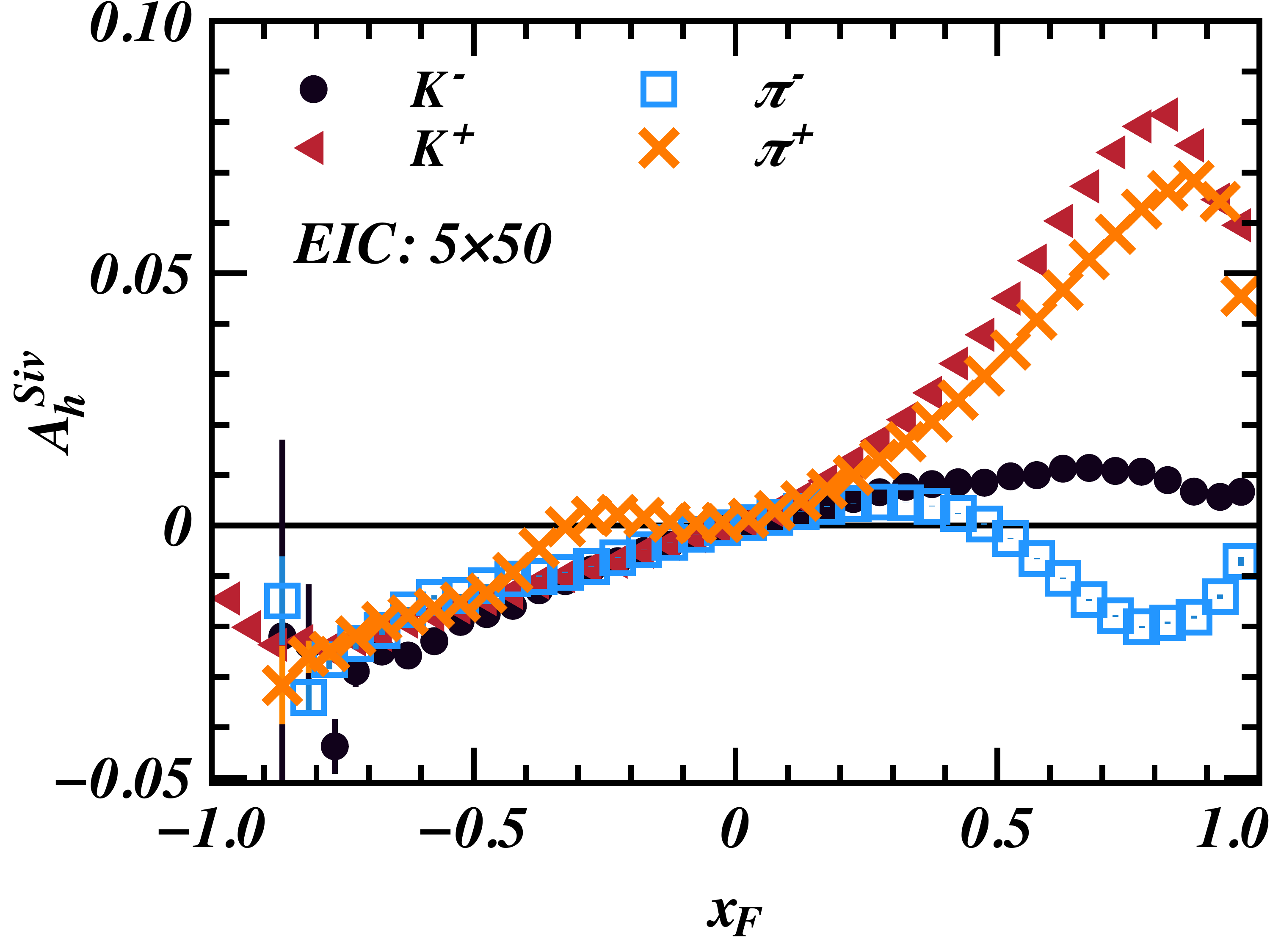}
\end{center}
\caption{\small Projections for Sivers asymmetry measurements as a function of $x_F$ at JLab (left) and future EIC (right) for positive pion
production, using 4 GeV electrons and 60 GeV protons (100 days at 10$^{34}$ cm$^{-2}$ sec$^{-1}$)~\cite{Matevosyan:2015gwa}.}
\label{fig-sivers-xf}
\end{figure}

The measurements of the SSAs for hadrons produced in the Target Fragmentation Region (TFR) will allow one to study the structure of
the nucleon through the fracture functions. These objects, though more complicated than the ordinary PDFs
and FFs, will provide important new information. An updated version of the PYTHIA, mPYTHIA was used to look at correlations between
different kinematical regions~\cite{Matevosyan:2015gwa}. Lund string model used in PYTHIA differs from the usual QCD factorized approach that describes the hadron production
in the Current Fragmentation Region (CFR) with a convolution of PDFs and FFs and in the TFR using two additional independent fracture functions.
Studies based on mPYTHIA, accounting only the correlation between the nucleon's transverse polarization and the transverse momentum of the struck
quark revealed sizable signal in the in TFR, comparable in size to that in the CFR (see Fig.~\ref{fig-sivers-xf}). Experimental measurements
of Sivers SSAs in both CFR and TFR will be important to reveal underlying correlations.

\begin{figure}[ht!]
\begin{center}
\vspace{0.0cm}
\includegraphics[width=0.42\textwidth]{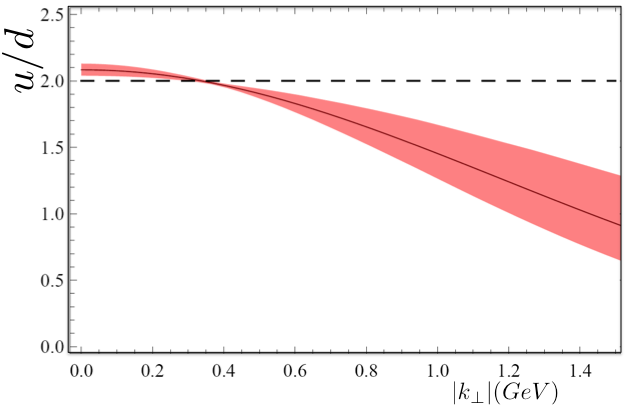}
\hspace{0.5cm}
\includegraphics[width=0.415\textwidth]{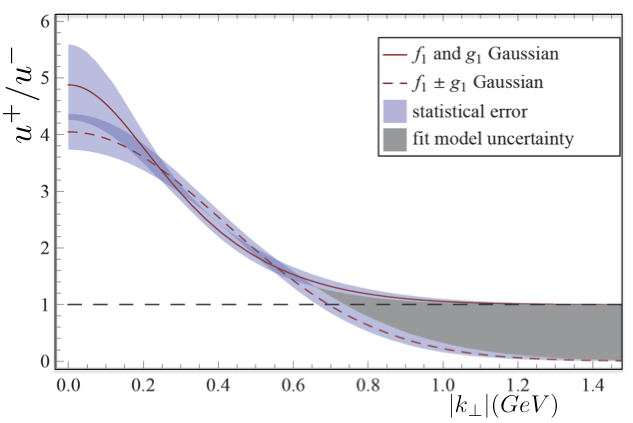}
\end{center}
\caption{Lattice calculations for $k_T$-dependence of ratios of $u/d$ quark distributions (left) and $u^+/u^-$-distributions (right)~\cite{Musch:2010ka}.}
\label{fig-lattice}
\end{figure}

\begin{figure}[ht!]
\centering
\includegraphics[width=0.6\textwidth]{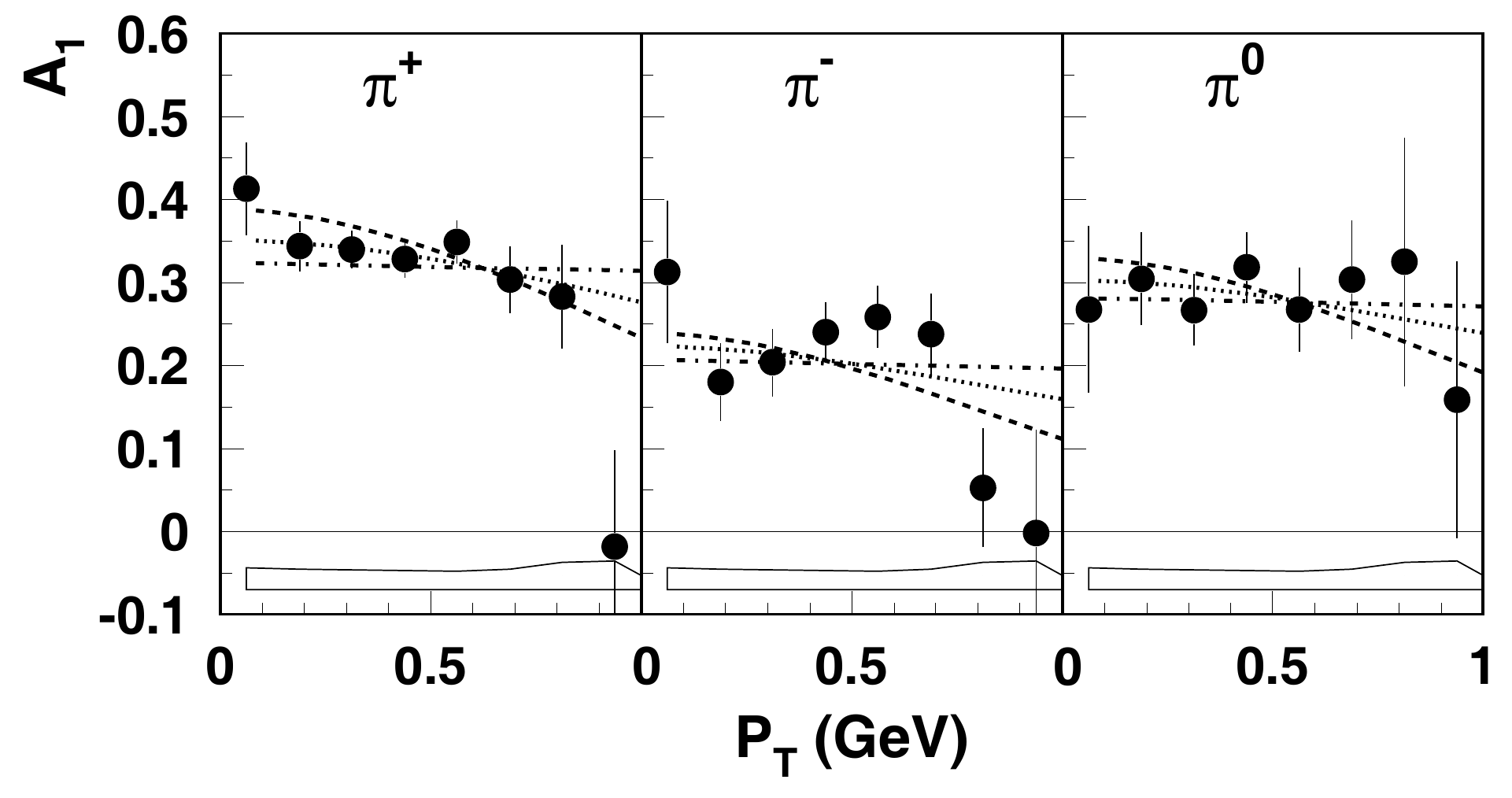}
\caption{\small
The double spin asymmetry $A_{1}$
 as a function of transverse momentum of hadrons, $\Phperp$,
averaged in the $0.4<z<0.7$ range. The empty band corresponds to systematic uncertainties. Three curves are calculations
for different transverse momentum widths (0.1, 0.17, 0.25) of $g_1$ at
a fixed width for $f_1$ (0.25)~\cite{Anselmino:2006yc}
}
 \label{a1pptdepvszq47x16}
\end{figure}

\begin{figure}[ht!]
\includegraphics[width=0.48\textwidth]{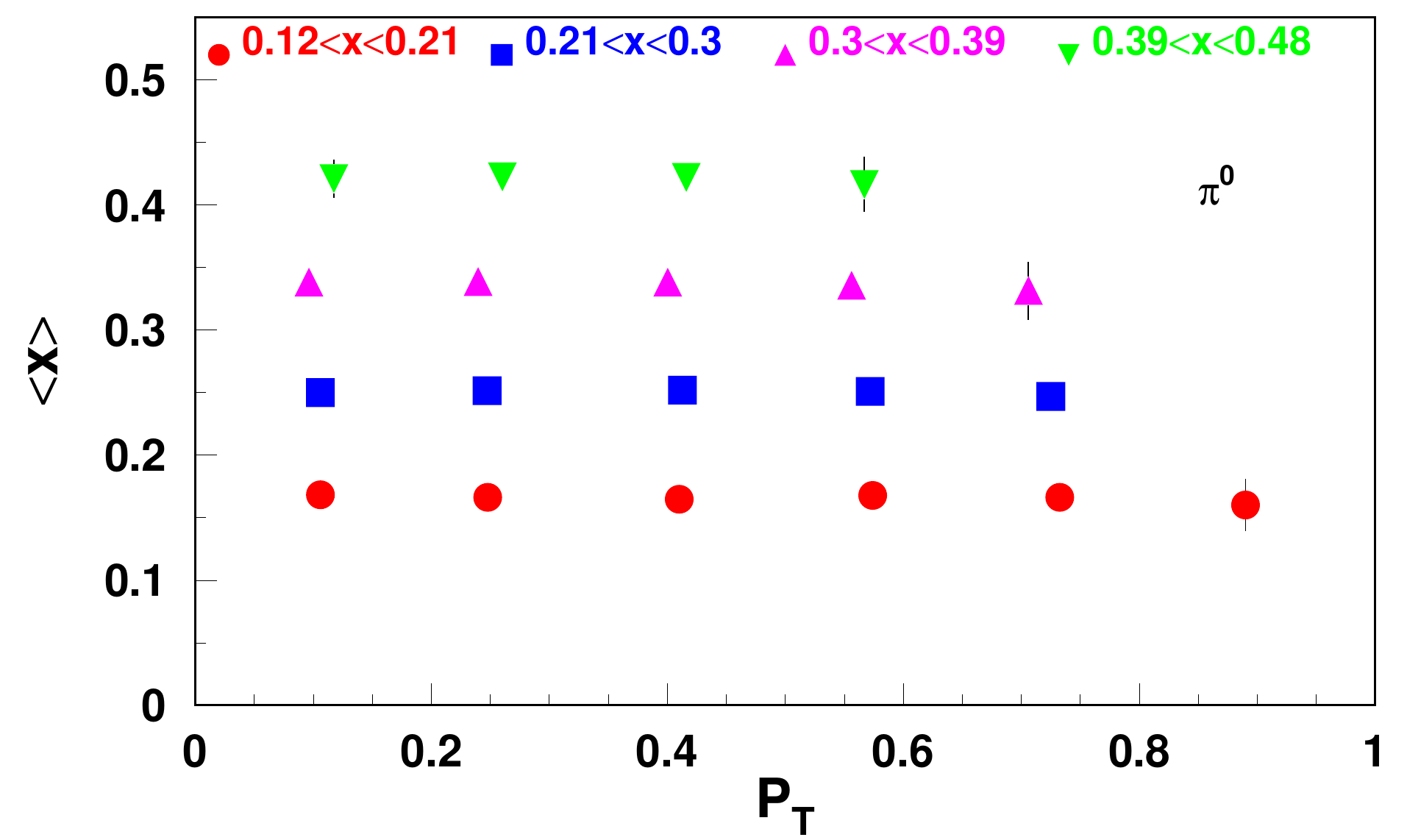}
\includegraphics[width=0.48\textwidth]{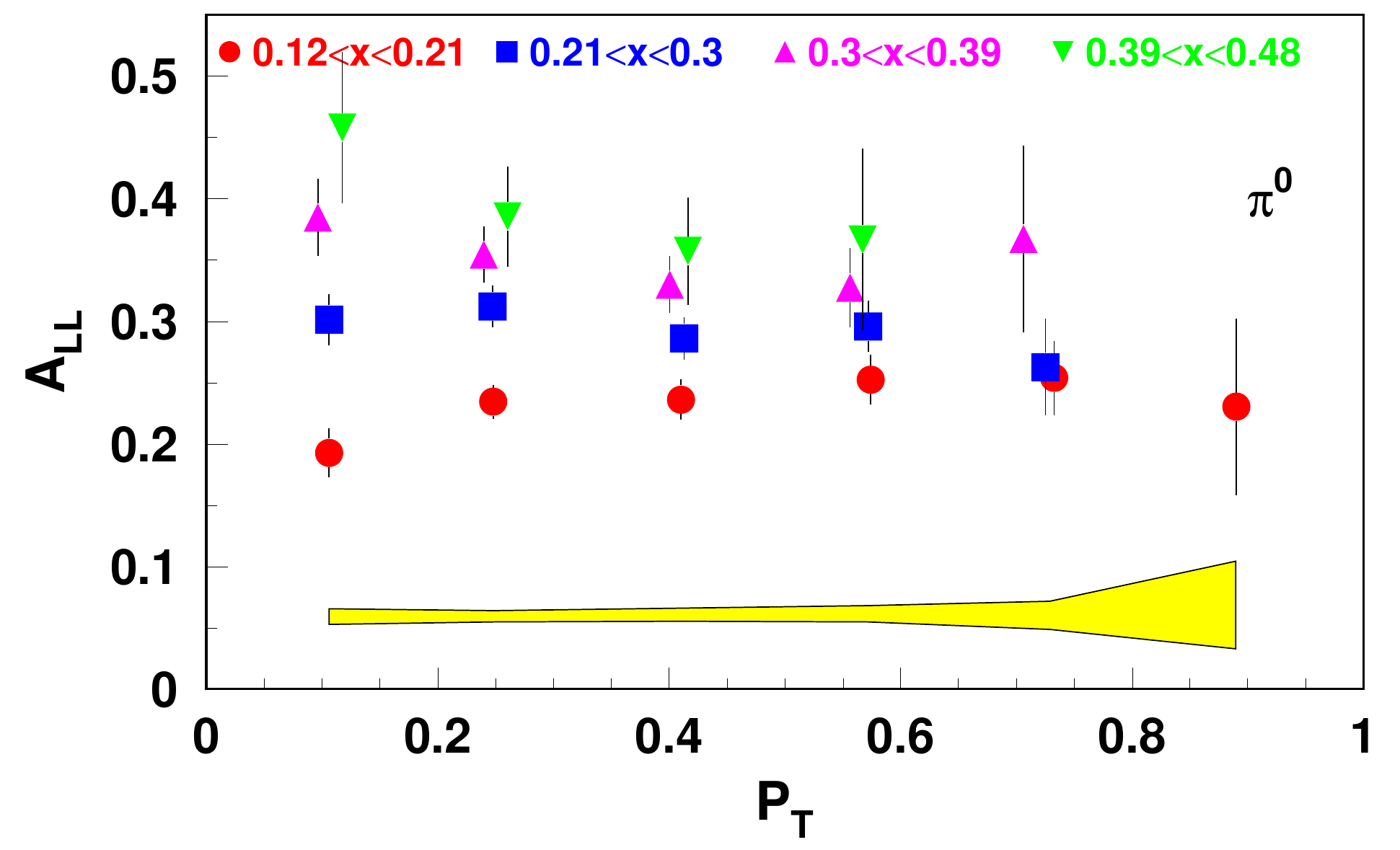}
\caption{The average $x$ versus $\Phperp$ for $ep\rightarrow e^\prime\pi^0X$ in CLAS kinematics (left), and the double spin asymmetry in semi-inclusive production of $\pi^0$ for different bins in $x$~\cite{Jawalkar:2017ube}.}
\label{fig-pi0all-pt-dep}
\end{figure}

\begin{figure}[ht!]
\includegraphics[width=0.48\textwidth]{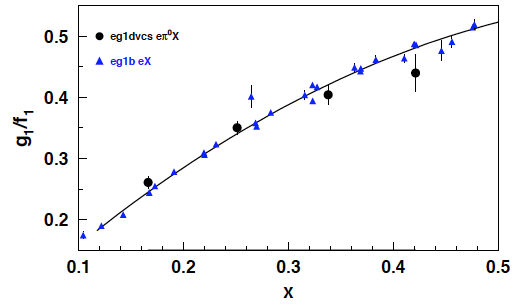}
\includegraphics[width=0.48\textwidth]{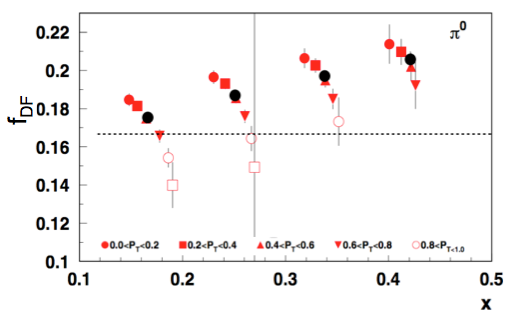}
\caption{The $A_{1}$ DSA in semi-inclusive production of $\pi^0$, compared to inclusive double spin asymmetry~\cite{Jawalkar:2017ube} (left) and dilution factor as a function of $x$ for different bins in $\Phperp$ (right).}
\label{fig-pi0all-x-dep}
\end{figure}

There have been many studies dedicated to model calculations of TMDs, see for
example~\cite{Ji:2002aa,Brodsky:2002cx,Brodsky:2002rv,Bacchetta:2003rz,Gamberg:2003ey,Meissner:2007rx,Gamberg:2007wm,Bacchetta:2007wc,Bacchetta:2008af,Gamberg:2008yt,Schweitzer:2010tt,Avakian:2010br,Boffi:2009sh,Pasquini:2010af,Pasquini:2008ax,Pasquini:2011tk,Pasquini:2014ppa}.
These models and calculations of asymmetries based on them could play a very important role as a first
step of description of the experimental observations, to give an intuitive way to
connect the physical observables to the dynamics of partons, and to provide key
inputs to unravel the partonic structure of the nucleon. Models provide clear way of
addressing fundamental questions, such as how the quark spin and its orbital
angular momentum contribute to the nucleon spin. Even though models do not contain full QCD dynamics, one may gain insight on full QCD by examining models. In addition, very exciting
results of TMDs have come from lattice QCD
calculations~\cite{Hagler:2009mb,Musch:2009ku,Musch:2010ka}, indicating, for instance, that
spin-orbit correlations could change the transverse momentum distributions of
partons. Lattice calculations suggested that transverse momentum distributions depend both on
flavor and the spin orientation of quarks (see Fig.\ref{fig-lattice}).
Measurements of the $\Phperp$-dependence of the double spin asymmetry $A_1$, performed at JLab, with longitudinally polarized NH$_3$ target~\cite{Avakian:2010ae}, suggest that widths of partonic distributions may indeed depend on the spin orientation (see Fig.\ref{a1pptdepvszq47x16}). The $\Phperp$-dependence of the $A_1$ DSA for positive and negative hadron productions measured recently by COMPASS~\cite{Parsamyan:2018ovx,Parsamyan:2018evv} and HERMES~\cite{Airapetian:2018rlq} appeared to be well compatible with a constant function.  This could indicate that transverse momentum widths of $g_1$ and $f_1$ are the same~\cite{Anselmino:2006yc} in the kinematics not dominated by valence quarks. The possible correlation between the $x$ and $\Phperp$ of the hadron in real experiments is one of the important issues to address in that kind of measurements. Such correlation tends to be much weaker for neutral pions.
The new set of measurements with much higher precision for neutral pions performed by
the CLAS collaboration supported the observed complex dependence of the double spin asymmetry
on the transverse momentum (see Fig.~\ref{fig-pi0all-pt-dep}). An important advantage of the $\pi^0$ data is the better uniformity and smaller variations of averages of $\Phperp$ with $x$ due to correlations between longitudinal and transverse momentum of quarks and hadrons.
Measurements performed with polarized nuclear targets (NH$^3$), require detailed account of significant nuclear background, and very careful treatment is needed to estimate the dilution factor, which defines the fraction of events originating from polarized quasi free protons (Fig.~\ref{fig-pi0all-x-dep}).
The double-spin asymmetries in DIS and $\pi^0$ SIDIS, in simple parton model, at large $x$, where the sea contribution is negligible, are expected to be roughly the same. CLAS measurements of both asymmetries indicate that already at 6 GeV, they are in good agreement (See Fig.~\ref{fig-pi0all-x-dep}).


\begin{figure}[ht!]
\begin{center}
\includegraphics[width=0.55\textwidth]{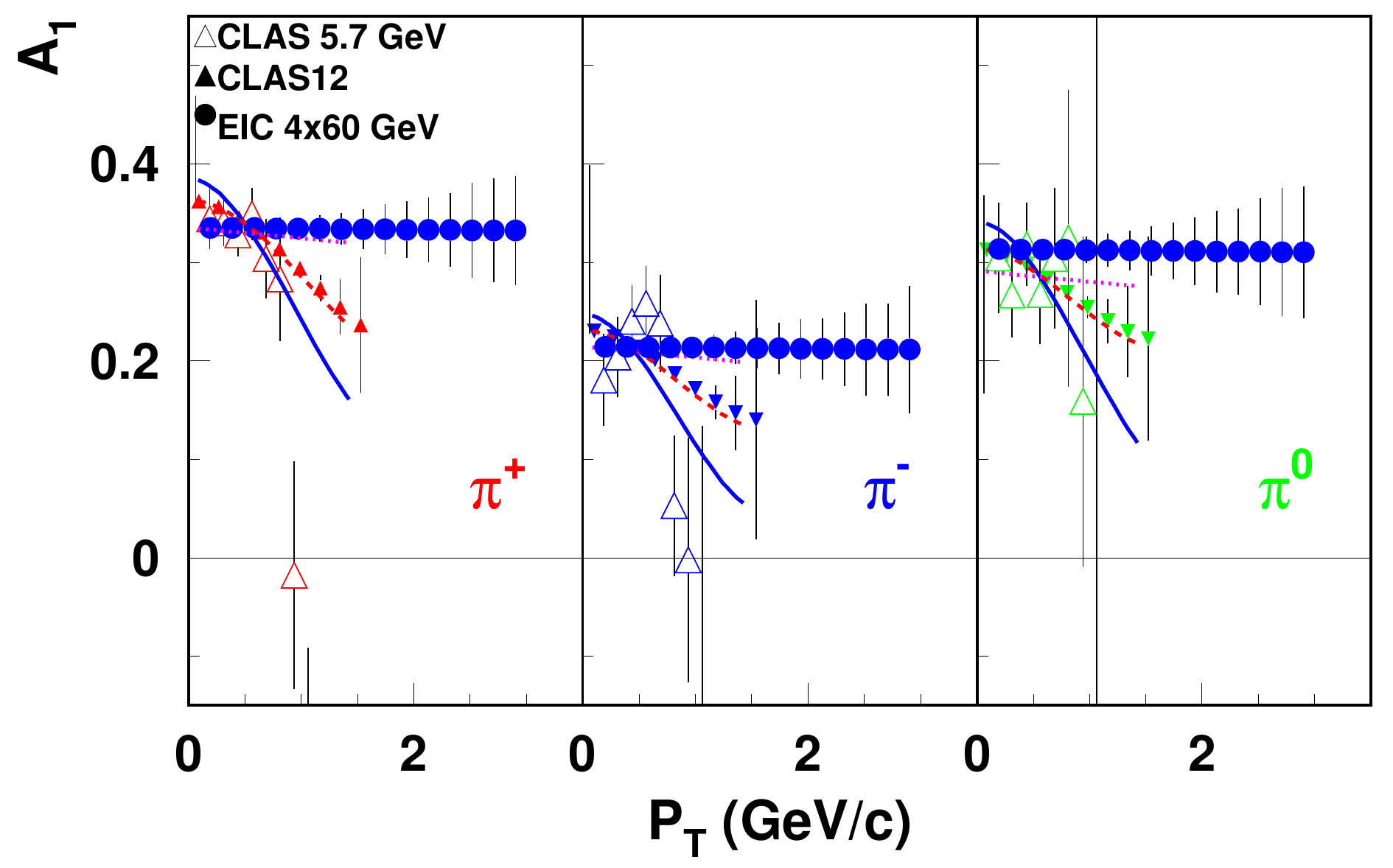} \hfill
\end{center}
\caption{Double-spin asymmetry, $A_{1}$, for pion production, using the
 EIC\cite{Accardi:2012qut} configuration with 4 GeV electrons and 60 GeV
 protons (100 days at 10$^{34}$ cm$^{-2}$ sec$^{-1}$), as a function of
 $\Phperp$, compared to published data from CLAS~\cite{Avakian:2010ae} and
 projected CLAS12 measurements~\cite{E12-09-009}. }
\label{Allpt_dep}
\end{figure}

Precision measurements using the upgraded CLAS12 detector with polarised
NH$_3$ and ND$_3$ targets will allow to access the $\kperp$-distributions of $u$
and $d$-quarks aligned and anti-aligned with the spin of the nucleon.
Projections for the resulting $\Phperp$-dependence of the double spin
asymmetries for all three pions are shown in Fig.~\ref{Allpt_dep} for a NH$_3$
target~\cite{E12-07-107,E12-09-009}. Integrated over transverse momentum, the
data will also be used to extract the $\kperp$-integrated standard PDFs.
Two proposals have been approved to study SSAs with longitudinally polarised target using
SoLID detector~\cite{E12-11-007} and Super-Bigbite spectrometer with polarised
$^3$He targets. The later one with kaon identification using a RICH detector.

The $A_{LT}^{\cos(\phi_h-\phi_S)}$ DSA provides access to the convolution of the ordinary unpolarized FF with the $g_{1T}$ TMD PDF which parameterizes the distribution of longitudinally polarized quarks in a transversely polarized nucleon:
\begin{eqnarray}
F_{LT}^{\rm cos(\phi_h-\phi_S)}\xzpt & = & {\cal C} \bigg[\frac{\bfhp\cdot\kt}{M_N} g_{1T}\xkts D_1\zpts \bigg].
\end{eqnarray}

\begin{figure}[ht!]
\centering
\includegraphics[width=0.8\textwidth]{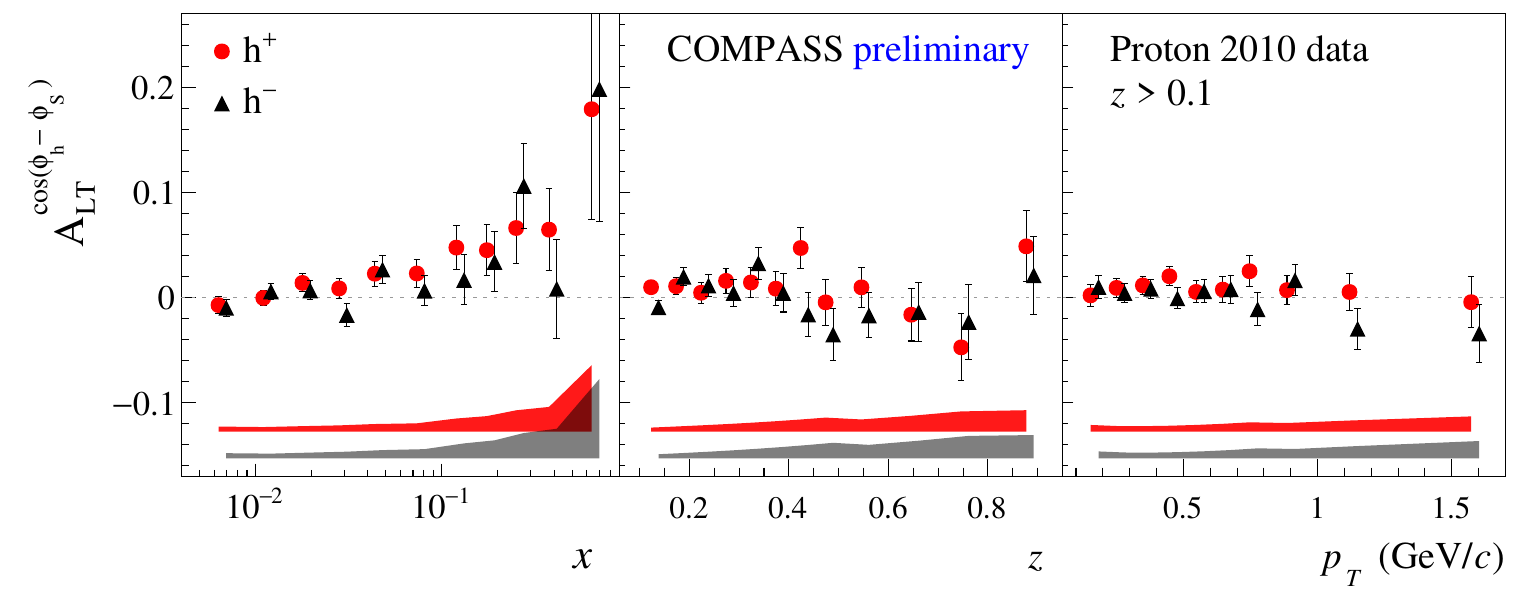}
\caption{The $A_{LT}^{\cos (\phiH -\phiS )}$ asymmetry extracted by COMPASS~\cite{Adolph:2016dvl,Parsamyan:2018evv}. Colored horizontal bands represent the systematic uncertainties.}
\label{fig:ALT_cmps}    
\end{figure}

\begin{figure}[ht!]
\centering
\includegraphics[width=0.7\textwidth]{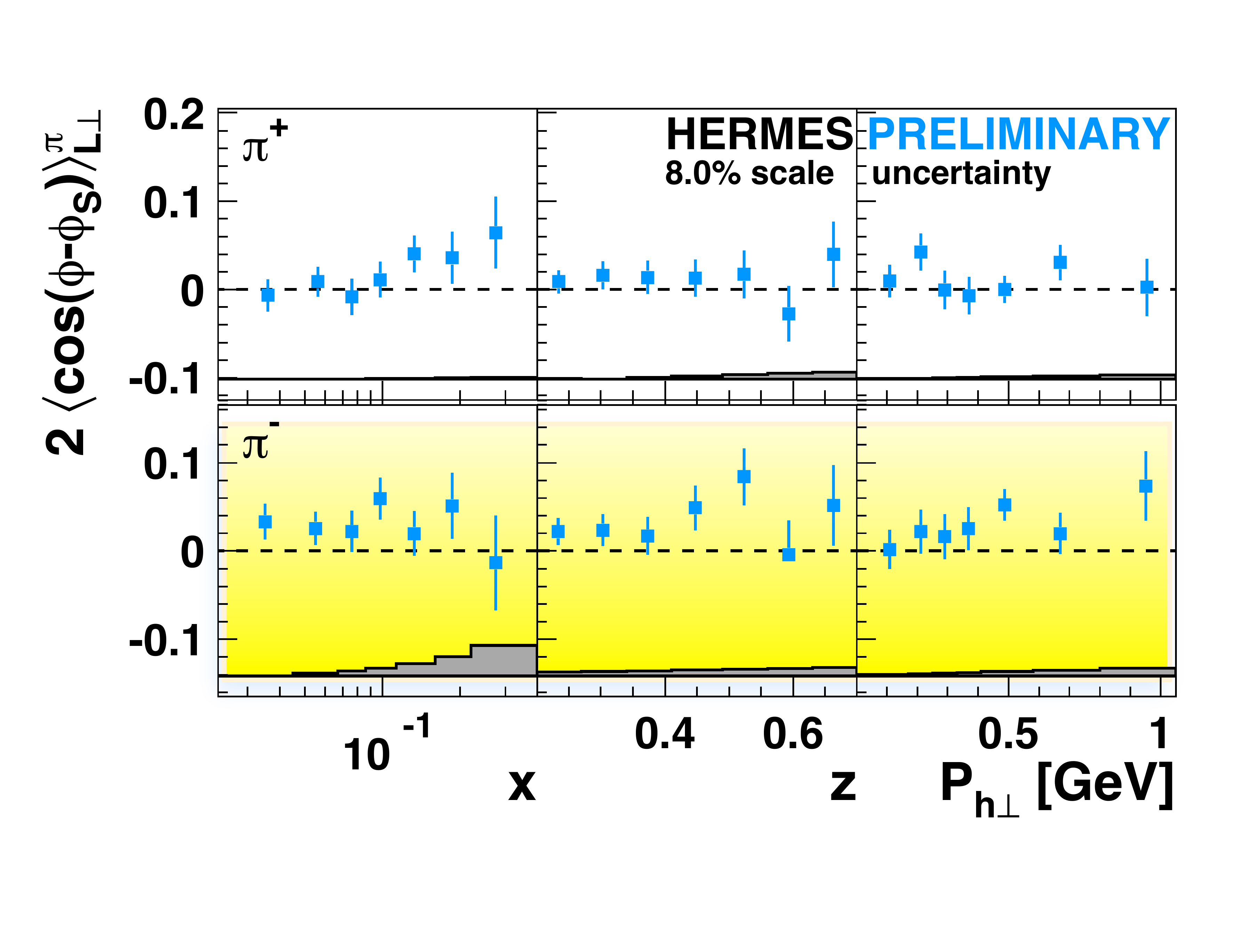}
\caption{The preliminary $A_{LT}^{\cos (\phiH -\phiS )}$ asymmetry extracted by HERMES~\cite{Pappalardo:2010zza}. Horizontal bands represent the systematic uncertainties.}
\label{fig:ALT_hms}    
\end{figure}

The $g_{1T}$ \textit{twist-2} chiral-even TMD PDF is the imaginary part of the interference terms between S and P wave components. It can be linked (through Lorentz invariance relation) to the \textit{twist-3} $g_2(x)$ PDF, which in its turn can be linked to \textit{twist-2} helicity PDF $g_1(x)$ using the so-called Wandzura-Wilczeck approximation~\cite{Kotzinian:2006dw}.
Measurements of the ${\cos(\phi_h-\phi_S)}$ amplitude by COMPASS and HERMES collaborations indicate, also that correlation may be significant at large $x$ (see Figs.~\ref{fig:ALT_cmps},\ref{fig:ALT_hms}).

In particular at COMPASS, for positive hadrons, the asymmetry is clearly positive increasing up to $\approx0.1$ in the interval of relatively large $x$ ($x>0.01$). For negative hadrons the effect is less prominent due to larger statistical uncertainties. Observed behaviour and the magnitude of the effect are well in agreement with the available model calculations~\cite{Kotzinian:2006dw} and with observations made by HERMES.
This modulation will be also measured by CLAS12 in the valence region with beam energy 10.6 GeV and transversely polarized HD-Ice target~\cite{C12-11-111}.

\section{Higher twist observables}
\label{Sec:highertwist}
Twist-3 TMDs, shown in Table.~\ref{tab:TMD-tables}, contribute to various observables in SIDIS. They describe quantum mechanical quark-gluon correlation
functions and thus do not have simple partonic interpretation as probability densities.
It is interesting to notice that higher twist asymmetries, especially those not receiving contributions from leading twist structure functions, were measured and in most of the cases were found not only to be not compatible with zero, but very significant. Good examples are the $\cos\phi_h$ moment of the unpolarized cross section, $F_{UU}^{\cos \phi_h}$, first measured by the EMC collaboration, back in the 80's~\cite{Aubert:1983cz,Arneodo:1986cf}, the $\sin\phi_h$ moment depending on the longitudinally target polarization, $F_{UL}^{\sin\phi_h}$, measured by HERMES collaboration in the 90s~\cite{Avakian:1999rr,Airapetian:1999tv,Airapetian:2001eg,Airapetian:2002mf}, and the $\sin\phi_h$ moment depending on the longitudinal polarization of the beam, $F_{LU}^{\sin\phi_h}$, measured at JLab~\cite{Avakian:2003pk,Aghasyan:2011ha,Gohn:2014zbz}. All those measurements were repeated and confirmed later by HERMES, COMPASS and JLab.

The gluon radiation recoil in SIDIS was predicted to lead to observable $\cos\phi_h$ modulations of the SIDIS cross section,
 and this effect was proposed by Georgi and Politzer in the 70s~\cite{Georgi:1977tv} as a clean test of perturbative QCD.
The gluon radiation indeed leads to azimuthal
dependencies in the semi-inclusive DIS cross section, but its contribution is significant
mainly at large values of $\Phperp$.

In 1978, Cahn~\cite{Cahn:1978se}  discussed origin of $\cos\phi_h$ modulation arising from non-zero intrinsic transverse momenta of partons. Although, suppressed by $\Phperp/Q$ that modulation (known as the Cahn effect),
appeared to be significant and dominating in the $\Phperp \sim $1 GeV range. The same mechanism contributes also to the $\cos 2\phi_h$ moments at twist-4 level. Significant azimuthal
modulations observed in various experiments indicate the importance of high twist effects. Additional contribution to $\cos \phi_h$ and
$\cos 2\phi_h$ moments could come from processes when the final meson
is produced at short distances via hard-gluon exchange, as proposed by Berger in 1980~\cite{Berger:1979xz}, and may also be
significant in the kinematic regime where the ejected meson carries most of
the virtual photon momentum. It appeared that the interplay between the parton transverse
momentum and spin, the so-called Boer-Mulders effect~\cite{Boer:1997nt}, in addition to leading-twist
contribution to the $\cos 2\phi_h$ can also generate sub leading contribution to the $\cos\phi_h$ amplitude.

The cross-section modulation $F_{UU}^{\cos \phi_h}$ originates from contributions only
at sub-leading twist level and is suppressed by $M/Q$.
As far as the structure function $F^{\cos \ph}_{UU}$ is higher-twist structure function it can only be accessed at moderate values of $Q$. Higher-twist observables are a key for understanding
long-range quark-gluon dynamics, for instance
they can be interpreted in terms of
the average transverse forces acting on a quark after
it absorbs the virtual photon~\cite{Burkardt:2008vd}.

In order to simplify the discussion, one can use the so-called Wandzura-Wilczeck-type (WW-type) approximation neglecting all quark-gluon-quark correlators.
Generically one can decompose higher-twist TMDs into leading-twist terms, current-quark mass terms and the so-called pure interaction-dependent (“tilde”) terms. This is accomplished by employing
equations of motion (EOM) and reveals that tilde-terms are not probability
densities but quark-gluon correlation functions.
Neglecting the tilde- and mass terms is sometimes referred
to as Wandzura-Wilczek approximation~\cite{Wandzura:1977qf}.
This step can be helpful in phenomenology to disentangle the
many contributions to twist-3 SIDIS
observables~\cite{DeSanctis:2000fh,Ma:2000ip,Efremov:2001cz,Efremov:2002ut},
and can in certain cases be a numerically useful
approximation~\cite{Avakian:2007mv,Accardi:2009au}.
Recently the authors of Ref~\cite{Bastami:2018xqd} have performed a comprehensive phenomenological study of the cross-section for the production of unpolarized hadrons in SIDIS, computing all twist-2 and twist-3 structure functions within Wandzura-Wilczek-type approximations and compared calculations to the existing experimental data.
For the $F_{UU}^{\cos \phi_h}$ structure function one obtains the following result: %
\begin{eqnarray} 
\label{Eq:cosphi}
F_{UU}^{\cos \phi_h} & \simeq & \frac{2M}{Q}\ C\left[-\frac{\bfhp\cdot\kt}{M_N} xf^\perp D_1 +
\frac{\bfhp\cdot\pt}{z m_h}
xh H_1^\perp \right]
\end{eqnarray}
where the first term is related to the Cahn effect~\cite{Cahn:1978se}, the second term, strictly speaking is 0 due to the sum rule~\cite{Bacchetta:2006tn}
\begin{eqnarray}
	x\,h^q(x) = 0\,.
\end{eqnarray}

\begin{figure}[ht!]
\centering
\includegraphics[width=0.75\textwidth]{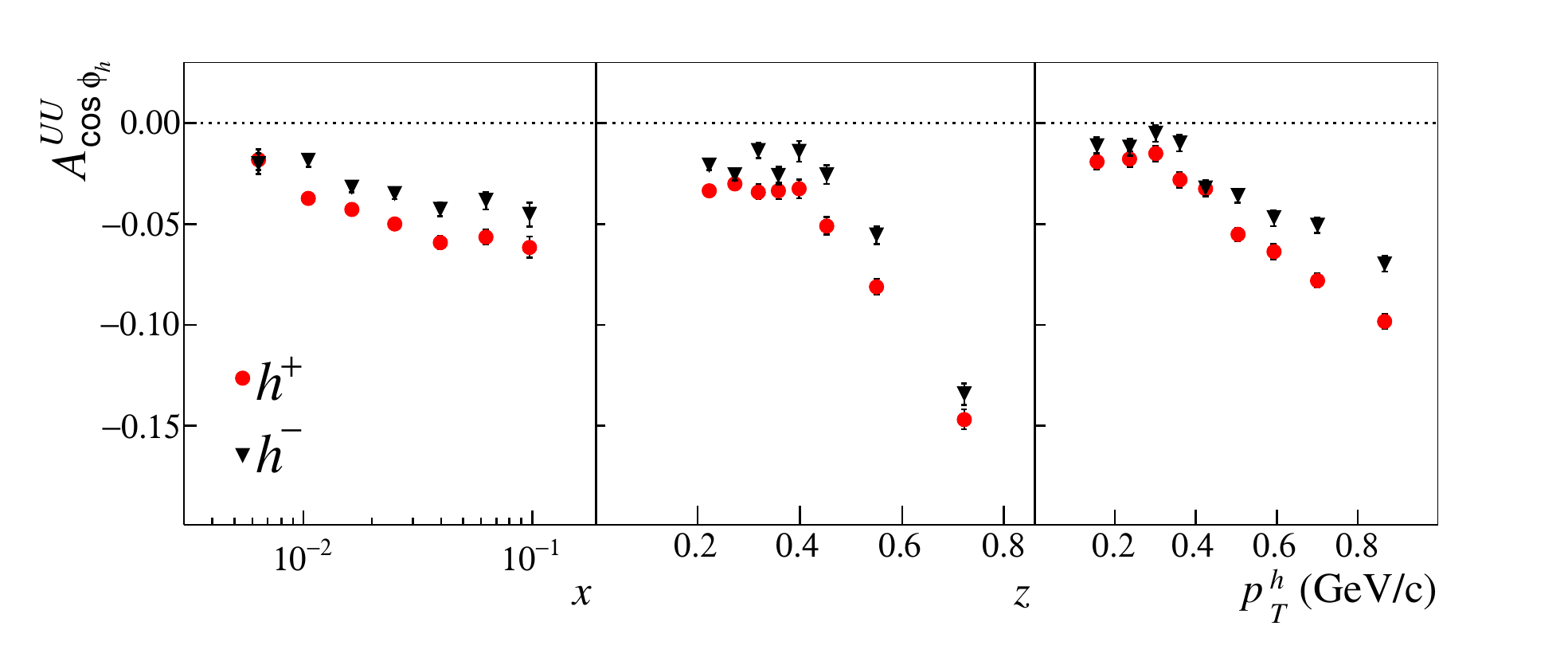}
\caption{The $A_{UU}^{\cos(\phiH)}$ asymmetry for positive and negative hadron productions as extracted by COMPASS~\cite{Adolph:2014pwc}.}
\label{fig:Cahn_cmps}
\end{figure}

Several measurements of $\cos \phi_h$ and $\cos 2\phi_h$ modulations in SIDIS experiments
has been published in the
past~\cite{Aubert:1983cz,Arneodo:1986cf,Adams:1993hs,Breitweg:2000qh}.
The CLAS collaboration measured non-zero cosine modulations for positive pions produced
by $6$ GeV/c electrons scattering off the proton~\cite{Osipenko:2008rv}. The
HERMES experiment have measured cosine modulations of hadrons produced in the
scattering of $27.5$ GeV/c electrons and positrons off pure hydrogen and deuterium
targets, where the lepton beam scatters directly off neutrons and protons (with
only negligible nuclear effects in case of
deuterium)~\cite{Airapetian:2012yg}. These modulations were
determined in a four-dimensional kinematic space for positively and negatively
charged pions and kaons separately, as well as for unidentified hadrons. At
COMPASS, positive and negative hadrons produced by the 160 GeV/c muon beam
scattering off a $\rm ^6LiD$ target have been measured in a three-dimensional
grid of the relevant kinematic variables $x$, $z$ and
$\Phperp$~\cite{Adolph:2014pwc}. In Fig.~\ref{fig:Cahn_cmps} COMPASS results are presented in one-dimensional representation, \textit{i.e.} as a function of $x$, $z$ or $P_{hT}$, while integrating over the other variables.

\begin{figure}[ht!]
\begin{center}
\includegraphics[width=0.46\textwidth]{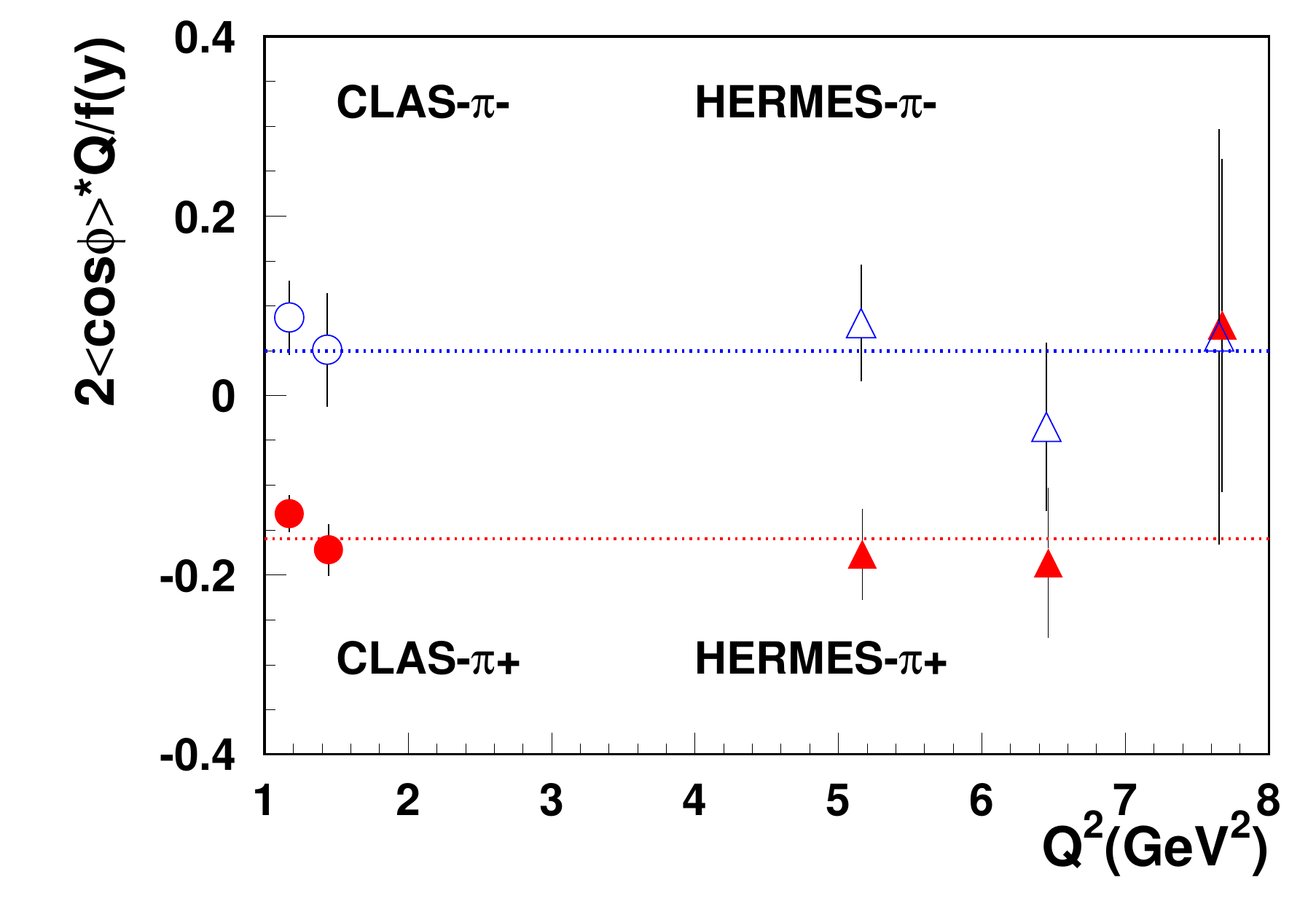}
\includegraphics[width=0.45\textwidth]{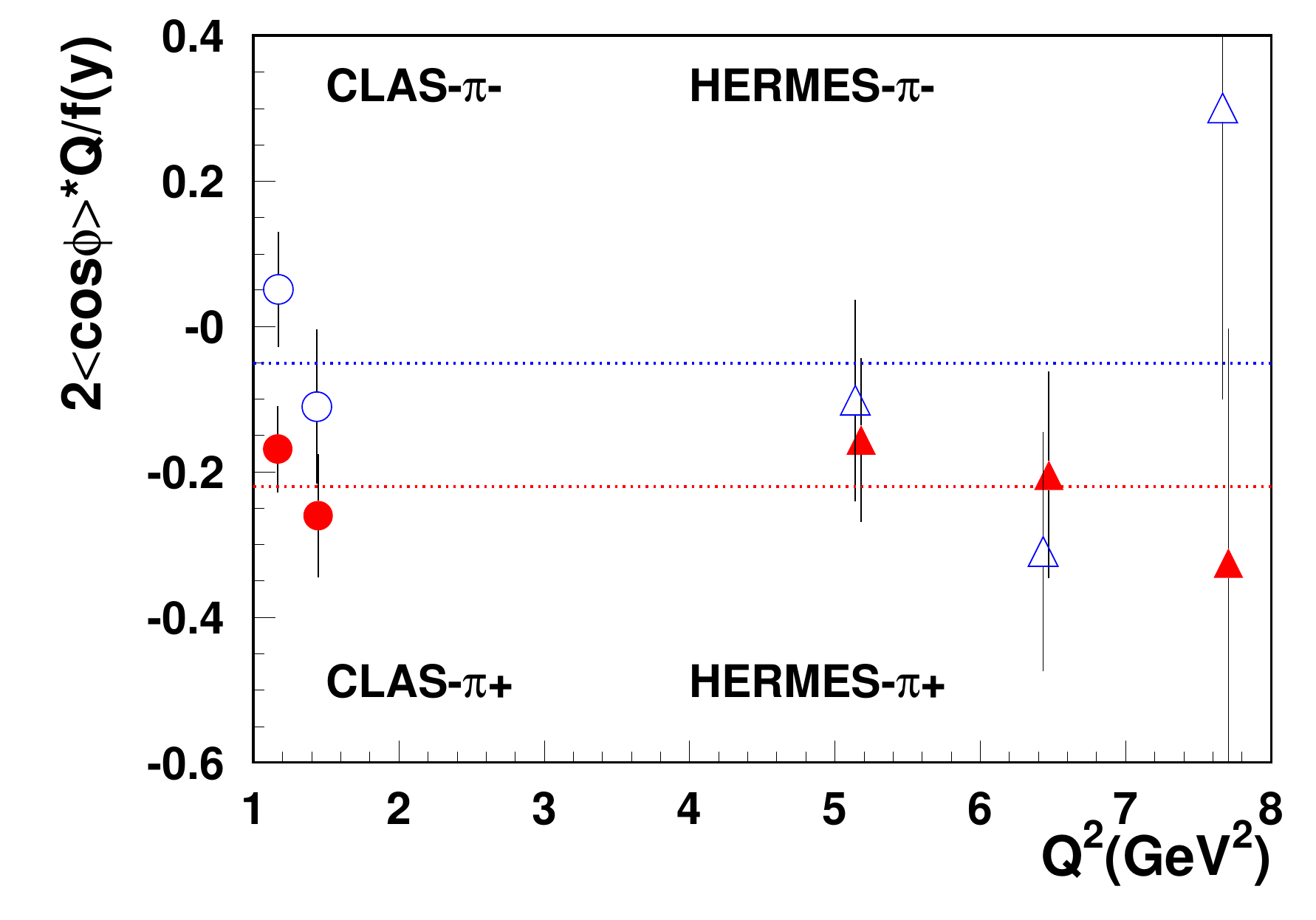}
\end{center}
\caption{(Color online) The $\cos\phi_h$ modulations in $\pi^\pm$ SIDIS plotted vs $Q^2$ for a bins with $x \approx 0.19$, and
$z\approx 0.35, \Phperp\approx 0.6$ GeV (left) and $z\approx 0.43, \Phperp\approx 0.8$ GeV (right).
Filled symbols are for $\pi^+$ and open symbols for $\pi^-$.
The triangles CLAS and open symbols are for HERMES~\protect\cite{Airapetian:2012yg}. $f(y)=\sqrt{2\epsilon(1+\epsilon)}$.}
\label{fig:clas-hermes}
\end{figure}

The azimuthal modulations have been studied phenomenologically, for instance Ref.~\cite{Boglione:2011wm} investigated effects due to the phase space limitations due to finite beam energies of real experiments as the cosine modulations are very sensitive to the corrections due to limitation of the phase space in experiments.
The twist-3 nature of $\cos\phi_h$ modulations could be tested by examining their $Q^2$dependence. In Fig.~\ref{fig:clas-hermes} CLAS measurements are compared with corresponding measurements from HERMES experiment~\cite{Airapetian:2012yg}, after taking into account the kinematic factors in the expression
of the $\cos\phi_h$ modulation and $\phi_h$ independent terms.
The CLAS and HERMES measurements are found to
be consistent with each other in a wide range of $Q^2$, as shown in Figs.~\ref{fig:clas-hermes}, indicating that at energies as low as 5-6 GeV, the
behavior of azimuthal modulations are similar to each other.
For comparison, the lowest $\xbj$ bin from CLAS and highest $\xbj$ bins from HERMES were used with equal average value of $\xbj\approx$ 0.19, $z\approx$ 0.45 and $\Phperp\approx$ 0.42 GeV.
The CLAS data provides
significant improvements in the precision of azimuthal moments for the kinematic region where the two data
sets overlap, and they extend the measurements to the large $\xbj$ region not accessible at HERMES, providing an important input for studies of higher-twist effects.
The $Q^2$-dependence of the $\cos\phi_h$ modulation is consistent with
the twist-3 nature of the contribution and within statistical uncertainties it is consistent with measurements performed at much higher energies and $Q^2$. Higher precision data from 12 GeV upgrade of JLab will provide essential information on to the $Q^
2$-dependence of observables in more details.

Large spin-azimuthal asymmetries observed at JLab for a longitudinally
polarised beam~\cite{Avakian:2003pk}, which have been
interpreted in terms of higher twist contributions, are also consistent with corresponding higher energy measurements at HERMES
\cite{Airapetian:2005jc} and COMPASS~\cite{Adolph:2014pwc}.
Within the same approximation as used in Eq.\ref{Eq:cosphi}, the expressions for $\sin\phi_h$ modulations in case of polarized beam or target can be written as
\begin{eqnarray}
F_{LU}^{\sin\phi_h} & \simeq & \frac{2M}{Q}\ {\cal C}\left[\frac{\bfhp\cdot\kt}{M_N} xg^\perp D_1
+\frac{\bfhp\cdot\pt}{z m_h}
xe H_1^\perp \right]
\end{eqnarray}

\begin{eqnarray}
F_{UL}^{\sin\phi_h} & \simeq & \frac{2M}{Q}\ {\cal C}\left[\frac{\bfhp\cdot\kt}{M_N} xf_L^\perp D_1
+\frac{\bfhp\cdot\pt}{z m_h}
xh_L H_1^\perp \right]
\end{eqnarray}
Distribution $e$ and $h_L$ are twist-3 TMD distribution functions that couple to chiral-odd Collins fragmentation function $H_1^\perp$, contributing to Collins-type terms in $F_{UL}^{\sin\phi_h}$ and $F_{LU}^{\sin\phi_h}$ and can be written in the following
way~\cite{Mulders:1996mp, Bacchetta:2006tn}
\begin{align}
x e &=x \tilde e + \frac{m}{M}\,f_1, \phantom{\frac{m^2}{M}}
\\
x h_L &= x {\tilde h}_L + \frac{\Phperp^2}{M^2}\, h_{1L}^{\perp} +\frac{m}{M}\,g_{1L}.
\end{align}

The Sivers-type contributions (terms that contain $D_1$) involve $f_L^\perp$ and $g^\perp$ TMDs, which
couple to the leading twist unpolarized fragmentation function $D_1$. Some of those functions can be studied in jet production, for instance, the T-odd twist-3 TMD $g^{\perp}$ gives
rise to a $\sin\phi_h$ azimuthal asymmetry in the production
of jets~\cite{Bacchetta:2004zf} in DIS with polarized lepton beams.
The higher twist TMDs, attracted a lot of theoretical attention, since first SSA was observed by HERMES~\cite{Avakian:1999rr}.
Some initial model calculations of unpolarized higher-twist TMDs were discussed in Refs.~\cite{Jaffe:1991ra,Signal:1996ct,
Jakob:1997wg,Schweitzer:2003uy,Wakamatsu:2003uu,Avakian:2010br}. A detailed list of recent calculations is presented in Ref.\cite{Lorce:2014hxa}. The function $g^{\perp}$ is an interesting object for theoretical studies. It was shown in Ref~\cite{Gamberg:2006ru} that $g^{\perp}$ has uncanceled light cone divergence and thus in principle TMD factorization at twist-3 fails. Additional theoretical studies of such functions as $g^{\perp}$  that arise at twist-3 level are needed to explore twist-3 factorization in detail and determine whether the factorization can be established or not.

\subsection{Model calculations}
\label{Sec:models_highertwist}
An important question that one could ask is whether twist-3 functions that we discuss are different from zero or not. Model calculations can certainly shed light on this matter. In fact many model calculations indicate sizable twist-3 functions, see for example Ref.~\cite{Lu:2015wja} where twist-3 fragmentation functions were studied.

WW approximation that we discussed in the previous subsection is certainly useful for numerical estimates but it removes the richness of the largely unexplored
but attractive non-perturbative physics of quark-gluon
correlations. This richness is precisely the important motivation to study
subleading-twist effects~\cite{Jaffe:1983hp,Jaffe:1989xx}.

Higher-twist TMDs and parton distribution functions of quarks are
expressed in terms of hadronic matrix elements of bilinear quark-field
correlators which
can be studied in quark models~\cite{Jaffe:1991ra}.
Quark models have been shown to give a useful description of
leading-twist TMDs and related SIDIS observables, provided one
applies them carefully within their range of applicability.

Quark models with interactions allow one to model
also the interaction-dependent tilde-terms.
\begin{figure}[ht!]
\centering
\includegraphics[width=0.27\textwidth]{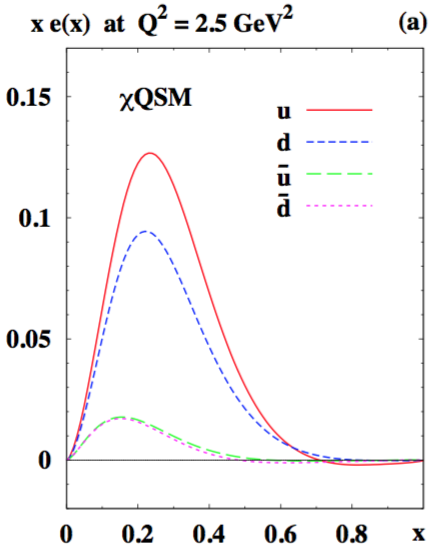}
\hspace{1cm}
\includegraphics[width=0.27\textwidth]{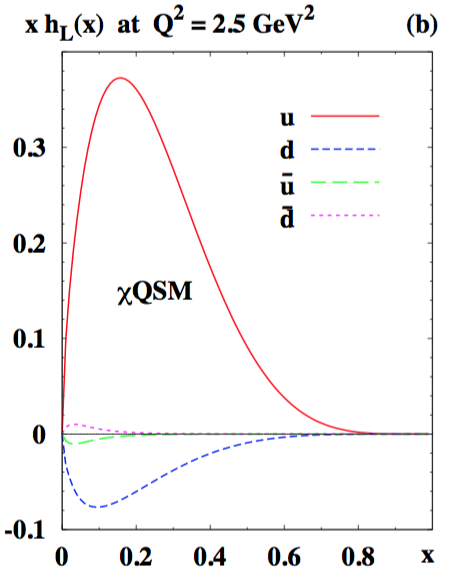}
\caption{The parton distribution functions $xe(x)$ and $xh_L(x)$
from the chiral quark soliton model ($\chi$QSM)~\cite{Cebulla:2007ej,Schweitzer:2001sr}
at $Q^2=2.5\,{\rm GeV}^2$.
\label{fig:twist3-e-hL-CQSM-2.5GeV2}}
\end{figure}
There are several model calculations of the twist-3 PDFs $e$ and $h_L$:
MIT bag model~\cite{Jaffe:1991ra,Signal:1996ct,Avakian:2010br},
diquark spectator model~\cite{Jakob:1997wg},
instanton QCD vacuum calculus~\cite{Balla:1997hf,Dressler:1999hc},
chiral quark soliton model~\cite{Wakamatsu:2000fd,Schweitzer:2003uy,Wakamatsu:2003uu,Ohnishi:2003mf,Cebulla:2007ej},
and the perturbative light-cone Hamiltonian approach
with a quark target~\cite{Burkardt:2001iy,Mukherjee:2009uy}.
In these calculations there are no
contributions from either strange or sea quarks, except for the chiral quark soliton model.
Fig.~\ref{fig:twist3-e-hL-CQSM-2.5GeV2} shows the parton distribution functions
$e(x)$ and $h_L(x)$ calculated in the chiral
quark soliton model~\cite{Cebulla:2007ej,Schweitzer:2001sr} at $Q^2=2.5\,{\rm GeV}^2$.
The bag model has given several powerful results and predictions of PDFs as well as
TMDs. It is a relativistic
model where quarks and antiquarks are excitations inside the confined bag.
It is generally assumed that the proton wave function is invariant under the $SU(6)$ spin-flavor
symmetry. In the case of two-body problems, this symmetry leads to proportionality between the
different flavor components. The contribution to $e(x)$ in the bag is entirely due to the bag
boundary, and therefore to the quark-gluon-quark correlation. The result of the model calculation
of the twist-3 $e^u(x), \, h_L^u(x)$, as well as of the unpolarized distribution $f_1^u(x)$,
is shown in Fig.~\ref{fig:twist3_bag}. On the other hand, the function $h_L(x)$ contains twist-2 and pure twist-3 contributions.
\begin{figure}[ht!]
\centering
\includegraphics[width=0.5\textwidth]{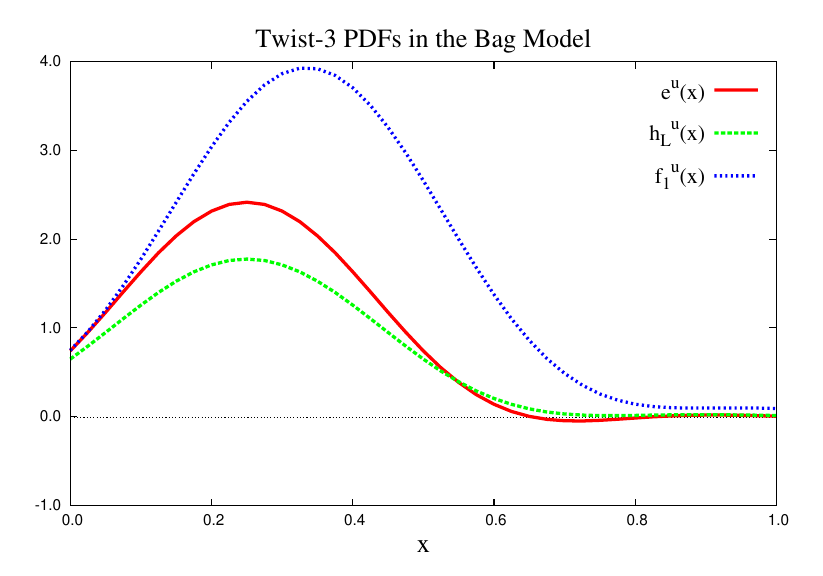}
\caption{Functions $e^u(x),\, h_L^u(x),$ and $f_1^u(x)$ calculated in the bag model~\cite{Jaffe:1991ra}.}
\label{fig:twist3_bag}
\end{figure}

\begin{figure}[ht!]
\centering
\includegraphics[width=0.49\textwidth]{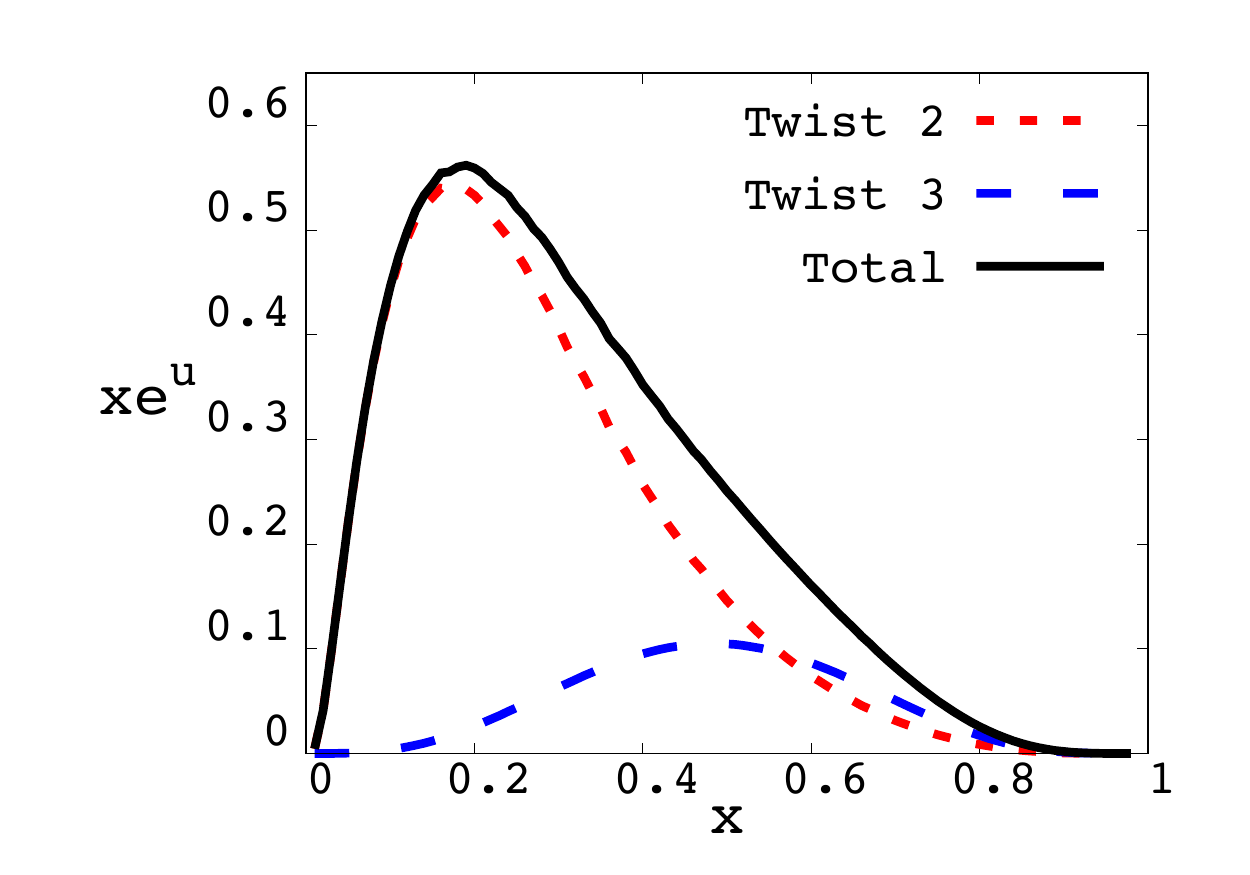}
\includegraphics[width=0.49\textwidth]{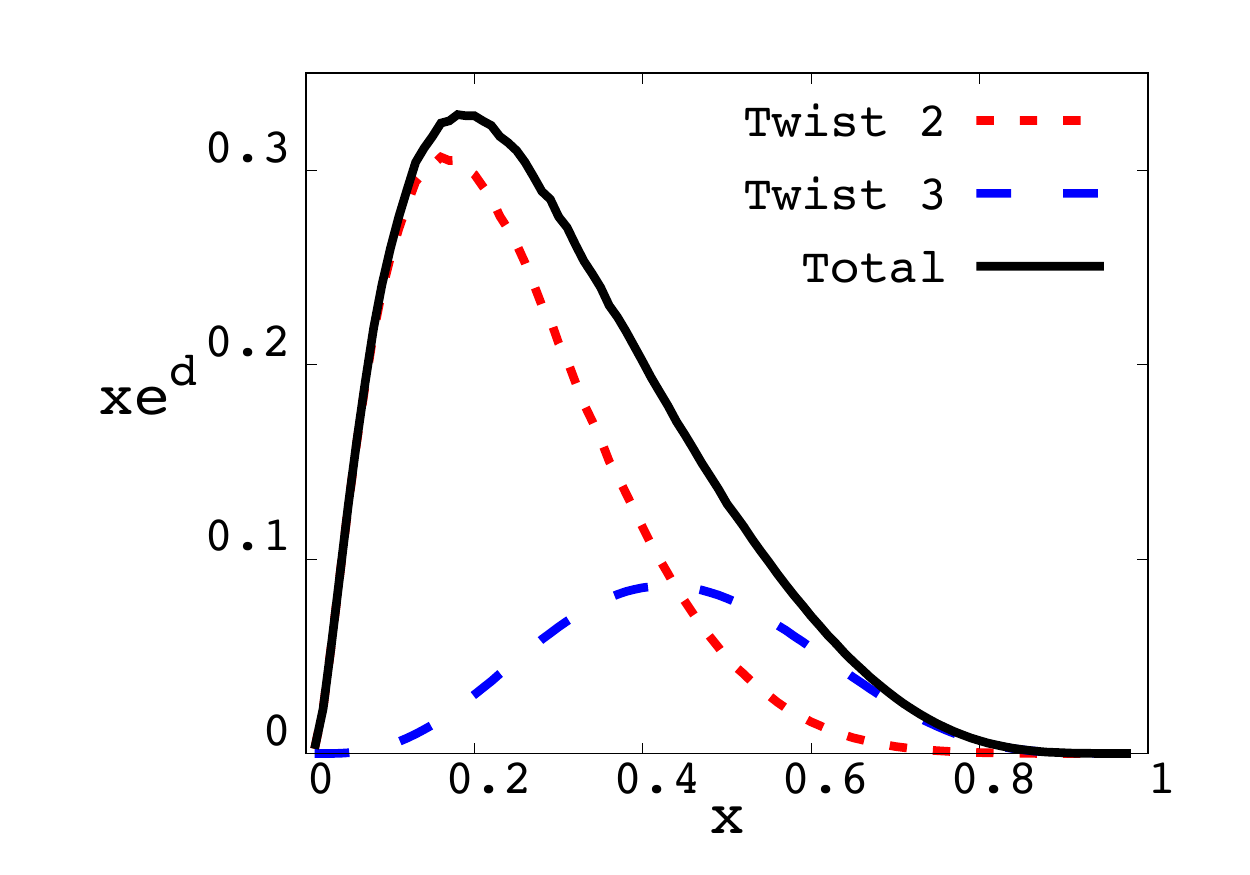}
\caption{Results for the PDF $xe(x)$ as function of $x$ for the up  (left
panel) and down (right panel) quark~\cite{Pasquini:2018oyz}. Red short-dashed curve: twist-2
contribution ($m/M f_1$) ; blue long-dashed curve:  pure twist-3
contribution ($\tilde e$); solid curve: total results, sum of the twist-2
and twist-3 contribution.}
\label{fig:twist3-e-pasquini}
\end{figure}

Although it is a popular assumption that pure twist-3 (and mass) terms are small~\cite{Tangerman:1994bb,Mulders:1995dh,Avakian:2007mv,Metz:2008ib}, this has rarely been justified by theoretical calculations.
Indeed, recent calculations using light-front wavefunctions (LFWFs)~\cite{Pasquini:2018oyz},
 taking into account the contribution from both the
  three valence quark ($qqq$)  and  three-quark plus one gluon ($qqqg$)
Fock-state of the nucleon, indicate the pure twist-3 contributions can be very significant in certain kinematics.
The LFWFs  are modeled using a parametrization derived from the  proton
distribution amplitudes, with parameters fitted to
the available phenomenological information on the unpolarized
leading-twist quark and gluon collinear  parton distributions.
The Fig.~\ref{fig:twist3-e-pasquini} presents the light-front model results for  the twist-2
contribution ($m/M f_1$), the pure twist-3 terms ($\tilde e$) and the
total results, for both the up and down quark.

\begin{figure}[ht!]
\centering
\includegraphics[width=0.5\textwidth]{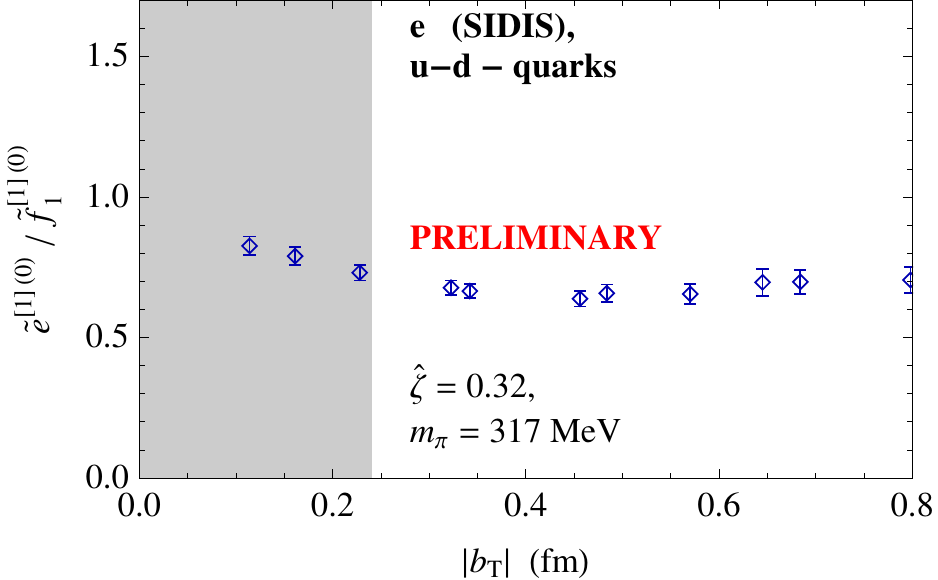}
\caption{The ratio of isovector $u-d$ combination of quark flavors for $e$ and $f_1$~\cite{Engelhardt:QCD-evol}. The shaded area represents the region in which discretization
effects may be significant.}
\label{fig:twist3_lattice}
\end{figure}

The twist-3 distributions have been recently also studied in lattice QCD calculations. The ratio of Fourier transformed to conjugate quark separation $b_T$-space twist-three TMD $e$, integrated over momentum fraction $x$ and
over the unpolarized
TMD $f_1$, likewise integrated and Fourier-transformed is shown in Fig.\ref{fig:twist3_lattice}. The framework for these calculations is described comprehensively in Refs.\cite{Musch:2011er,Engelhardt:2015xja}.
Data were obtained from a Lattice QCD calculation of the proton
matrix elements defining these TMDs, where the staple-shaped
gauge link in the relevant quark bilocal operator extends in the
direction appropriate for the SIDIS process. The Collins-Soper-type
parameter $\hat{\zeta } = v\cdot P /(|v| |P|)$ characterizes the rapidity
difference between the proton momentum $P$ and the direction of
the staple legs $v$. The phenomenologically most relevant range of
$\hat{\zeta } $ lies appreciably above the value accessed in this
calculation, which was moreover performed at an artificially high
pion mass. While Lattice QCD calculations closer to the physical case
remain to be performed, extraction on lattice is consistent in sign and magnitude with simple model calculations discussed above and presented in Fig.\ref{fig:twist3_bag}. One can see that lattice computations suggest that $e$ has almost the same relative size as twist-2 $f_1$.

With several functions contributing to the same observable one faces a difficult task when dealing
with sub-leading twist in SIDIS, but it is important
to stress that each of them provides a different and
independent view of the quark-gluon dynamics in the
nucleon or in the fragmentation process.
The importance of these observables for spin physics and QCD is
very high, indeed these were the first instances that
single spin phenomena in SIDIS have been measured and
this has triggered important theoretical developments.
Both structure functions, $F_{LU}^{\sin\phi_h}$ and $F_{UL}^{\sin\phi_h}$, were subject to numerous theoretical
and phenomenological studies
\cite{Kotzinian:1999dy,Boglione:2000jk,Efremov:2000za,
Anselmino:2000mb,Ma:2000ip,Oganessian:2000um,Wakamatsu:2000fd,
Efremov:2001cz,Efremov:2001ia,Ma:2001ie,Bacchetta:2002tk,
Efremov:2002sd,Efremov:2002ut,Ma:2002ns,Efremov:2003tf,
Efremov:2003eq,Schweitzer:2003yr,Efremov:2004hz},
see also~\cite{Afanasev:2003ze,Yuan:2003gu,Bacchetta:2004zf,Metz:2004je,Bacchetta:2006tn}.
Nevertheless there is presently no satisfactory understanding
which are the functions of origin of these modulations from the point of view of quantifying contribution of underlying functions.

Other sub-leading structure functions have also been
studied~\cite{Ji:1993vw,Levelt:1994np,Oganesian:1997jq}
though less extensively, since there is far less data
available though measurements of some subleading structure
functions were reported~\cite{Parsamyan:2007ju,Pappalardo:2010zza,Parsamyan:2018evv}.
The TMDs $e$ and $g^{\perp }$ are pure twist-3
interaction-dependent quark-gluon correlators, i.e.\
$e=\tilde{e}$ and $g^{\perp}=\tilde{g}^{\perp}$
up to current quark mass terms, and hence vanish in
the Wandzura-Wilczek-type approximation discussed above.
This means that in this approximation the entire
$F_{LU}^{\sin\phi_h}$ would vanish, while in experiment a
clearly non-zero effect is seen \cite{Avakian:2003pk,Airapetian:2006rx}.

A detailed study of non-perturbative properties of sub-leading twist TMDs has been performed recently by Lorce and collaborators~\cite{Lorce:2014hxa}. The formalism to describe unpolarized higher-twist TMDs in the light-front framework based on a Fock-space expansion of the
nucleon state in terms of free on-shell parton states has been developed, and some numerical results in
a practical realization of this picture performed by the light-front constituent quark model. The results from the light-front constituent quark model were also compared to available phenomenological information, showing a satisfactory agreement.

\begin{figure}[ht!]
\centering
\includegraphics[width=0.75\textwidth]{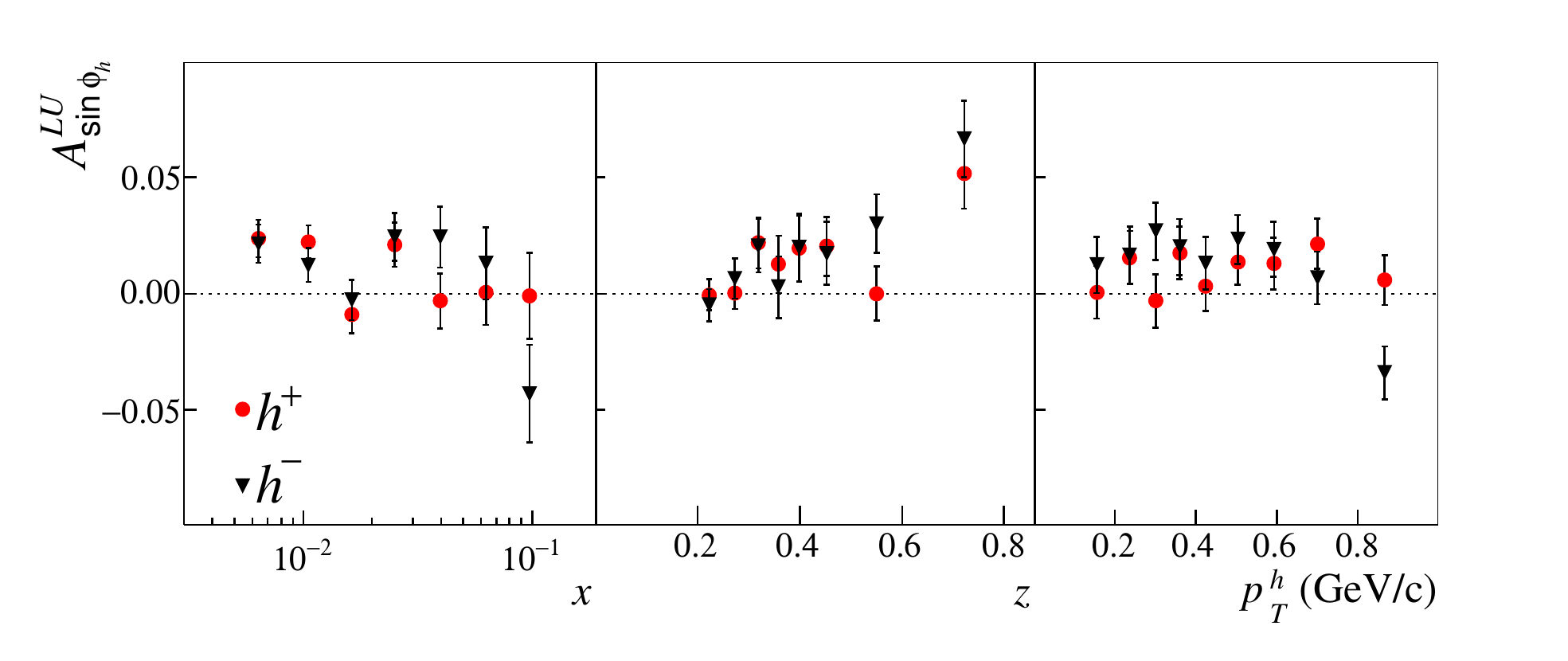}
\caption{The $A_{LU}^{\sin(\phiH)}$ asymmetry extracted for positive and negative hadron productions by COMPASS~\cite{Adolph:2014pwc}.}
\label{fig:ALU_cmps}
\end{figure}

\begin{figure}[ht!]
\centering
\includegraphics[width=0.75\textwidth]{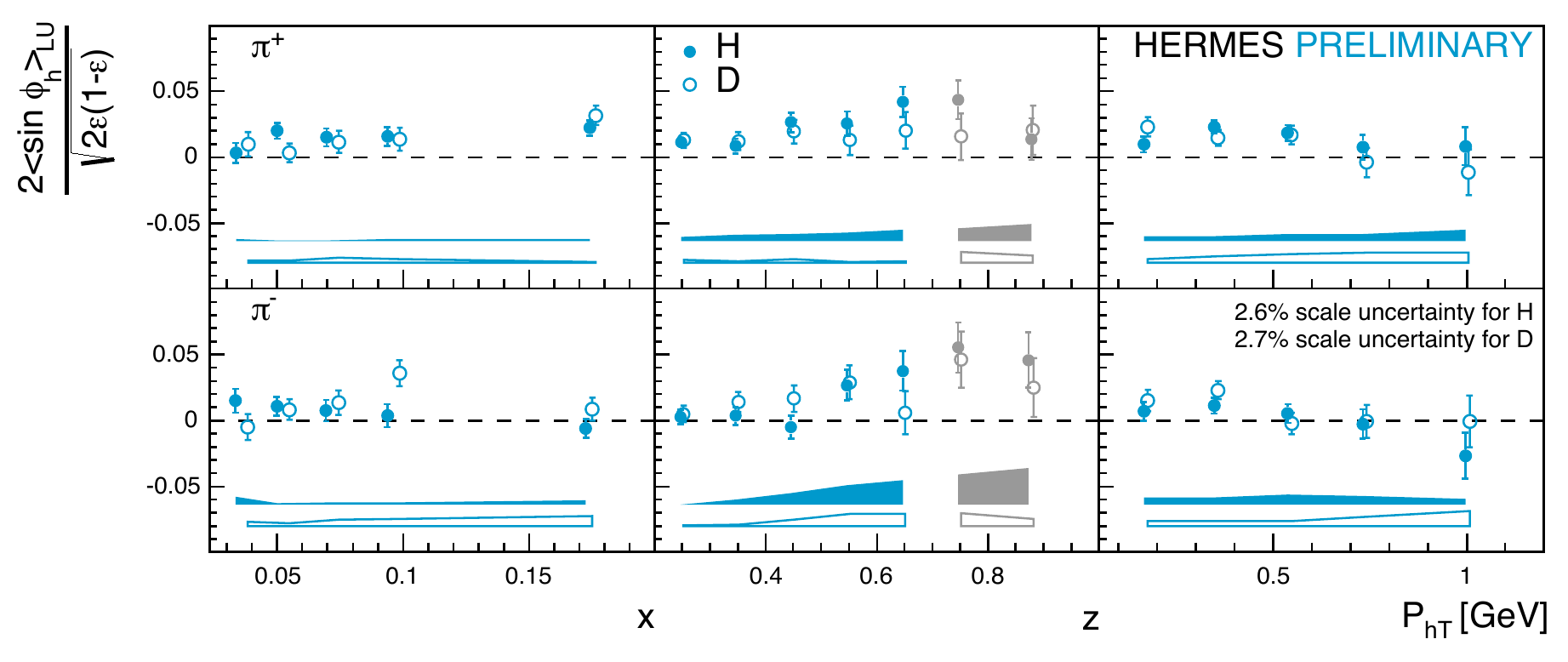}
\caption{The preliminary $A_{LU}^{\sin(\phiH)}$ amplitudes for charged
pions extracted from data on unpolarized hydrogen (H) and deuterium
(D) targets by HERMES~\cite{Marukyan:2018sde}. The asymmetries are corrected for the depolarization factor.}
\label{fig:ALU_hms}
\end{figure}

In Figs.~\ref{fig:ALU_cmps},\ref{fig:ALU_hms} COMPASS (deuteron) and recent HERMES (deuteron, proton) results for $A_{LU}^{\sin(\phiH)}$ asymmetries  are shown as a function of $x$, $z$ and $P_{hT}$.

\begin{figure}[ht!]
\centering
\includegraphics[width=0.41\textwidth]{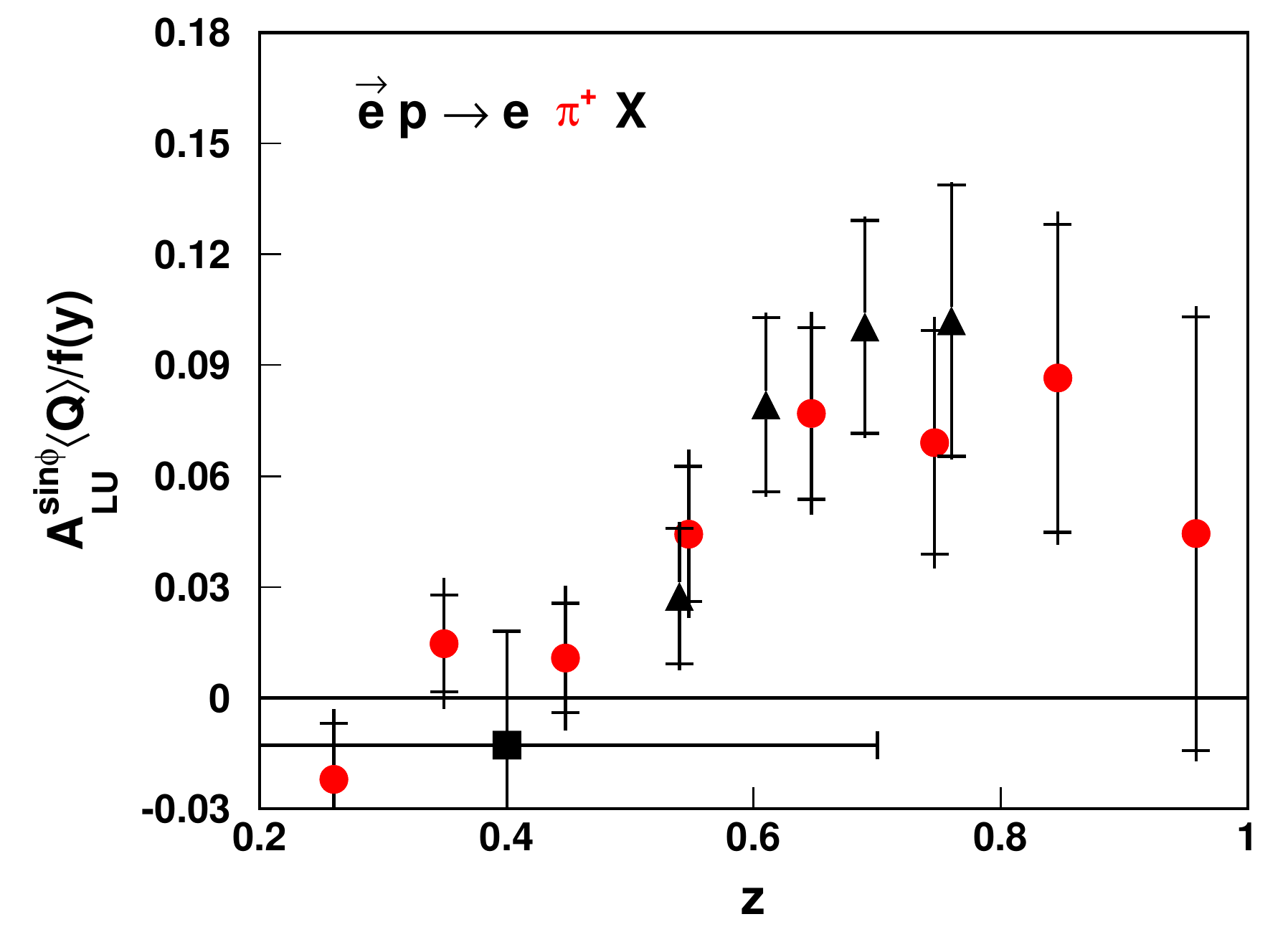}
\includegraphics[width=0.45\textwidth]{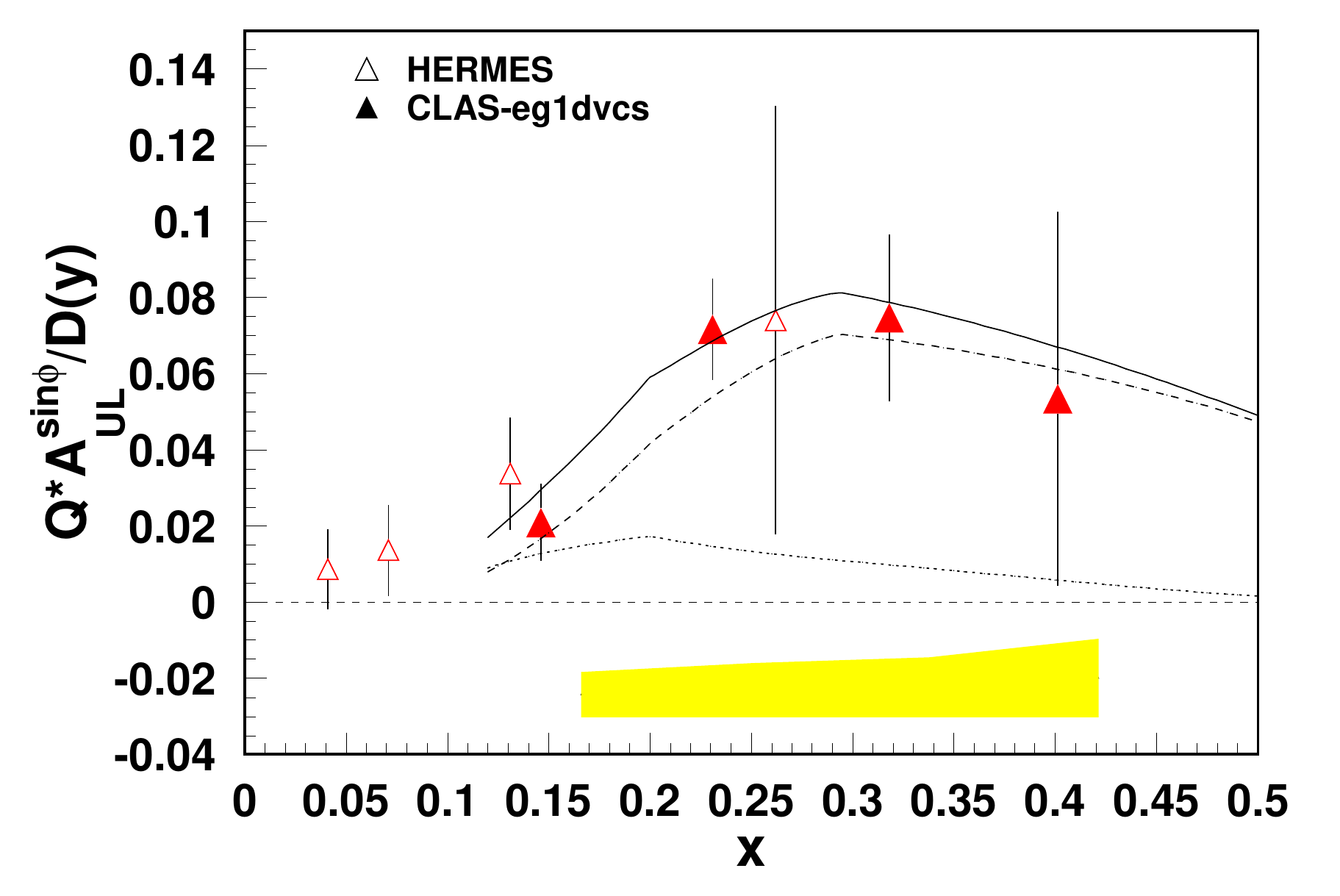}
\caption{The comparison of SSA  $A_{LU}^{\sin\phi}$ as a function of $x$ for neutral pions at HERMES~\cite{Airapetian:2006rx} and CLAS for similar average $\Phperp$ ~\cite{Aghasyan:2011ha} (left panel). Both Asymmetry moments are multiplied by the kinematic factor $Q/f(y)$. The right panel shows the comparison of SSA $A_{UL}^{\sin\phi}$ for neutral pions at HERMES and CLAS~\cite{Jawalkar:2017ube} (right). The factor $D(y)$ for longitudinally polarized case is given in Eq.\ref{eq:x_secSIDIS}. The dashed and dotted lines are twist-3 calculations from from Sivers and Collins type terms \cite{Mao:2012dk,Lu:2016gtj}, respectively.}
\label{fig:alu-ul-her-clas}
\end{figure}

\begin{figure}[ht!]
\centering
\includegraphics[width=0.7\textwidth]{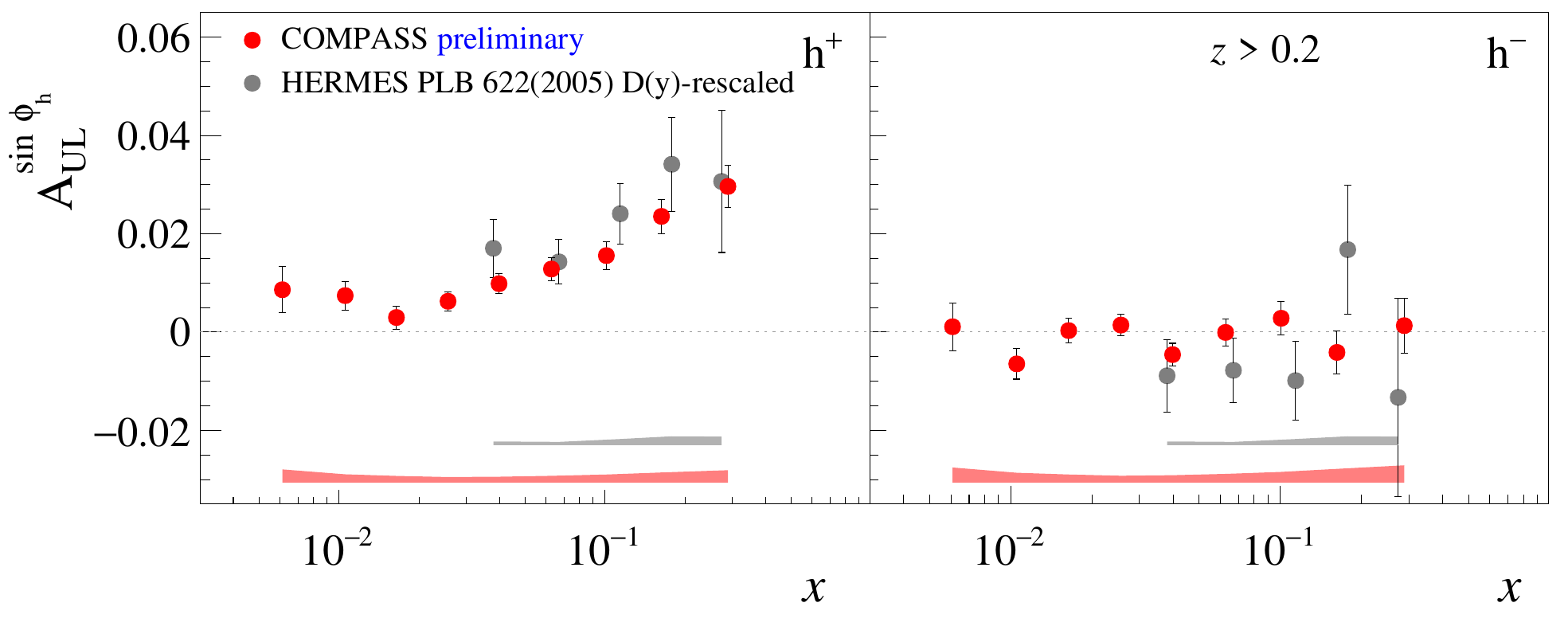}
\caption{The preliminary results for $A_{UL}^{\sin(\phiH)}$ obtained by COMPASS~\cite{Parsamyan:2018ovx,Parsamyan:2018evv} are shown together with the HERMES results~\cite{Airapetian:2005jc}. Presented results are not rescaled for $1/Q$. HERMES results are rescaled for $D(y)$.}
\label{fig:AUL1_cmps_hms}
\end{figure}

\begin{figure}[ht!]
\centering
\includegraphics[width=0.42\textwidth]{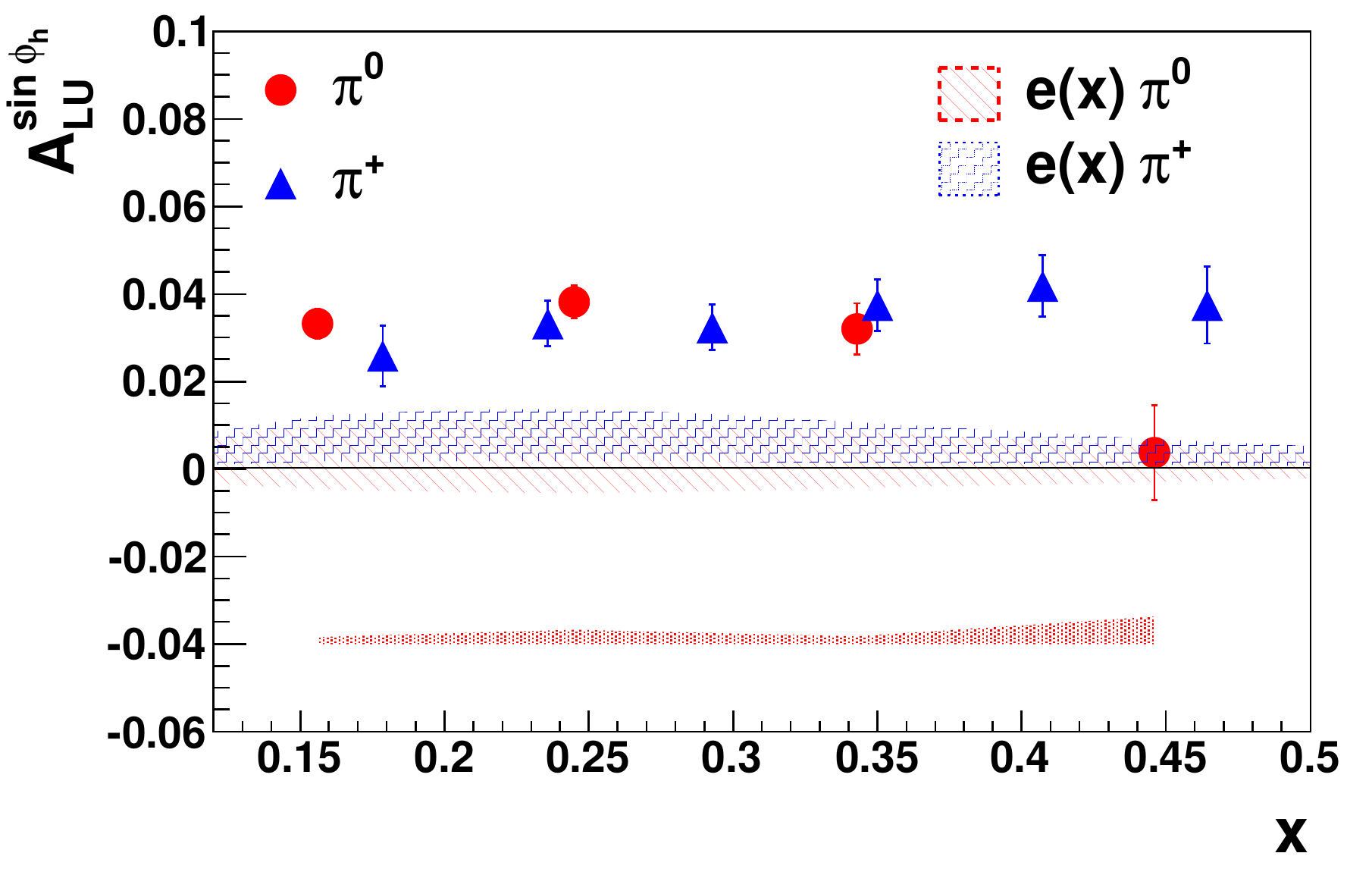}
\includegraphics[width=0.57\textwidth]{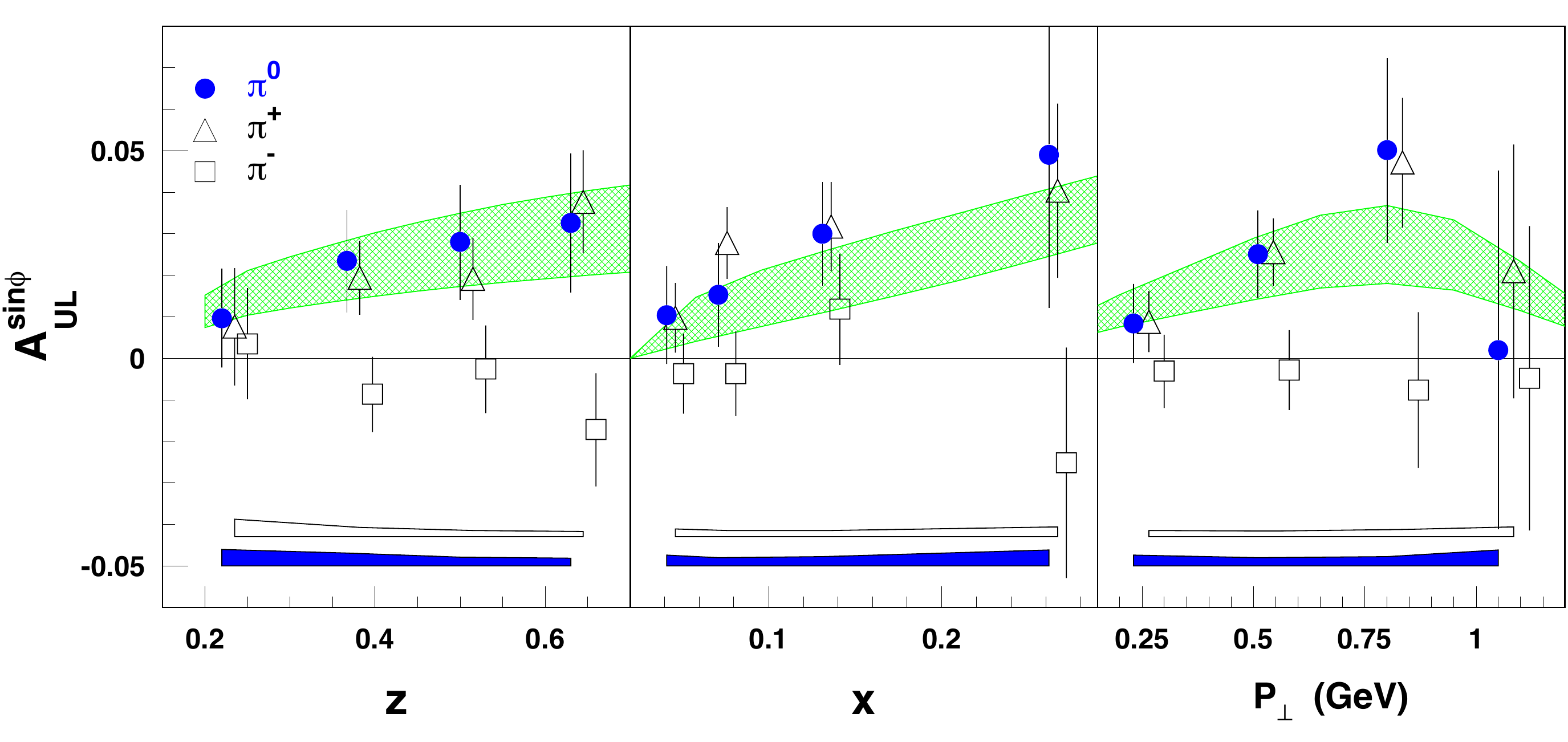}
\caption{
Left panel: comparison of $A_{LU}^{\sin(\phiH)}$ for $\pi^+$ and $\pi^0$ from CLAS data~\cite{Aghasyan:2011ha} for the same average $\Phperp=0.38$ GeV and $0.4<z<0.7$  (left).
The error bars correspond to statistical and bands to systematic uncertainties. The hatched band shows  model calculations involving only the Collins effect for $\pi^0$.
Right panel: $A_{UL}^{\sin(\phiH)}$    for different flavors using HERMES data~\cite{Airapetian:2001eg}. Error
bars include the statistical uncertainties only. The filled and open bands at the bottom of the panels represent the systematic
uncertainties for neutral and charged pions, respectively. The shaded areas show a range of predictions of a model calculation \cite{Oganessian:1998ma,DeSanctis:2000fh}
applied to the case of $\pi^0$ electroproduction.}
\label{fig:clasalu-ul}
\end{figure}

Since the structure functions $F_{LU}$ and $F_{UL}$ contain kinematical terms depending on the beam energy for given kinematics, as well as additional $1/Q$ suppression factor, direct comparison of ratios of structure functions involved in those observables between different experiments requires accounting for those terms. After corrections the data seem to be consistent also between CLAS and HERMES $A_{LU}^{\sin\phiH}$, see Fig.~\ref{fig:alu-ul-her-clas}. Recent high precision measurements of $A_{UL}^{\sin\phiH}$ performed at COMPASS are also consistent with similar measurements at HERMES, see Fig.\ref{fig:AUL1_cmps_hms}. Both asymmetries exhibit similar kinematical behaviour and, more importantly similar flavor dependence.  Accounting for difference in energies and average $y$ in all comparisons was done by dividing by the kinematic factors defined in Eq.\ref{eq:x_secSIDIS} (for $A_{UL}^{\sin\phiH}$ the $D(y)=\sqrt{2\epsilon(1+\epsilon)}$). Comparison of $A_{LU}^{\sin\phiH}$ and $A_{UL}^{\sin\phiH}$ for $\pi^+$ and $\pi^0$ (see Fig.\ref{fig:clasalu-ul}) indicates, that in both cases they are consistent with each other. Latest measurements of  $A_{UL}^{\sin\phiH}$ by CLAS~\cite{Jawalkar:2017ube} for all pion flavors is consistent with HERMES measurements, confirming that $\pi^+$ and $\pi^0$  show similar behaviour both for $A_{LU}^{\sin\phiH}$ and $A_{UL}^{\sin\phiH}$.

\begin{figure}[ht!]
\begin{center}
\includegraphics[width=0.4\textwidth]{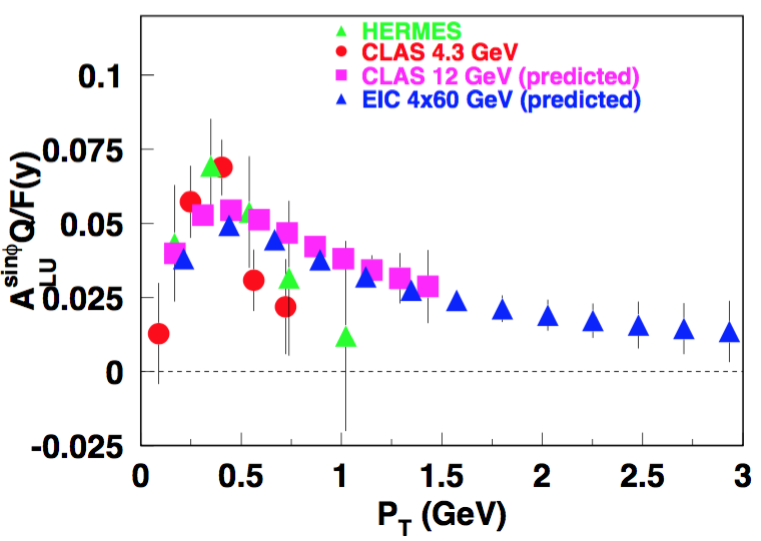}
\includegraphics[width=0.44\textwidth]{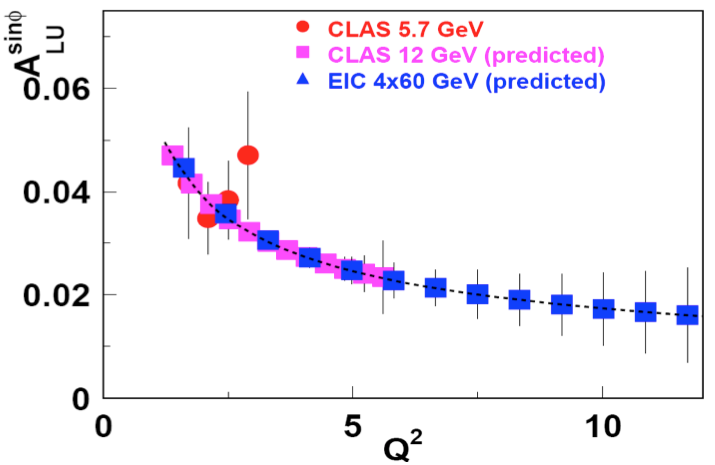}
\end{center}
\caption{\small Projections for higher-twist lepton spin
asymmetry $A_{LU}^{\sin\phi}$ for positive pion
production, using 4 GeV electrons and 60 GeV protons (100 days at 10$^{34}$ cm$^{-2}$ sec$^{-1}$), as a
function of $\Phperp$ (left) and $Q^2$ (right) compared to published data
from CLAS~\cite{Avakian:2003pk} and
HERMES~\cite{Airapetian:2006rx} and projected CLAS12~\cite{PACauu} in one
$x,z$ bin ($0.2<x<0.3$, $0.5<z<0.55$).}
\label{eicvsclas12}
\end{figure}

Understanding of quark-gluon dynamics is crucial for interpretation of upcoming SIDIS data from
Jefferson Lab 12 GeV upgrade, where studies of TMDs are one of main driving forces.
Significantly higher, compared to JLab12, $\Phperp$ range accessible at EIC would allow for studies
of transverse momentum dependence of various distribution and fragmentation functions as well as
transition from TMD regime ($\Phperp/z \ll Q$) to collinear perturbative regime ($\Phperp/z \sim Q$).
Measurements of spin and azimuthal asymmetries as a function of the final hadron transverse
momentum at EIC will extend (see Fig. \ref{eicvsclas12}) measurements at JLab12~\cite{PACauu}
to significantly higher $\Phperp$ and lower $x$ and will provide access to studies of TMDs beyond the valence region.
Much higher $Q^2$ range accessible at EIC would allow for
studies of $Q^2$-dependence
of different higher twist spin-azimuthal asymmetries (Fig \ref{eicvsclas12}),
which, apart from providing important
information on quark-gluon correlations are needed for understanding of possible corrections from
higher twists to leading twist observables.

Both Jefferson Lab12 and future EIC will measure also other azimuthal moments arising due to different quark-gluon-quark interactions. Many of them have been
already measured by HERMES and COMPASS in corresponding kinematical regions, including $A_{LL}^{\cos\phi}$ and $A_{UT}^{\sin\phi_S}$, see Figs~\ref{fig:ALL1_cmps},\ref{fig:AUT1_cmps},\ref{fig:AUT1_hms}. A common feature of all higher twist asymmetries is the increase at large $x$.

The $F_{LL}^{\cos\phi_h}$ structure function can be presented as follows:

\begin{eqnarray}
F_{LL}^{\cos\phi_h} & \simeq & \frac{2M}{Q}\ {\cal C}\left[-\frac{\bfhp\cdot\kt}{M_N} xg_L^\perp D_1-\frac{\bfhp\cdot\pt}{z m_h} xe_L H_1^\perp \right]
\end{eqnarray}

\begin{figure}[ht!]
\centering
\includegraphics[width=0.5\textwidth]{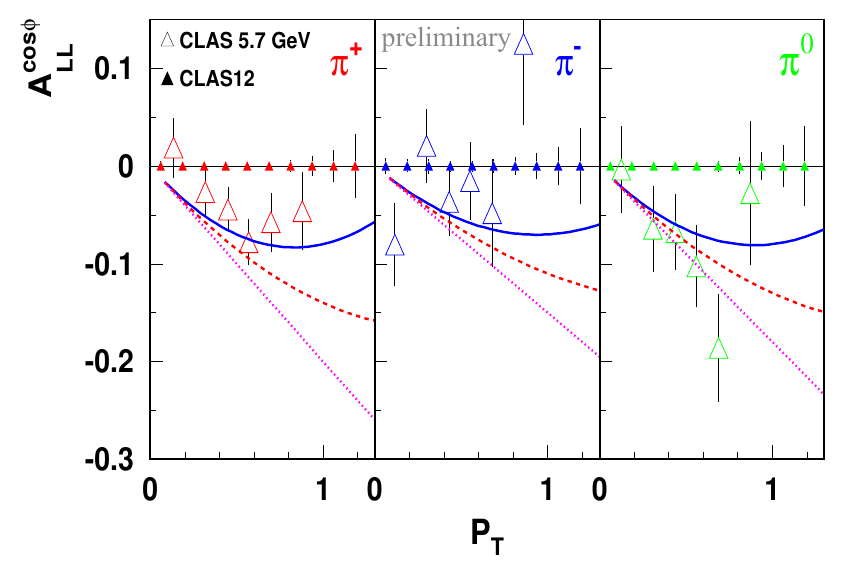}
\caption{The $A_{LL}^{\cos(\phiH)}$ as a function of the hadron transverse momentum. The filled triangles show projected uncertainties from approved JLab12 measurements~\cite{E12-07-107} compared to   5.7 GeV  preliminary result (open triangles).  The lines correspond to model calculations from Ref.~\cite{Anselmino:2006yc}.}
\label{fig:ALL1_JLab}
\end{figure}

\begin{figure}[ht!]
\centering
\includegraphics[width=0.7\textwidth]{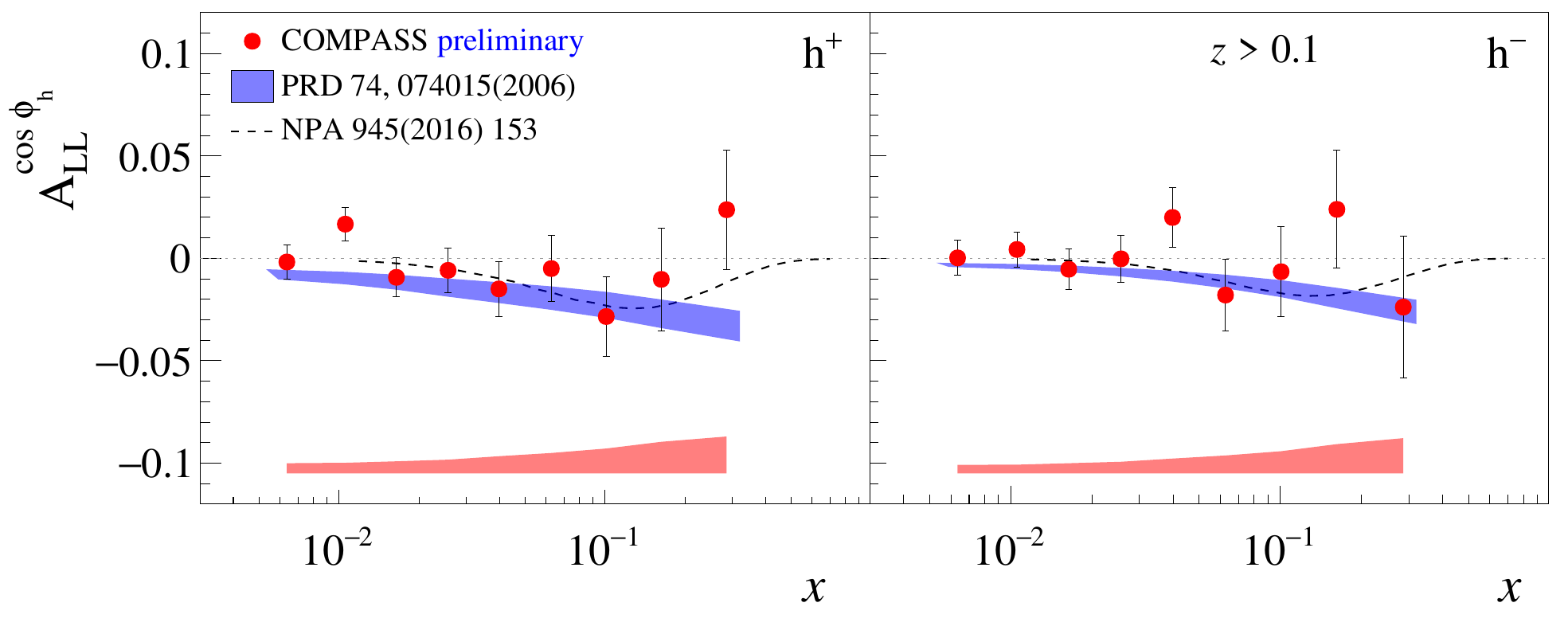}
\caption{The preliminary results for $A_{LL}^{\cos(\phiH)}$ obtained by COMPASS~\cite{Parsamyan:2018ovx,Parsamyan:2018evv} are shown together with available model predictions~\cite{Anselmino:2006yc,Mao:2016hdi}.}
\label{fig:ALL1_cmps}
\end{figure}

\begin{figure}[ht!]
\centering
\includegraphics[width=0.8\textwidth]{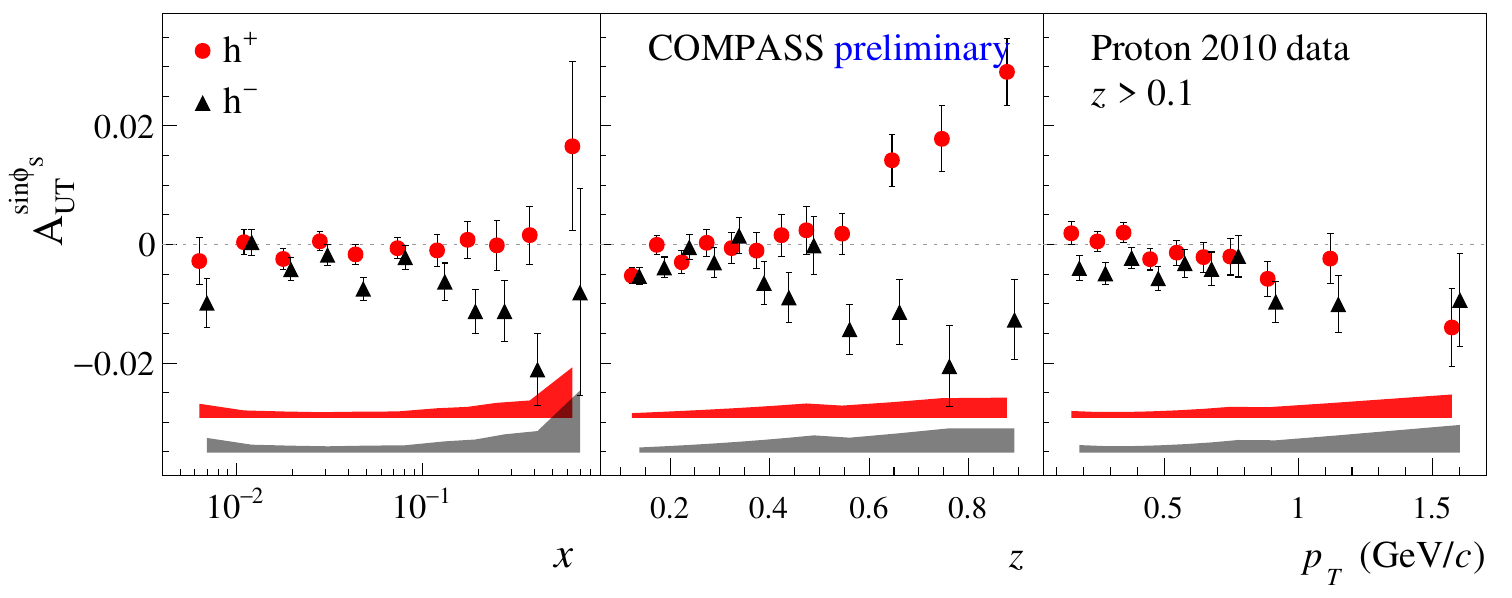}
\caption{The $A_{UT}^{\sin(\phiS)}$ asymmetry extracted by COMPASS~\cite{Adolph:2016dvl,Parsamyan:2018evv}.}
\label{fig:AUT1_cmps}
\end{figure}

\begin{figure}[ht!]
\centering
\includegraphics[width=0.6\textwidth]{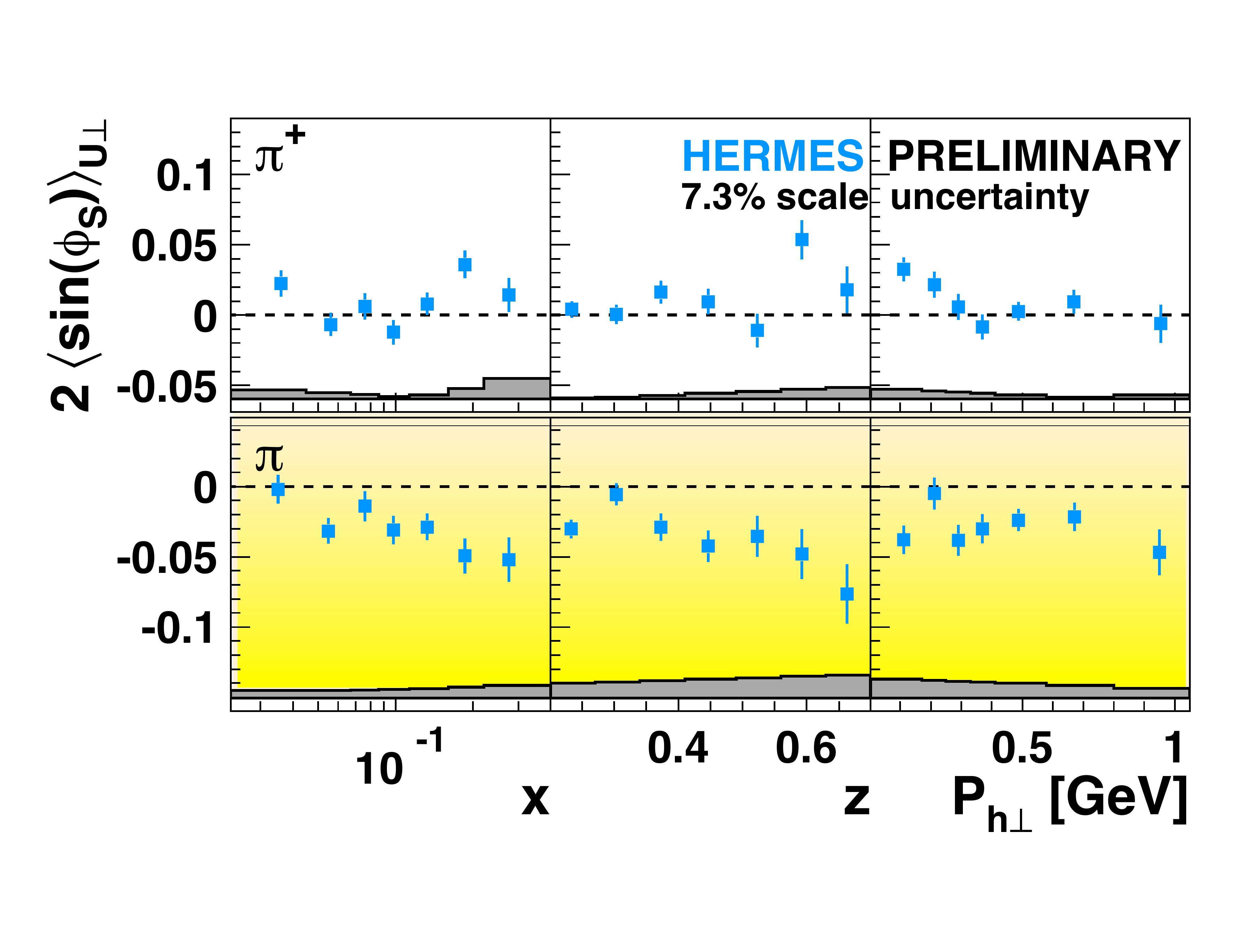}
\caption{The preliminary $A_{UT}^{\sin(\phiS)}$ asymmetry extracted by HERMES~\cite{Pappalardo:2010zza}.}
\label{fig:AUT1_hms}
\end{figure}

Within WW-type approximation the $e_L$ vanishes and we remain with $xg_L^\perp\simeq g_{1L}$ PDF. Corresponding double spin asymmetry is an analogue of the aforementioned Cahn effect, but for the polarized (helicity) cross section~\cite{Anselmino:2006yc}.
In Figs~\ref{fig:ALL1_JLab},~\ref{fig:ALL1_cmps} JLab and COMPASS results for this DSA are shown.

The $A_{UT}^{\sin\phiS}$ subleading-twist SSA is a peculiar term, since within the WW-type approximation it can be interpreted as a combination of Sivers and Collins constituents (within WW-type approximation). The corresponding structure function reads:

\begin{eqnarray}
F_{UT}^{\sin\phi_S} & \simeq & \frac{2M}{Q}\ {\cal C}
\Bigg\{xf_{T} D_1+\frac{\kt\cdot\pt}{2M_Nz m_h}\left[xh_T   - xh_T^\perp   \right] H_1^\perp\Bigg\}
\end{eqnarray}

If one employs WW-type approximation, then $xf_{T}\simeq f_{1T}^\perp$ is a Sivers type and $xh_T   - xh_T^\perp  \simeq -2 h_1$ is a Collins type contribution. COMPASS and HERMES results for this SSA are presented in Figs.~\ref{fig:AUT1_cmps},\ref{fig:AUT1_hms}. In both cases the asymmetry exhibits a non-zero signal at large $z$, in particular, for negative hadrons.

\section{Flavor dependence of spin-azimuthal asymmetries}
\label{Sec:flavor}
One of the important challenges in phenomenological fits is to disentangle the flavor dependence of the TMD PDFs from the measurements for a limited set of final particle types. The difficulty is that in the corresponding cross section the TMD PDFs are convoluted with the fragmentation functions to the specific hadrons and these are summed over the flavor of the parton. Thus having SSA measurements for as large a number of final hadron types as possible would allow, removing some of the assumptions about the  flavor dependence of the corresponding TMDs, Sivers PDFs, in particular.
Measurements with different final state hadrons are crucial for understanding of dynamics and sorting out contributions due
to spin-orbit correlations occurring in the initial distributions and in final stage of hadronization.

For example, electroproduction of neutral pions has several important advantages: 1) suppression of higher-twist contributions at large hadron energy fraction $z$~\cite{Afanasev:1996mj}, which are particularly important at JLab energies where small $z$ events are contaminated by target fragmentation; 2) the absence of $\rho^0$ production which complicates the interpretation of the charged single-pion data; 3) the fragmentation functions for $u$ and $d$ quarks to $\pi^0$ are the same in first approximation; and 4) suppression of spin-dependent fragmentation for $\pi^0$s, due to the roughly equal magnitude and opposite sign of the Collins fragmentation functions for up and down quarks~\cite{Airapetian:2010ds,Alekseev:2010rw,Abe:2005zx,theBABAR:2013yha,Ablikim:2015pta}. In addition it was shown that the longitudinal photon contribution, is suppressed in exclusive neutral pions production~\cite{Defurne:2016eiy} with respect to the transverse photon contribution, which is higher twist, suggesting that longitudinal photon contribution to SIDIS $\pi^0$ will also be suppressed.
At large $x$, where the sea contribution is negligible, $\pi^0$ multiplicities and double spin asymmetries will provide direct info on the fragmentation function of $u$ and $d$-quarks to $\pi^0$.
These factors simplify the interpretation of $\pi^0$ observables, such as single and double-spin asymmetries.
Furthermore, neutral pions are straight-forward to identify with little background using
the invariant mass of two detected photons.
Similar, important feature, like equal for $u$ and $d$ fragmentation functions exhibit also the production of charged pion pairs.

\section{Dihadron production}
\label{Sec:dihadrons}
Measurements of Collins asymmetries, $A_{UT}^{\rm \sin(\phi_h+\phi_S)}$, at HERMES and COMPASS, confirm a non-zero Collins fragmentation functions, which
are likely  generated due to correlation of the transverse spin of fragmenting quarks and the orbital motion of $q\bar{q}$ pairs~\cite{Artru:1990zv}. That means that the leftover pair of the $q\bar{q}$ will end up in the second hadron. Hadron pair production allow for a
possibility of extraction of transversity distribution in a complementary to single hadron production case and was used in extraction of transversity in an analysis with combined electron-proton and proton-proton data \cite{Radici:2018iag}.
In order to extract
the transversity distribution from single hadron production, the Collins
function should be determined through the measurement of azimuthal asymmetries in
the distribution of two almost back-to-back hadrons in $e^+e^-$
annihilation~\cite{Boer:1997mf}. The Belle and BaBar collaborations measured this
asymmetry~\cite{Abe:2005zx,Seidl:2008xc,BABAR:2013yha},
making the first-ever extraction of $h_1^q$
possible from a simultaneous analysis of $e p^{\uparrow} \to e' \pi X$ and
$e^+ e^- \to \pi \pi X$ data~\cite{Anselmino:2007fs}.

In spite of this breakthrough, some questions still hinder
the extraction of transversity from single-particle-inclusive measurements.
The most crucial issue is the treatment of evolution effects, since the measurements were
performed at very different energies. The convolution
$h_1^q \otimes H_1^{\perp\, q}$ involves the transverse momentum of quarks.
Hence, its evolution should be described in the framework of the
transverse-momentum-dependent
factorization~\cite{Collins:1981uk,Ji:2004wu}. Quantitative explorations in this direction suggest
that neglecting evolution effects could lead to overestimating transversity~\cite{Boer:2008fr}. The so-called TMD framework has been extensively studied \cite{Aybat:2011zv}(e.g. Bessel weighting treatment of cross sections~\cite{Boer:2011xd}). The first extraction of transversity with TMD evolution was presented in Refs.~\cite{Kang:2014zza,Kang:2015msa} and was found to be compatible with earlier studies without full account of TMD evolution from Refs.~\cite{Anselmino:2007fs,Anselmino:2008jk} thus confirming an idea that in ratios evolution effects generically cancel to certain extent.

In this context, it is of paramount importance to extract transversity in an independent way,
requiring only standard collinear factorization where the above complications are absent
(see, e.g, Refs.~\cite{Collins:1989gx,Brock:1993sz} and references therein). Results of the newest analysis from Ref~\cite{Radici:2018iag} are compatible with Refs.~\cite{Kang:2014zza,Kang:2015msa,Anselmino:2007fs,Anselmino:2008jk} and with extraction of transversity that takes into account lattice QCD calculations of tensor charge done in Ref.~\cite{Lin:2017stx}, even though some tension still remains among various results.

\begin{figure}[ht!]
\centering
\includegraphics[width=0.6\textwidth]{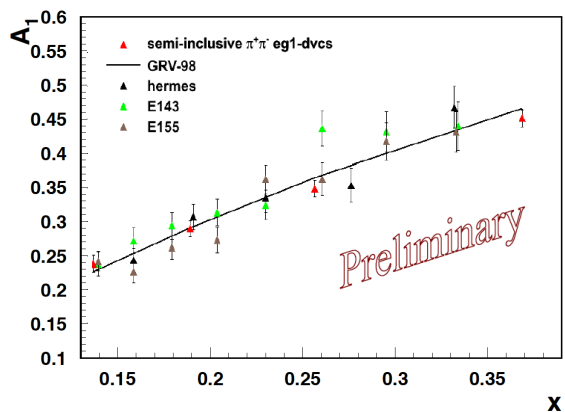}
\caption{The double spin asymmetry in semi-inclusive production of $\pi^\pm$ pairs, compared to inclusive DIS measurements
from HERMES~\cite{Airapetian:1998wi}, CLAS~\cite{Prok:2014ltt} and SLAC E143\cite{Abe:1994cp} and E155~\cite{Anthony:2000fn} experiments.}
\label{fig-dihad_A1_all}
\end{figure}

The semi-inclusive deep-inelastic production of two charged pions with
small invariant mass,%
\begin{equation}
\ell(l) + N(P) \to \ell(l') + H_1(P_1)+H_2(P_2) + X
, ,\nonumber
\end{equation}
which was suggested as a complementary source of information on the quark-gluon dynamics, bears similar features.
The double-spin asymmetry of pion pair production has been measured at JLab and shown to be consistent with inclusive DIS asymmetry.
Although, dihadron production in SIDIS requires higher energies and $Q^2$, than single hadron SIDIS,
measurements of double-spin asymmetries at CLAS (see Fig.~\ref{fig-dihad_A1_all}) are already at 5.7 GeV compatible with simple leading twist predictions for equality of double spin asymmetries in $eX$, $e\pi^+\pi^-X$, and $e\pi^0X$, assuming the sea quark contributions
are negligible at large $\xbj$ and fragmentation functions sum of charged pions are flavor independent.

CLAS measurement of the double spin asymmetry from inclusive DIS (also from HERMES and SLAC) are consistent with CLAS measurements
of double spin asymmetry in
charged pion pair production. At low energies the multiplicities are low and one has to apply cuts on the missing mass of the final state to avoid contributions from exclusive states.

\begin{figure}[ht!]
\centering
\includegraphics[width=0.45\textwidth]{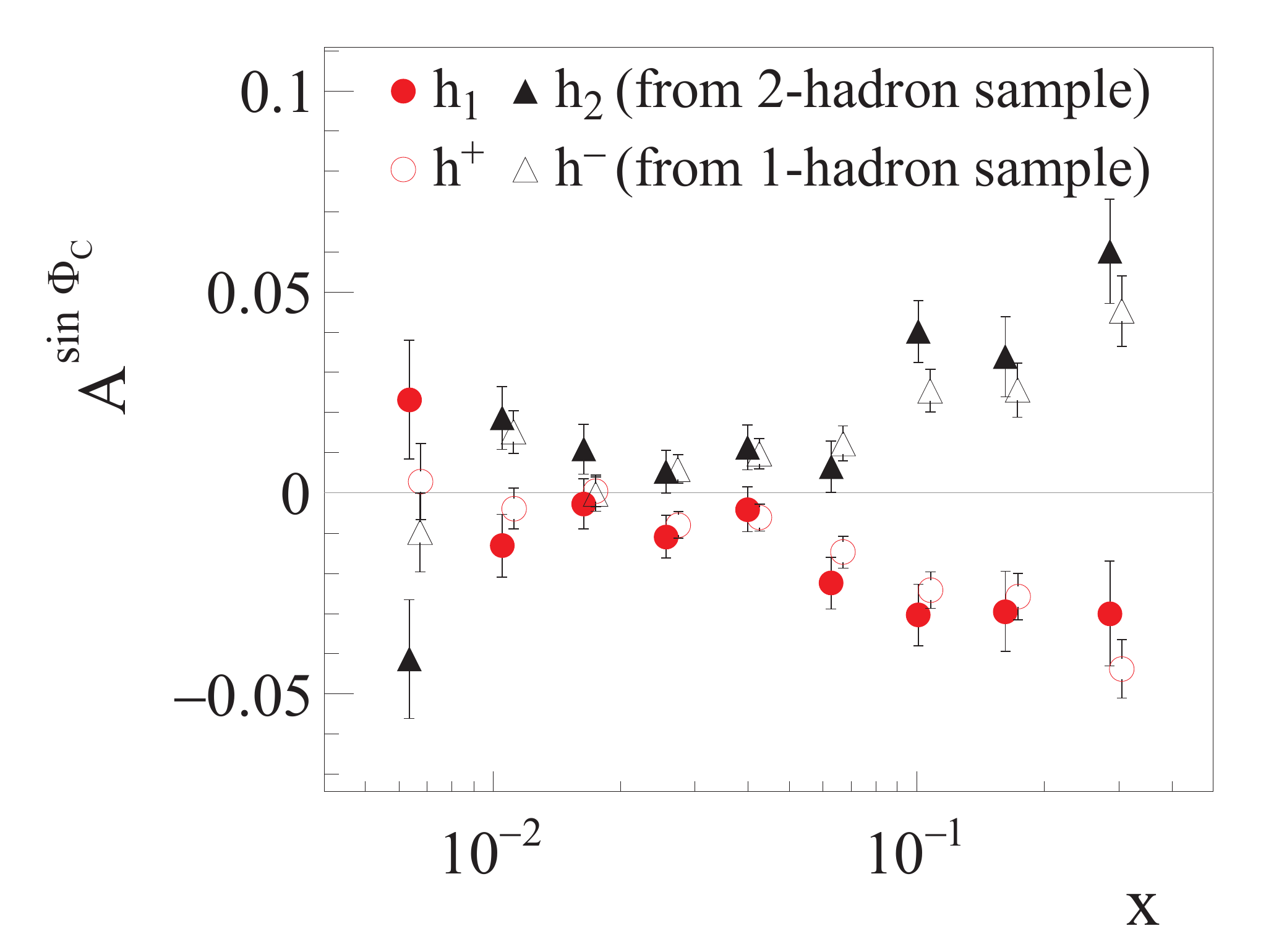}
\caption{(Color online) Comparison of the Collins asymmetries for single (full red circles) and di-hadrons (full black triangles)
 from COMPASS measured as function of x~\cite{Adolph:2015zwe}.}
\label{fig-dihadr-compass}
\end{figure}

\begin{figure}[ht!]
\includegraphics[width=0.5\textwidth]{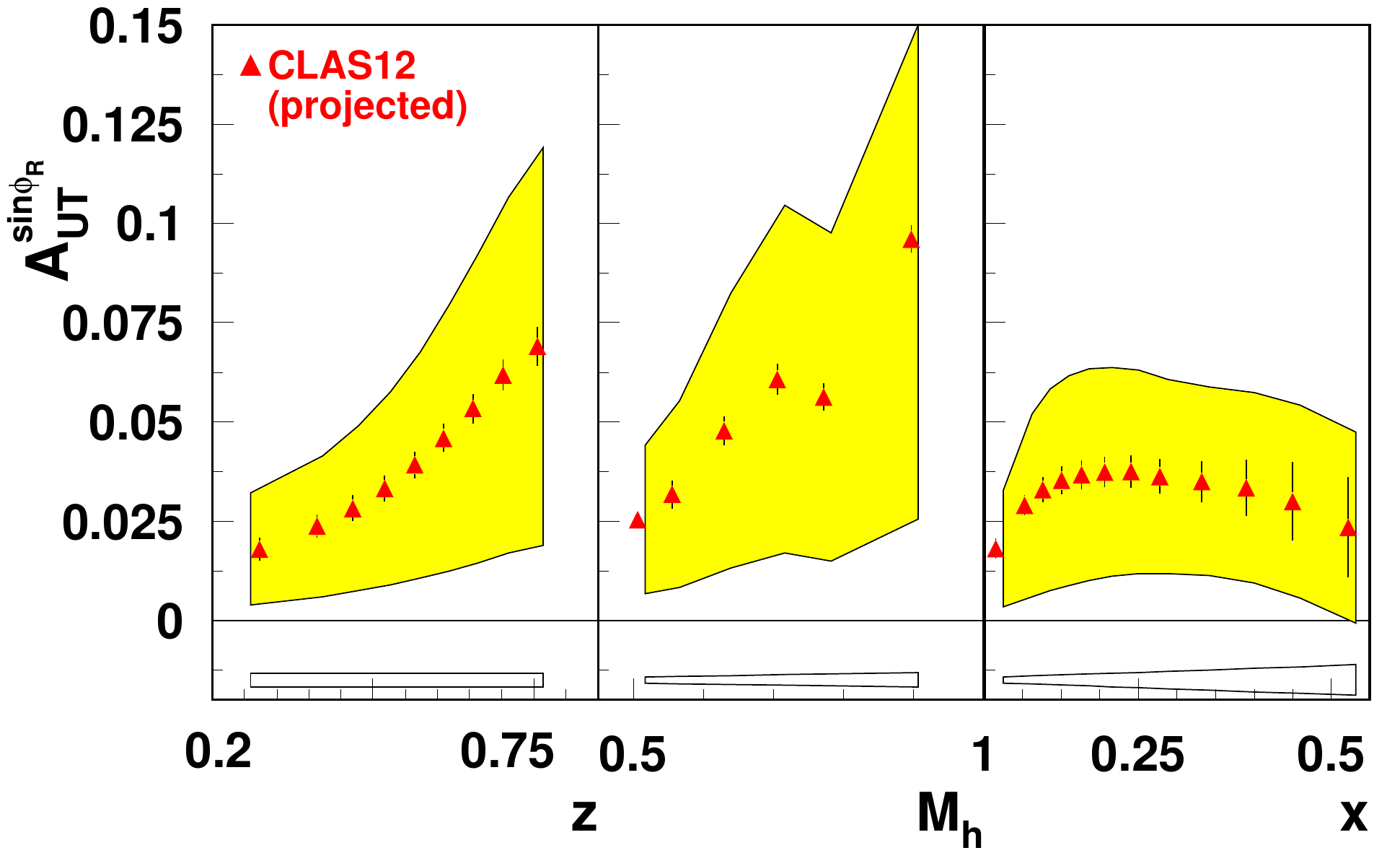}
\includegraphics[width=0.5\textwidth]{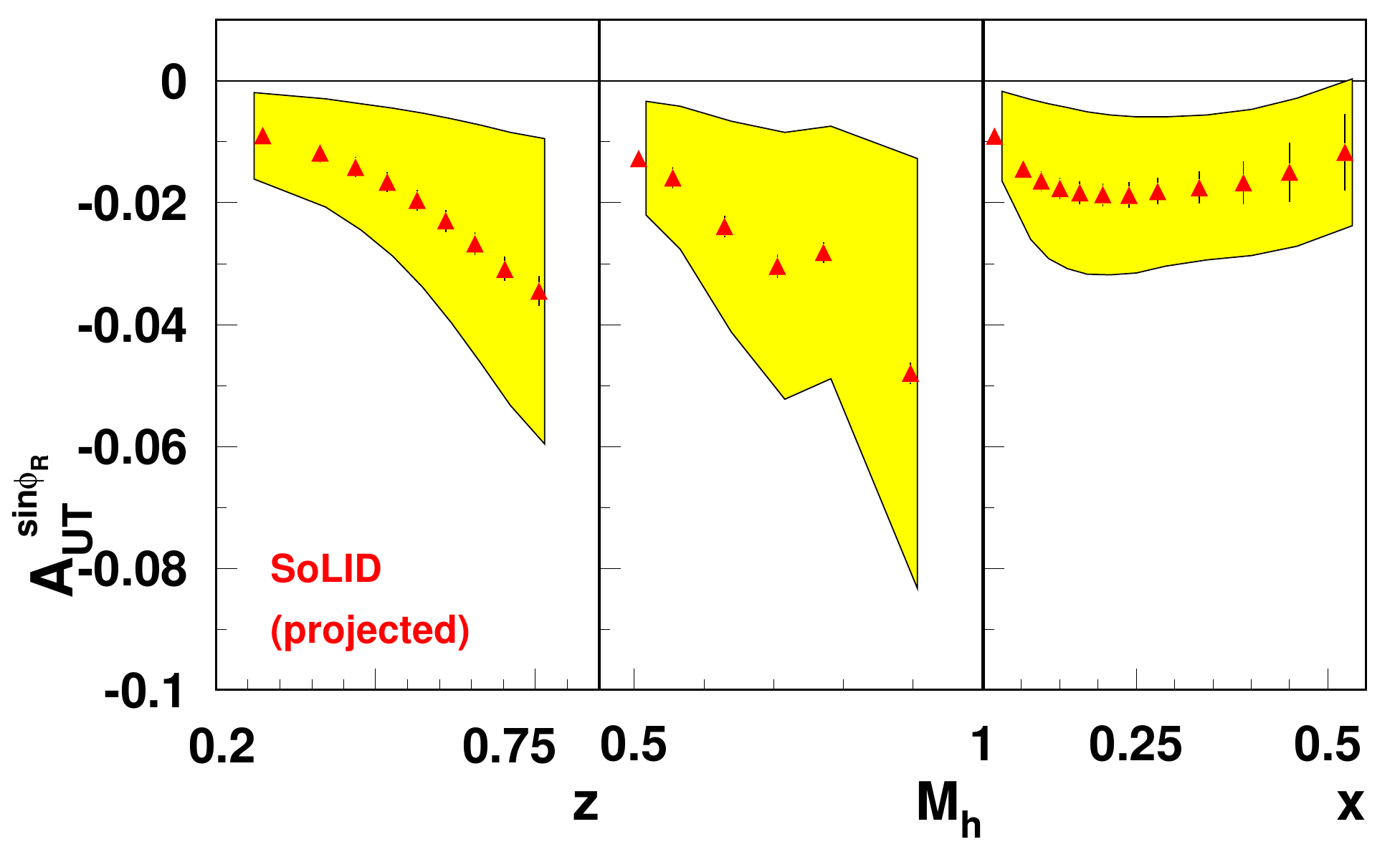}
\caption{The transverse spin-dependent azimuthal asymmetry, $A_{UT}^{\sin\phi_R}$ predicted in approved measurements by clas12 on proton\cite{C12-12-009} (left) and
SoLID on neutron\cite{E12-10-006A} (right).}
\label{fig-dihadr-all}
\end{figure}

Studies of integrated values of Collins asymmetries in the single-hadron sample were shown to be the
same as the Collins-like asymmetries of di-hadron sample which in turn are related with the integrated values
of di-hadron asymmetry (see Fig.\ref{fig-dihadr-compass}), suggesting that both, the single hadron and di-hadron transverse-spin
dependent fragmentation functions are driven by the same underlying mechanism~\cite{Adolph:2015zwe}.

Two experiments have been approved at JLab to study dihadrons with transversely polarized targets. Combination of hydrogen measurements
with CLAS12~\cite{C12-12-009} and $^3$He with SoLID~\cite{E12-10-006A} (see Fig.\ref{fig-dihadr-all}) would allow to perform complementary to TMD method
flavor decomposition of the transversity distribution.

\begin{figure}[ht!]
\includegraphics[width=0.35\textwidth]{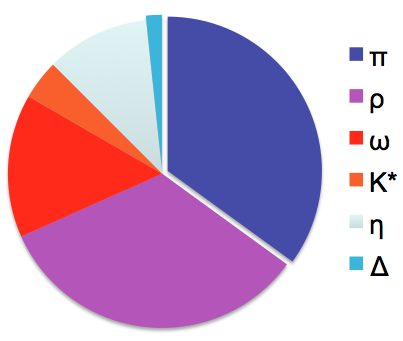}
\hspace{1cm}
\includegraphics[width=0.45\textwidth]{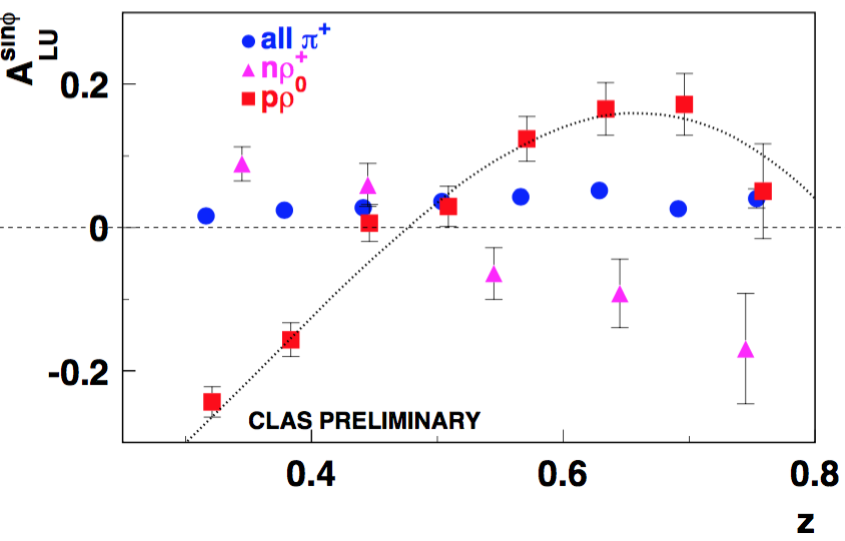}
\caption{Left panel: The fraction of pions in $ep\rightarrow e'\pi X$ in HERMES kinematics (27.5 GeV), produced directly from the string in PYTHIA MC (shown as $\pi$) and from different decay channels ($\rho,\omega, K^*,\eta$ and $\Delta$). Right panel: The $A_{LU}^{\sin\phi}$ asymmetry measured by CLAS on hydrogen target, for final state $\pi^+$ originating from exclusive $\pi^+\pi^-$ events dominated by $\rho^0$ (red squares),  from exclusive $\pi^+\pi^0$ events dominated by $\rho^+$ (magenta triangles) and integrated over all sources (blue circles)~\cite{Avakian:2004cw}.}
\label{fig-pi0contr}
\end{figure}

The interpretation of di-hadron production, as well as interpretation of single-hadron production is intimately related to contributions to those samples from vector mesons. The general procedure for SIDIS analysis, so far, was requiring
estimates of contribution of diffractive $\rho^0$, so theoretical studies can account for their contribution.
 Since the spin dependent fragmentation (Collins function)
of rho mesons is very different from pions, in fact predicted to have an opposite sign~\cite{Czyzewski:1996ih}, the final interpretation
of pion asymmetries will be very sensitive to relative fraction of pions coming from vector meson decays. The fraction of pions coming from correlated di-hadrons in general, and the rho decays, in particular, produced in the region of $\Phperp/z<Q$, contribute to the region $\Phperp/z>Q$, due to much smaller $z$ of decay pions, making the interpretation of data more complicated.
The PYTHIA based MC, suggests that the dominating fraction of pions are indeed
coming from vector meson decays (see Fig.\ref{fig-pi0contr}). Measurements of SSA performed by CLAS indicate that there is a
significant asymmetry in single-pions sample originating from rhos, which is also very different for different vector mesons.
The size of the asymmetry reaches $\sim$20\% and is opposite for $\pi^+$ originating from exclusive $\rho^+$ and $\rho^0$ decays~\cite{Avakian:2004cw}.

Significant single beam spin asymmetries were predicted in di-hadron production for several angular correlations.
The usual $sin(\phi_R)$ single beam spin asymmetry~\cite{Bacchetta:2003vn} involves subleading-twist functions in the so-called "collinear kinemaitcs" that involves  the measurement of only the relative transverse momenta of the hadron pair. Recently, a new leading-twist single beam spin asymmetry involving the helicity-dependent DiFF $G_1^\perp$ has been proposed~\cite{Matevosyan:2018dea}, in analogy with the target single spin asymmetry  in Refs.~\cite{Matevosyan:2017liq}, which requires the measurement of both total and relative transverse momenta of the pair. Yet another leading twist beam spin asymmetry involves the so-called fracture functions~\cite{Anselmino:2011bb}, where one of the  hadrons is produced in the current region by the fragmenting parton, while the second one is produced in the target region by the target remnant.
CLAS preliminary measurements~\cite{Pisano:2014ila} indicated a very significant non-zero beam-spin asymmetry $A_{LU}$
both on $^2$H and NH$_3$ targets.
Non zero single-spin asymmetries ($A_{LU}$) have also been observed in back-to-back pion and proton electroproduction~\cite{Avakian:2016zos},
indicating that spin-orbit correlations between hadrons
may be very significant. While this opens a new avenue for studies of the complex nucleon structure
in terms of quark and gluon degrees of freedom, it also suggests, that there are likely no uncorrelated hadrons in SIDIS.

\section{Precision measurements of TMDs}
\label{Sec:precision}
 The main mission of experimental exploration of 3D momentum and spatial structure of the nucleon by Jefferson Lab, and other facilities in the world,  is providing scientific community with unbiased data on observables that can be uniquely interpreted in theoretical and phenomenological models, i.e. cross-sections, multiplicities, asymmetries, structure functions. It is paramount that experimental results are not influenced by theoretical and/or phenomenological expectations and in this sense a blind analysis will be the most desirable one. Of course in order to achieve such a degree of unbiased measurements one will need to carefully study all pitfalls and limitations of data analysis itself. Experimental Halls of Jefferson Lab are likely to work together on issues related to interpretation and extraction of experimental data.

Although the interest in TMDs has grown enormously, we are still in need
of a consistent theoretical and phenomenological description spanning
the full kinematic regime covered by the (un)polarized world-data.
Some TMDs have been already phenomenologically extracted,
mainly from analyzing azimuthal distributions of single hadrons in SIDIS.
To obtain a full picture about the 3D momentum structure of the partons
in the nucleon from high to low $x$, it is important to connect the
theoretical approaches to extract TMDs including evolution.
The studies of 3D PDFs in general, and TMDs in particular, require a
lot more attention to uncertainties due to input parametrizations, than in 1D case, as
more degrees of freedom and bigger number of input parameters may generate
biased model uncertainties.
For example it was argued, based on studies using the mPYTHIA MC generator, that the current extractions
of the Sivers PDF might be significantly underestimated~\cite{Matevosyan:2015gwa}. Thus, improvements in estimating the true relative size of
the Sivers term in the cross section would be needed in future precision studies.

\begin{figure}[ht!]
\centering
\includegraphics[width=0.6\textwidth]{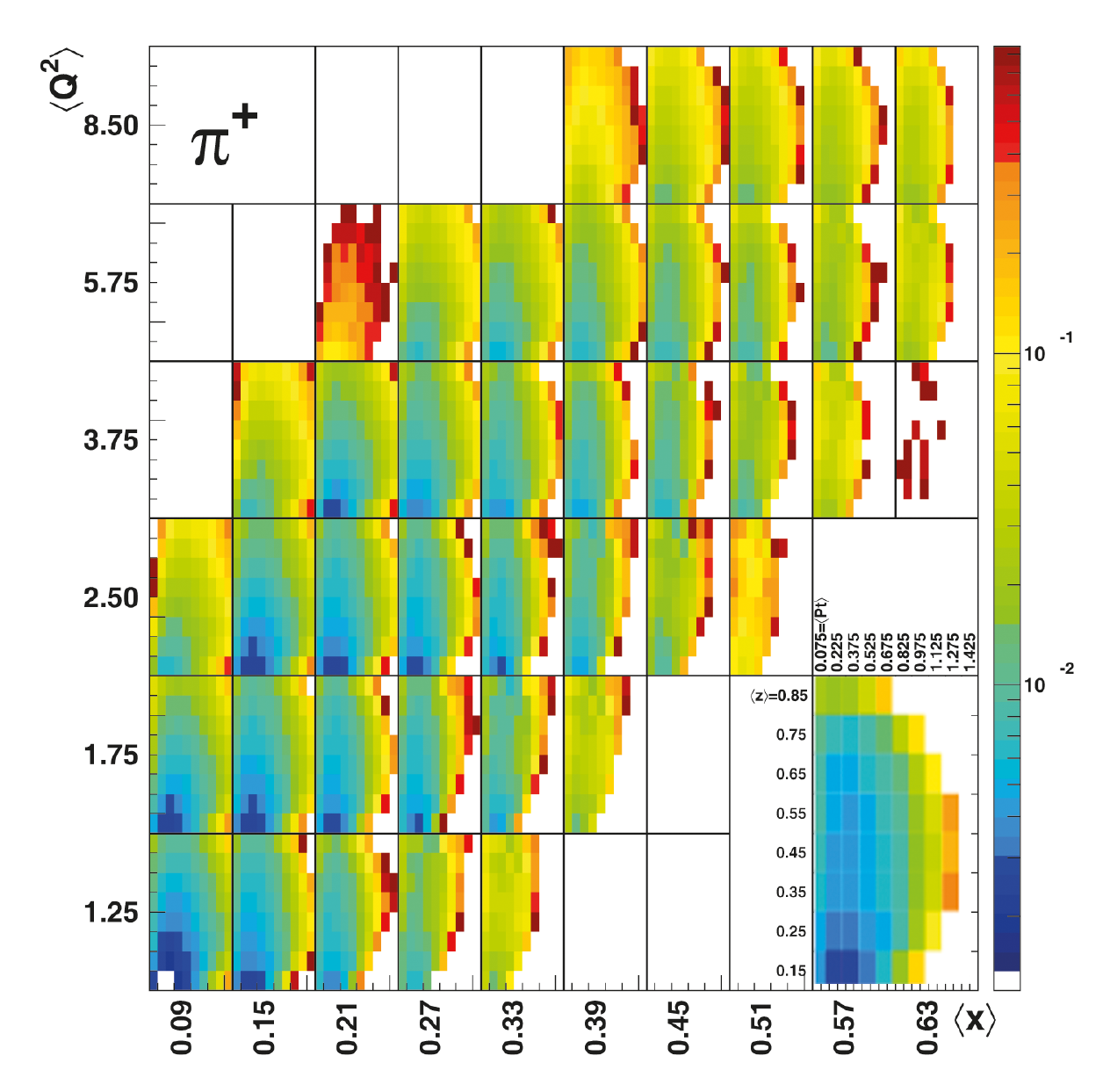}
\caption{The CLAS12 4D kinematical acceptance. The small insert shows an example of 2D bin in $z$ vs $\Phperp$ in each of  2D bins in $x$ and $Q^2$}
\label{clas12-4D-accept}
\end{figure}

The 3D partonic structure accessible in hard scatterings is rich and complex.
The understanding of the contributions to the final transverse-momentum
dependence of different azimuthal moments in the cross
section will require detailed studies.
Monte-Carlo event generators accounting for spin-orbit correlations
will be crucial to study the dependence on different model inputs,
as well as sensitivity of the extraction of underlying TMDs on various
experimental uncertainties including acceptances, resolutions and
radiative corrections. The acceptances of wide angle detectors are already themselves pretty complicated (see Fig.\ref{clas12-4D-accept}).

All that makes the development of a
framework for testing different extraction procedures a high priority
task for the community involved in studies of 3D PDFs.
Development of realistic Monte-Carlo generators accounting for TMD
evolution effects, spin-orbit and quark-gluon correlations will be a
great support for a multifaceted effort to study the fundamental 3D
structure of matter.

The future 3D nucleon structure extraction
framework, should include in addition to extraction procedures
a library of 3D PDFs~\cite{Hautmann:2014kza, Alekhin:2014irh,Hautmann:2014uua}
and MC simulation frameworks using that library
as input and allowing to check the quality of extracted PDFs for
specific experimental conditions in a full range of accessible kinematics.

\subsection{Radiative moments in unpolarized SIDIS}

The corrections due to radiative effects in SIDIS are on the interface of the theory and experiment, and require joint focused
activities on the both theoretical and experimental parts. While the theory effort addresses theoretical uncertainties of Radiative Corrections (RC) constrained by lack of knowledge of hadronic structure, the experimental approach is on the methodology of data analysis.
Currently there are no exact calculations of RC to SIDIS for arbitrary polarizations, with all available calculations representing various kinds of approximations. Model independent RC can be and must be accounted exactly in modern measurements in SIDIS within a new procedure of RC of experimental data using sets of semi-inclusive and exclusive  Structure Functions (SFs).
A self consistent approach is currently available only for the unpolarized target.
Possible (reasonable) approximations used in literature and data analysis practice include: i) soft photon approximation, ii) leading log approximation, iii) peaking approximation, iv) Compton-peak approximation, and v) using different Monte Carlo generators (e.g., Radgen). The soft photon approximation is very convenient because RC is factorized at the Born cross section and completely cancels in spin asymmetries, but is wrong because of hard photon emission that can not be neglected even in experiments with much poorer accuracy comparing to the SIDIS experiments in JLab. The leading log (and peaking) approximation could estimate RC from hard photon with semi-inclusive process but is not applicable for exclusive radiative tail. The Compton-peak approximation was good for elastic radiative tail in DIS, but estimates show that it will not work even for exclusive radiative tail. The most popular Monte-Carlo-based approximation uses the generator Radgen~\cite{Akushevich:1998ft} for generating radiative photons and evaluating RC based on simulated samples (e.g., with and without using Radgen). Unfortunately this approach is not applicable for analysis of azimuthal asymmetries in SIDIS, because key parts of the cross section responsible for $\cos(\phi_h)$ and $\cos(2\phi_h)$ do not appear in Radgen. Therefore, correct cross section of radiated photon cannot be reconstructed using Radgen.

\begin{figure}[ht!]
\centering
\includegraphics[width=0.48\textwidth]{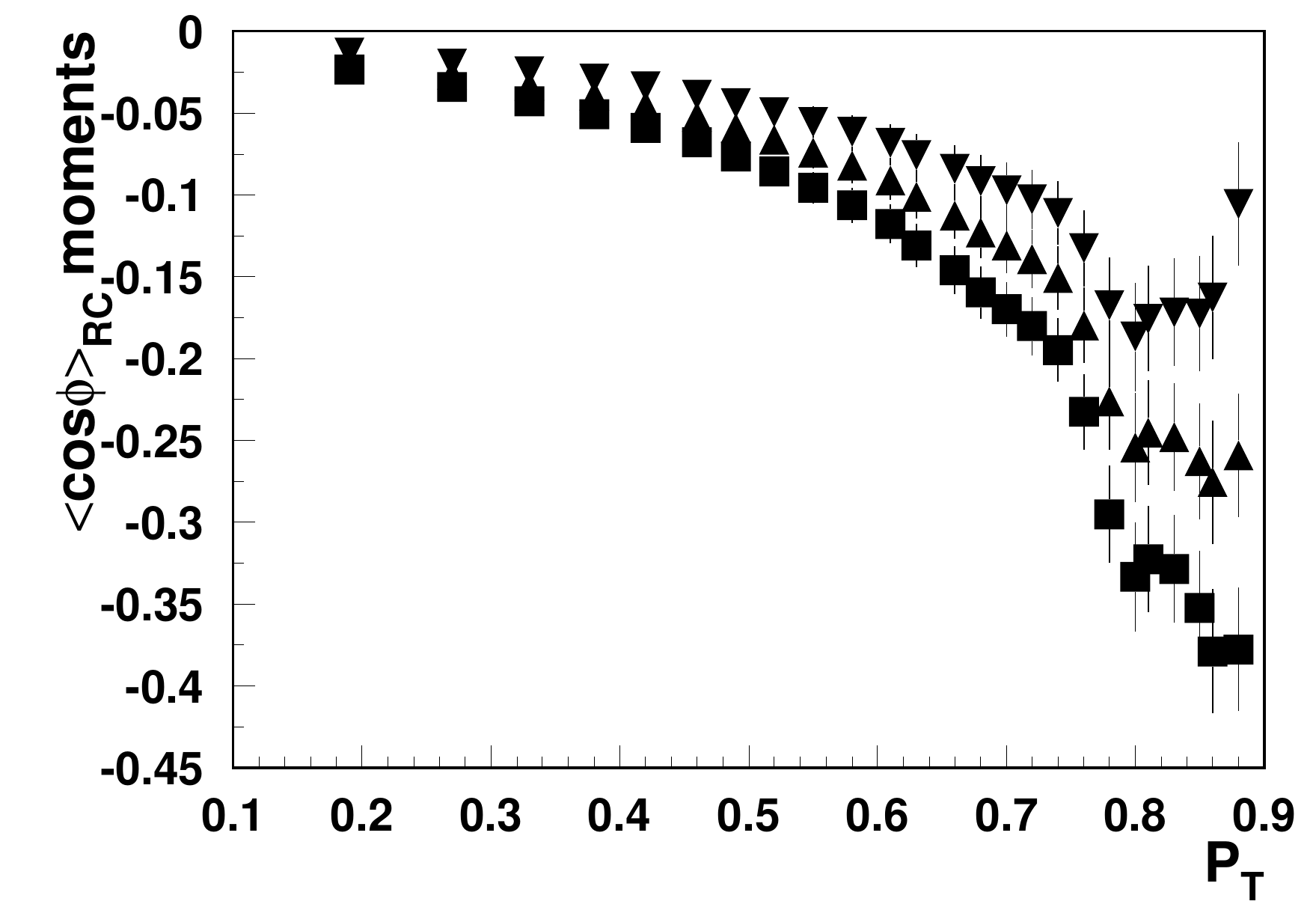}
\includegraphics[width=0.48\textwidth]{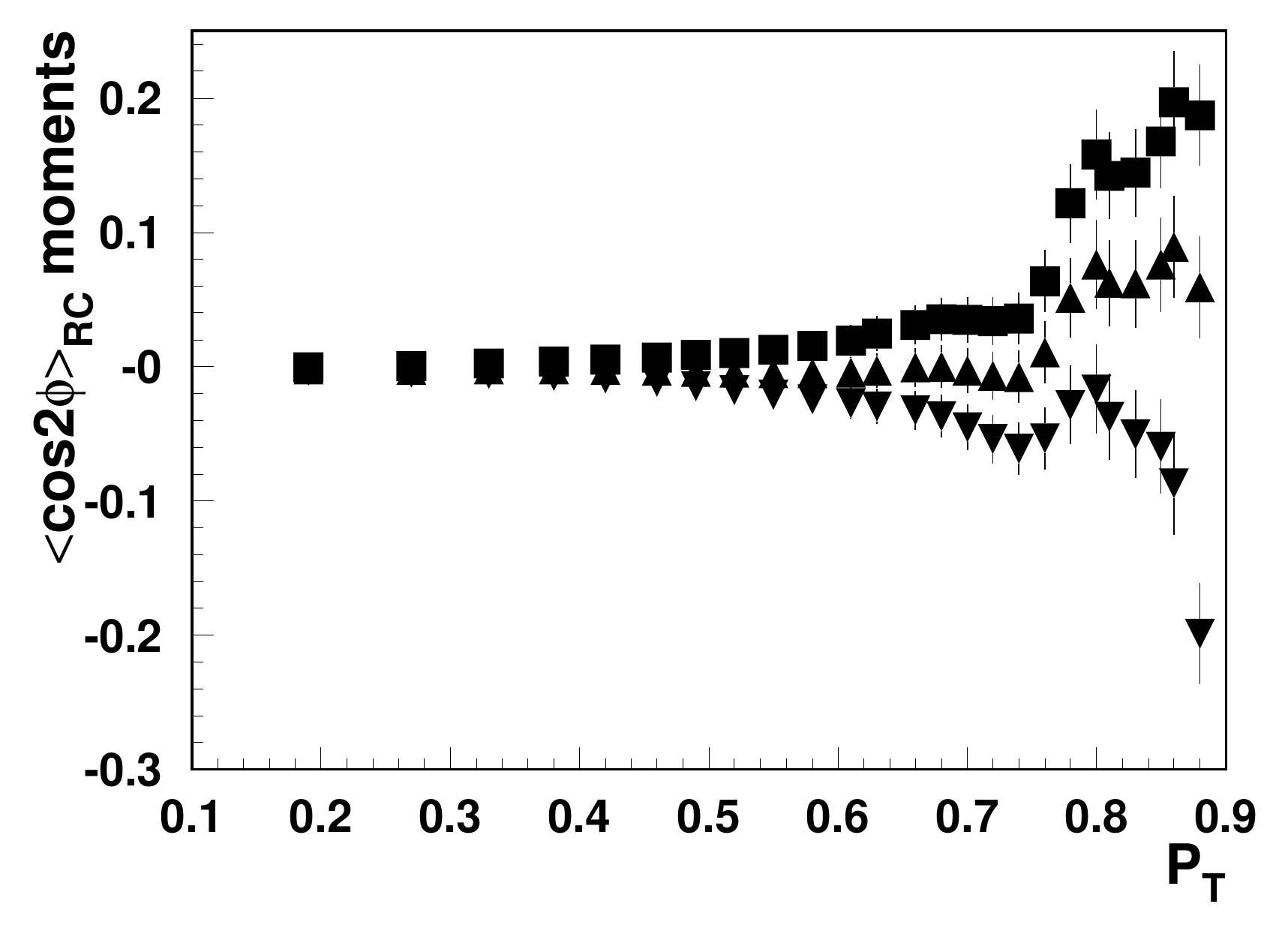}
\caption{Left panel:
Sensitivity of the $\cos\phi$-moment generated by radiative effects to $\phi$-dependence of three different structure
functions with $\cos\phi$ input amplitudes equal to -5\% (squares), -10\% (triangles up), and -15\% (triangles down) calculated from
haprad2.0~\cite{Akushevich:2007jc} as a function of the transverse momentum of a pion in SIDIS in JLab12 kinematics
at $x=0.3$ and $z=0.3$.
Right panel: The same for the $\cos 2\phi$-moments generated by radiative effects.
}
\label{fig-radcor}
\end{figure}

Preliminary analyses (based on HAPRAD 2.0 and SIRAD) are: i) $x$ and $Q^2$-dependences are similar to what we have in DIS; ii) RC goes down with increasing $z$, e.g., RC factor can change from 1.05 to 0.85 between $z$=0.2 and 0.8 for the same $x$ and $Q^2$; the $z$-dependence of RC is generated by decreasing the phase space of radiated photon with increasing $z$; iii) $\Phperp$-dependence is strong: RC can increase by a factor of 2 or more for very high $\Phperp$: both semi-inclusive and exclusive processes have large RC for large $\Phperp$; iv) RC to $\phi$-dependence can be large (RC generate new $\phi$-dependence and therefore new observables like $ \langle \cos(3\phi_h)\rangle$ that are exactly zero in SIDIS); and v) RC from exclusive radiative tail has its own dependence on kinematical variables and can give a high contribution especially as small missing mass of the $e^\prime\pi X$, $M_x^2$ (e.g., 0.95 and 1.4 without and with exclusive radiative tail for $M_x^2$=1.5 GeV$^2$ or 1.05 and 1.3 for $M_x^2$=3.0 GeV$^2$) and for high $\Phperp$.
Radiative correction in the polarized case are largely unknown. The effect in SIDIS may be significant with strongly dependence on the model for structure functions. The strong model dependence can be partly addressed within the RC iteration procedure of experimental data.
An illustration of possible effect of RC in unpolarized case shows that terms from SFs $F_{UU}^{\cos\phi}$  and  $F_{UU}^{\cos 2\phi}$ can significantly contribute to the base term,
so  higher harmonics generated by RC may be essential. For example, for kinematical point $E_{beam} = 10$ GeV , $Q^2$ = 2.5 GeV$^2$, x = 0.25, z= 0.2, and $\Phperp$= 0.7 GeV , the estimate of $\langle \cos3\phi_h\rangle$ is about 5\% , and corrections
to the calculated $\phi_h$-integrated cross section only due to presence of $\phi$-dependent SFs varies from ~3-30\% depending on different assumptions. Some results obtained from program HAPRAD~\cite{Akushevich:2007jc}, specifically developed to calculate the radiative corrections in SIDIS are presented in Fig.~\ref{fig-radcor}. Radiative moments can be large at large values of $\Phperp$ and, in addition some new moments can be generated due to radiative effects.
A self-consistent methodology of extraction of TMDs in SIDIS has yet to be developed. The strategy of RC can be developed by generalizing the RC procedure for DIS. The RC procedure of experimental data should involve an iteration procedure in which the fits of SFs of interest are re-estimated at each step of this iteration procedure. This procedure could be defined with and without involving Monte Carlo generator. Independently on whether MC is involved the procedure has to include the following steps: i) the fits of SFs are constructed to have the model in the region covered by the experiment; ii) constructing the models in the regions of softer processes, resonance region, and exclusive scattering using experimental data or theoretical models, iii) checking that the constructed models provide correct asymptotic behavior when we go to the kinematical bounds (Regge limit, QCD limit); iv) jointing all the models to have continuous function of all four variables in all kinematical regions necessary for RC calculation; v) implementing this scheme in a computer code and define the iteration procedure; vi) implementing the procedure of separation SFs in data and model each of them if several SFs are measured in an experiment; vii) constructing the models for other SFs if necessary (e.g., unpolarized SFs when spin asymmetries are measured); viii) paying specific attention to exclusive SFs, because the radiative tail from exclusive peak is important (or even dominate) in certain kinematical regions; and ix) paying specific attention to $\Phperp$ dependence, because RC is too sensitive for $\Phperp$ model choice.

\subsection{Possible effects due to experimental errors}

\begin{figure}[ht!]
\centering
\includegraphics[width=0.41\textwidth]{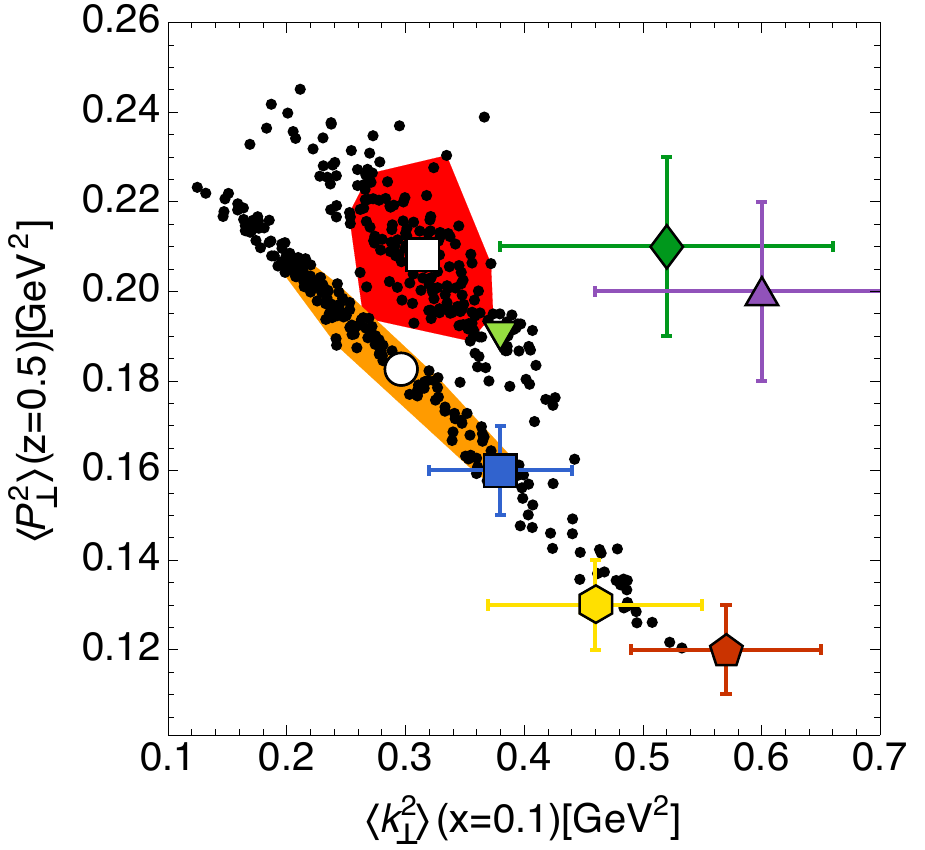}
\includegraphics[width=0.49\textwidth]{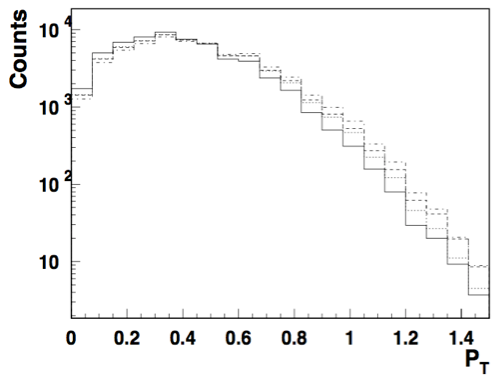}
\caption{Extraction of average $\Phperp$ and $k_T$ from experiments~\cite{Signori:2013mda}, and variations of the $\Phperp$-distributions for 4 different combinations of $\Phperp$ and $k_T$ ($\akperp/\apperp=0.2/0.20, 0.2/0.24, 0.4/0.16,0.4/0.2$ GeV$^2$) at CLAS12 kinematics for the
$z$-bin around 0.5.}
\label{fig-extr-sign}
\end{figure}

Traditionally SIDIS experiments in most cases were extracting different azimuthal moments
with assumption, supported by simplified test, that other moments in the cross section have a negligible effect on the moments of interest.
Studies of transverse momentum distributions using the SIDIS multiplicities~\cite{Signori:2013mda} indicate wide spread of the values for average transverse momentum of quarks due to variation of experimental errors withing few percent (Fig.~\ref{fig-extr-sign}).
Better separation of different options require wide range in $\Phperp$, in particular large $\Phperp$, where the experimental acceptances are becoming more complicated, radiative corrections large  and cross sections drop.
Unaccounted contributions from higher twists, target mass corrections, target fragmentation, modeling, and other factors, may introduce more significant variations of extracted TMDs, making the validation process crucial.

\subsection{Energy of the beam and phase space limitations}
One of the important items in interpretation of precision data expected from JLab are the
phase space limitations due to finite beam energies of real experiments
\cite{Boglione:2011wm}. The cosine modulation in particular is most sensitive to transverse momentum distributions leading to significant corrections due to
limitation of the phase space in experiments.


Our studies using MC simulation of Cahn and Boer-Mulders contributions, assuming Gaussian distributions from Ref.\cite{Boglione:2011wm}, indicate that phase space limitations may have very different effects on different contributions. A detailed simulations of effects of accounting the maximum possible transverse momentum of
quarks (assuming on-shell quarks) on the magnitude of the $\cos\phi_h$-moment from two competing contributions suggest that azimuthal moments from different sources can change significantly due to limited phase space, which
suppresses the Cahn contribution~\cite{Boglione:2011wm}, and  can lead to relative enhancement of the Boer-Mulders contribution.  If the underlying assumptions could be validated, that will indicate significant suppression of azimuthal asymmetries generated at distribution level.


\subsection{Extraction framework: assumptions and modeling}

Recent studies have shown that especially in the region of large $x$,
where little or no direct experimental information is available,
the uncertainty related to the choice of parametrization and methodology
may be as large or larger than the statistical uncertainty
\cite{Aaron:2009aa,Jimenez-Delgado:2014xza}.
Various assumptions involved in modern extractions of TMDs from available data
rely on conjectures, in particular, on of the transverse momentum dependence of distribution and fragmentation functions ~\cite{deFlorian:2007aj,Anselmino:2005nn,
Amrath:2005gv,Bacchetta:2007wc,PhysRevD.83.074003,Hirai:2007cx,
Anselmino:2008sga,Matevosyan:2012ga,Matevosyan:2011vj,Casey:2012hg,Signori:2013mda}
 making estimates
of systematic errors due to those assumptions
extremely challenging.

The main goal of the Extraction and VAlidation framework (EVA) is to assist extraction of 3D PDFs, by testing different extraction procedures and estimating
systematics related to different assumptions and models used in the extraction procedure~\cite{Avakian:2015vha}.
The input of the extraction and validation framework will include also multiplicities and asymmetries, but the preferred input will be the table with Elementary Bin Counts (EBC).
The EBC table will contain counts in the smallest size multidimensional bins limited only by detector resolution in all relevant kinematical variables involved in a given process.
For example the single hadron SIDIS will have bins in $x,y,z,\Phperp,\phi$ and exclusive production of photons or mesons, correspondingly $x,y,t,\phi$. The values of helicities for the incoming lepton, $\lambda$, and target nucleon, $\Lambda$, as well as other relevant information like beam energy, target type, etc., should also be in the table. The counts should be corrected for acceptance, background and
other experimental factors,  with corresponding statistical and systematic uncertainties. Optionally the radiative correction factor, extracted from unfolding using some specific
procedure could be specified.
The resolution size EBC bins would also allow application of alternative data analysis, typically based on event-by-event procedures (ex. Bessel-Weighting method~\cite{Aghasyan:2014zma}).

The EBC table will have info on boundaries of bins and average values of kinematical variables within the elementary bin.
The simplest implementation of the EBC file could be by using JavaScript Object Notation (JSON) file format used for serializing and transmitting structured data.
A table of values of normalized counts in elementary bins has been extracted in 5-dimensional bins in $x,Q^2,z,\Phperp$ and $\phi$~\cite{Avakian:2016pqj}.
Full 5-dimensional table with counts in  elementary bins with all relevant information
(helicities, beam energies, target,...) would allow rebinning, proper integrations over other variables, web browsing and graphical presentations.
While keeping it human readable, the data will be machine readable (will need API).
Reduction of the size of the bins will be only limited by detector resolution and
available MC statistics for acceptance extraction chain.
Small size of the bins would allow precision calculations of all relevant variables.
Much wider bins, used in experiments, when azimuthal moments were extracted in the process of experimental analysis of the
collected data, will always loose some relevant information, limiting the usage of data to specific tasks.
More complex processes (like 2 hadron production) will have more variables.

The main challenge in extraction of $\phi_h$ distributions of hadrons in SIDIS is the handling of the detector acceptance, both geometrical acceptance (the location of active detector elements) and the efficiency in the active regions.
For a given kinematical bin, the acceptance is defined as the ratio of the number of reconstructed events (using a GEANT based simulation of the given detector) to the number of generated events.

\begin{figure}[ht!]
\begin{center}
\includegraphics[width=0.8\textwidth]{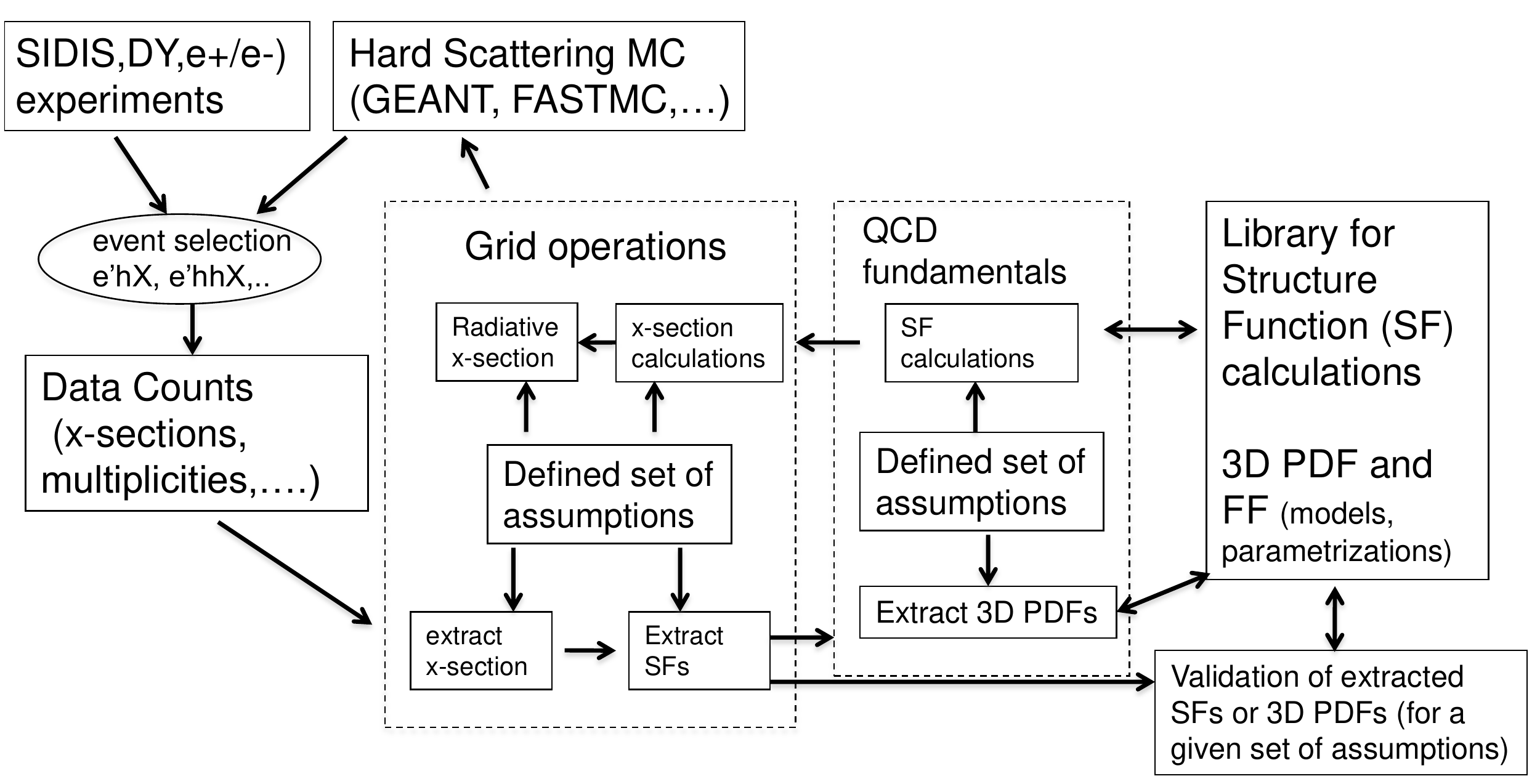}
\end{center}
\caption{The diagram for the extraction and validation framework (EVA).}
\label{fig-eva}
\end{figure}

Overall design of the extraction and validation framework with definitions of the services and data flow between them is shown in Fig.~\ref{fig-eva}.
The extraction part will take as input the EBC tables and extract SFs based on a given set of SFs from the existing library with radiative corrections applied using the same set. Extracted structure functions will be used to extract underlying TMD PDFs and FFs involved in calculation of corresponding SFs.
The validation process will start with calculation of relevant (for given observable) SFs using TMD PDFs and FFs from the library. Those SFs will be used to calculate the
Born cross section, which will be used to calculate the radiative cross section. The radiative cross section will be used to generate the process of
interest (ex. $ep\rightarrow e^\prime\pi^+X$). The generated events will go through the GEANT simulation and reconstruction chain of a given experiment, eventually providing the EBC table for a given process.
The TMD PDFs and FFs extracted from the EBC tables will be compared with input values to validate the process of extraction. The EVA chain would allow estimates of
systematic errors due to different assumptions (for instance extraction the Sivers TMD, ignoring the $\cos\phi_h$ modulation of unpolarized cross section or SF with $\sin\phi_S$, or ignoring the evolution of 3D PDFs in extraction, and so on).
All data flow between different blocks could go through JSON files.


Compared to PDFs, the status of TMD extractions is still in an early stage of development.
Phenomenological efforts have been summarized recently by introduction of a library
of fits and parametrizations for transverse-momentum-dependent parton distribution
functions (TMD PDFs) and fragmentation functions (TMD
FFs) together with an online plotting tool, TMDplotter~\cite{Hautmann:2014kza}.
Observables  constructed by taking ratios, such as asymmetries, are not ideal grounds for the study of TMD evolution effects. More effort should be made towards measuring properly normalized SIDIS and $e^+e^-$, and Drell-Yan cross sections (both unpolarized and polarized)

\section{Summary}
\label{Sec:summary}

In summary, spin and azimuthal moments, which may be related to quark-gluon correlations have been measured for all pion channels in a fully differential way and shown to be very significant and flavor dependent. They are indispensable part of SIDIS analysis,
and their understanding is important for interpretation of all kind of spin-azimuthal asymmetries.
Comparison of azimuthal moments measured at Jefferson Lab with HERMES and COMPASS measurements, supports the higher twist nature of the $\cos\phi_h$ and $\sin\phi_h$ moments.
Measurements of di-hadron single spin asymmetries indicate that final state hadrons tend to be correlated.
Several effects, including limited phase space for finite beam energies, higher twist contributions, radiative corrections and possible background from target fragmentation region should be considered as important elements for interpretation of systematic uncertainties in precision measurements in polarized SIDIS experiments, and in particular at Jefferson Lab. Understanding of the scale of contributions ($\sim M^2/Q^2$, $\sim \Phperp^2/Q^2$, Target/Current correlations, etc) will define the limits on precision needed to various aspects of TMDs, such as  evolution, higher twists, etc.
Sophisticated frameworks for calculation of 3D PDFs, such as TMDlib~\cite{Hautmann:2014kza} can be interfaced to extend the scope of available models, which can be used in validation of 3D PDFs extraction framework. The future Extraction and VAlidation framework will serve to help both experimental community and phenomenological community to test results and assure model independence of measured data.

In this review we have not covered hadron-hadron facilities such as RHIC~\cite{Aschenauer:2015eha}, and $e^+e^-$ facilities, such as BELLE. It is however, important to emphasize that hadron structure and TMDs in particular are to be studied in as many processes as possible. For instance, measurements in one process, such as SIDIS, cannot definitely answer a question on universality of TMDs. Only a comprehensive study of data coming from various experiments and various processes will allow to unravel the underlying parton landscape of the nucleon.
\section{Acknowledgements}
\label{Sec:Acknowledgements}

We thank  I. Akushevich,  M. Engelhardt, H. Matevosyan, B. Pasquini and P. Rossi for their help and discussions. This work was partially supported by the U.S.\
Department of Energy under Contract No.~DE-AC05-06OR23177 and by the National Science Foundation under Contract No. PHY-1623454, and Department of Energy under Contract within the framework of the TMD Topical
Collaboration.

\bibliographystyle{varenna}
\bibliography{3dstructure}
\end{document}